\documentclass[11pt,a4paper]{article}
\usepackage{jheppub}
\usepackage{amsmath,multirow}
\usepackage[usenames,dvipsnames,table]{xcolor}
\usepackage{comment,mathrsfs}
\usepackage{graphicx}
\usepackage{soul}
\usepackage{blindtext}
\usepackage{slashed}
\usepackage{subcaption}

\usepackage{tikz}

\title{Hydrodynamics, spin currents and torsion}

\author[a]{A.D. Gallegos,}
\author[a]{U. G\"ursoy,}
\author[b]{and A. Yarom}
\affiliation[a]{
Institute for Theoretical Physics, Utrecht University,
Leuvenlaan 4, 3584 CE Utrecht, The Netherlands}
\affiliation[b]{Department of Physics, Technion, Haifa 32000, Israel}

\abstract{We construct the canonical constitutive relations for a fluid description of a system with a spin current, valid in an arbitrary number of dimensions in the absence of parity breaking or time reversal breaking terms. Our study encompasses the hydrostatic partition function, the entropy current, Kubo formula, conformal invariance, and the effect of charge. At some stages of the computation we turn on a background torsion tensor which naturally couples to the spin current.}

\begin{document}

\maketitle

\section{Introduction}
\label{S:Introduction}

Hydrodynamics is a robust effective theory capable of describing a wide variety of phenomena, ranging from stellar evolution through fluid flow in a pipe to heavy ion collisions. The dynamics of fluid flow are captured by the conservation laws of the underlying many body theory at hand, supplemented by constitutive relations which parameterize the conserved currents as local functions of the dynamical variables. For example, energy can be parameterized by a local temperature field and energy conservation serves as a dynamical equation for it. 

In a relativistic theory, on a flat torsionless background, translation invariance implies that the energy momentum tensor, $T^{\mu\nu}$, is conserved. In relativistic hydrodynamics the constitutive relations for the energy momentum tensor are expressed in terms of a temperature field $T$ and a velocity field $u^{\mu}$ normalized such that $u^{\mu}u_{\mu}=-1$. 
As in any effective theory, the constitutive relations are the most general ones possible, compatible with the physical constraints of the problem. In hydrodynamics these include symmetries, unitarity, the second law of hydrodynamics and various Onsager relations.
Once the constitutive relations are available (usually in terms of a derivative expansion), the equations of motion for $T$ and $u^{\mu}$ are given by energy momentum conservation and form the relativistic version of the Navier-Stokes equations. 

Over the last few years there has been increasing interest in a theory of hydrodynamics in the presence of an independently conserved angular momentum density, $J^{\mu\nu\rho}$, or, alternately, a non vanishing spin current, $S^{\mu\nu\rho}$;
\begin{equation} 
\label{E:JandS}
	J^{\mu\nu\rho} = x^{\nu}T^{\mu\rho} - x^{\rho}T^{\mu\nu} - S^{\mu\nu\rho}.
\end{equation}
In the context of heavy ion collisions there is experimental evidence for correlations between the spin polarization of $\Lambda$-hyperons and the angular momentum of the quark gluon plasma in off center collisions \cite{STAR:2017ckg,Adam:2018ivw}. Likewise, an experimental realization of spin currents in liquid metals has been demonstrated in \cite{Takahashi}. A proper theoretical understanding of these phenomenon necessitates a consistent hydrodynamic theory of spin currents. While there has been a significant amount of work on the matter, see, e.g., \cite{Voloshin:2004ha,Abreu:2007kv,Becattini:2007sr,Betz:2007kg,Becattini:2013fla,Florkowski:2017dyn,Florkowski:2017ruc,Becattini:2018duy,Becattini:2020ngo,Becattini:2020sww,Becattini:2020riu,Singh:2020rht,Manes:2020zdd,Bhadury:2021oat,Florkowski:2021pkp,Floerchinger:2021uyo,Singh:2021man,Das:2021aar,Weickgenannt:2021cuo,Becattini:2021iol,Speranza:2021bxf,Goncalves:2021ziy,She:2021lhe,Yi:2021ryh,Wang:2021ngp,Hongo:2021ona,Lin:2021mvw,Chen:2021azy,Buzzegoli:2021wlg,Liu:2021nyg,Valle:2021nfv,Dong:2021fxn,Sonin:2010,PhysRevLett.87.187202,PhysRevLett.114.157203,PhysRevLett.112.227201,PhysRevB.90.094408,Hongo:2022izs}, 
a fully consistent theory of hydrodynamics of spin currents is lacking. This work aims at filling in this gap, expanding on the recent work of \cite{Gallegos:2021bzp}.

The first issue that needs to be addressed when constructing a theory of hydrodynamics with spin currents is the identification of a canonical spin current. In a relativistic theory, on a flat torsionless background, Lorentz invariance dictates that energy and momentum are conserved and, as a result, that angular momentum is conserved. Indeed, it is well known  (see \cite{BELINFANTE1940449,rosenfeld1940tenseur}) that in the absence of torsion, it is always possible to add an improvement term to the energy momentum tensor, $T^{\mu\nu}$, such that the stress tensor satisfies $T^{\mu\nu}=T^{\nu\mu}$, as an additional equation of motion. Then, angular momementum conservation  follows from energy-momentum conservation. Put differently, the spin current $S^{\mu\nu\rho}$ is ambiguous due to the possibility of adding improvement terms to the stress tensor, and, in particular it can be set to zero by judiciously choosing these terms.

One way to deal with this ambiguity is to couple the theory to an external spin connection, $\omega_{\mu}{}^{ab}$, which is independent of the vielbein $e^{a}{}_{\mu}$ due to non vanishing torsion. In this case, one can define a canonical stress tensor and spin current via
\begin{equation}
\label{E:variation}
	\delta S = \int d^dx |e| \left(T^{\mu}{}_{a} \delta e^{a}{}_{\mu} + \frac{1}{2} S^{\mu}{}_{ab} \omega_{\mu}{}^{ab} + E \cdot \delta \phi\right)
\end{equation}
where $|e|$ is the determinant of the vielbein, $E$ denotes the equations of motion, and $\delta\phi$ the variation of the dynamical fields associated with the action describing the constituents of the fluid. Here and in the remainder of this paper greek indices will denote spacetime indices and roman indices tangent bundle indices. As we discuss in depth in section \ref{S:Canonicalspincurrent}, the presence of torsion precludes an improvement term, leading to a unique spin current. Of course, after obtaining the stress tensor and spin current via \eqref{E:variation}, one can set the background vielbein to the Minkowski one and the spin connection to the metric compatible, torsionless, one. 

Since the stress tensor and spin current are well defined in the presence of torsion, one can generate a set of constitutive relations for them by introducing a hydrodynamic spin chemical potential $\mu^{ab}$ in addition to the temperature and velocity field. We compute these constitutive relations in two steps. In the first step, discussed in section \ref{S:hydrostatic}, we use the hydrostatic partition function introduced in \cite{Jensen:2012jh,Banerjee:2012iz} to construct the constitutive relations associated with hydrostatic equilibrium configurations. Then, in section \ref{S:nonhydrostatic} we add dissipative corrections to the constitutive relations and check compatibility of the resulting constitutive relations with the entropy current and Onsager relations. The constitutive relations obtained this way are the most general ones possible in a generic number of spacetime dimensions. For ease of reference we have collected our results in appendix \ref{A:fullconstitutive}. A full discussion of the constitutive relations which explicitly break parity or time reversal via the presence of the Levi-Civitta tensor is left for future work.

After obtaining the full set of constitutive relations for hydrodynamics with a spin current, we carry out a linearized analysis of the resulting equations of motion, allowing us to identify the various hydrodynamic modes of the problem and to express the appropriate coefficients in the constitutive relations in terms of Kubo relations. We do this in section \ref{S:linearized}. We then turn our attention to charged fluids in section \ref{S:charge} and to conformal invariant fluids in section \ref{S:conformal}. 

One of our main findings is that under the ``simplest'' possible assumptions, the spin chemical potential should be considered as a first order in derivative quantity. This is somewhat unusual since the temperature, velocity and chemical potential associated with a conserved charge are naturally evaluated as zeroth order in derivatives quantities. Not only is the spin chemical potential first order in derivatives, its equations of motion are algebraic, tying its value to the acceleration of the fluid, its vorticity and the background torsion. While somewhat puzzling at first, these features of the spin chemical potential are what allow for the recovery of standard hydrodynamics in the limit where the torsion tensor vanishes---with an appropriate choice of improvement term, the dynamics of the velocity field and temperature effectively decouple from that of the spin chemical potential. This feature of hydrodynamics with a spin current, and other effects associated with the spin current are discussed further in section \ref{S:discussion}.

\section{The canonical spin current}
\label{S:Canonicalspincurrent}

As discussed in the introduction, the first step in setting up a hydrodynamic theory of spin currents is to construct a well defined spin current. The challenge being the inherent ambiguity in the definition of the stress tensor constructed via Noether's procedure due to improvement terms. These improvement terms modify the structure of the angular momentum density 
and therefore the structure of the spin current, as defined in \eqref{E:JandS}. In what follows we will go over the origin of the ambiguity of the spin current and make two observations. The first is that the ambiguity in the spin current is removed once the theory is placed on a torsionfull background geometry. Thus, by coupling the theory to torsion one can obtain a well defined spin current. Our second observation is that the ambiguity in the spin current does not affect the combined dynamics of the stress tensor and spin current. Therefore, to obtain a well defined spin current we can couple our theory to a torsionfull background and take the torsionless limit to obtain physical results.   
While the material contained in this section is known, it is central to our construction, so we present it here for completeness, and to set up the notation for the rest of this work.

Given an action $S[\phi;\,\chi]$ with dynamical fields $\phi$ and couplings $\chi$, we define the variation
\begin{equation}
	\delta S [\phi;\,\chi] = \int  \left( C \cdot \delta \chi + E \cdot \delta \phi \right) d^dx
\end{equation}
such that $E$ denotes the equations of motion. If $\delta S=0$ and $\delta \chi=0$ for a particular $\delta \phi$, we say that $\delta \phi$ is an infinitesimal symmetry. If $\delta S = 0$ for a particular (non vanishing) $\delta \chi$ and $\delta \phi$ we say that $\delta \chi$ and $\delta \phi$ constitute a spurionic symmetry. If we parameterize an infinitesimal symmetry via $\delta \phi = \lambda \delta_{\lambda}\phi$ with $\lambda$ a constant then for a spacetime dependent $\lambda$ we neccessarily have
\begin{equation}
\label{E:defJ2}
	\delta S[\phi;\,\chi] = - \int J^{\mu} \nabla_{\mu} \lambda d^dx = \int E \cdot \lambda \delta_{\lambda} \phi \, d^dx
\end{equation}
which gives Noether's theorem, $\nabla_{\mu}J^{\mu} = 0$, under the equations of motion.
	
The transformation associated with $\delta \phi = \lambda(x) \delta_{\lambda}\phi$ is not a symmetry. But, if we manage to couple the action to a source $\chi_{\mu}$ such that a generic transformation of the action yields
\begin{equation}
	\delta S[\phi;\,\chi,\,\chi_{\mu}] = \int \left( J^{\mu} \delta \chi_{\mu} + C \cdot \delta \chi + E \cdot \delta \phi \right) \,d^dx
\end{equation}
then $\delta \phi = \lambda(x) \delta_{\lambda}\phi$, $\delta \chi = 0$ and $\delta \chi_{\mu} = \nabla_{\mu}\lambda$ will be a spurionic symmetry with the associated conservation law $\nabla_{\mu}J^{\mu}=0$. 
Conversely, suppose $\delta S =0$ for some spurionic symmetry $\delta \phi = F(\lambda(x))$, $\delta \chi = \chi(\lambda(x))$ and $\delta \chi_{\mu} = X_{\mu}(\lambda(x))$ where $F$, $X$ and $X_{\mu}$ are linear functions of $\lambda$ with (possibly) a finite number of derivatives acting on them. Then, $J^{\mu}$ defined in \eqref{E:defJ2} will satisfy a certain identity under the equations of motion. Often, this identity is referred to as a conservation equation even though, as should be clear from the above construction, it will result in $\nabla_{\mu}J^{\mu}=0$ only for $\delta \chi_{\mu} = \nabla_{\mu}\lambda$ and $\delta \chi = 0$. In what follows we will refer to identities associated with the above spurionic symmetry as conservation laws if they lead to a conservation equation of the form $\nabla_{\mu}J^{\mu}=0$ and to non conservation laws otherwise. Our entire discussion can be mapped to a quantum field theory (modulo a treatment of anomalies) where the non conservation laws are referred to as Ward identities and hold inside correlation functions as long as there is no overlap between $J^{\mu}$ and the  inserted operators.

Before applying this somewhat abstract construction to the vielbein and spin connection, let us first discuss it in a more familiar setting. Consider a set of dynamical fields $\phi$ coupled to an external metric $g_{\mu\nu}$ and an external gauge field, $A_{\mu}$ such that the action is coordinate reparameterization invariant and gauge invariant. In this case, the coordinate reparameterizations and gauge transformations are spurionic symmetries. A general variation of the fields and sources is given by
\begin{equation}
	\delta S[\phi;\,g_{\mu\nu},\,A_{\mu}] = \int \left(\frac{1}{2} T^{\mu\nu}\delta g_{\mu\nu} + J^{\mu}\delta A_{\mu} + E \cdot \phi \right) \sqrt{g} d^dx\,.
\end{equation}
Only the gauge field transforms under gauge transformations so $J^{\mu}$ will satisfy a conservation law, $\nabla_{\mu}J^{\mu}=0$. On the other hand coordinate reparameterizations will modify the metric and the gauge field leading to the non-conservation law
\begin{equation}
\label{E:Jouleheating}
	\nabla_{\mu}T^{\mu\nu} = F^{\nu\mu}J_{\mu}
\end{equation}
with $F^{\mu\nu}$ the field strength associated with the external gauge field $A_{\mu}$. 
Often, the right hand side of \eqref{E:Jouleheating} is referred to as a Joule heating term. It is interpreted as the work done on the system in the presence of an external field. We will see, shortly, that similar terms arise when placing a theory in a background with torsion.

Indeed, moving on to our main construction, consider a field theory with dynamical fields $\phi$, coupled to a vielbein $e^a{}_{\mu}$ and spin connection $\omega_{\mu}{}^{ab}$. The variation of the action associated with this theory is given in \eqref{E:variation}. Such an action can be made invariant under coordinate reparameterizations, $\xi^{\mu}$, and local Lorentz transformations, $\theta^a{}_{b}$, as long as the dynamical fields transform as appropriate tensors, e.g.,
\begin{equation}
	\delta Q^a{}_{\mu} = \pounds_{\xi} Q^a{}_{\mu} - \theta^a{}_{b}Q^b{}_{\mu}
\end{equation}
for some tensor $Q^a{}_{\mu}$,
and the vielbein and spin connection transform as
\begin{align}
\begin{split}
\label{E:deltaew}
	\delta e^{a}{}_{\mu} =& \pounds_{\xi} e^{a}{}_{\mu} - \theta^a{}_{b} e^{b}{}_{\mu} \\
	\delta \omega_{\mu}{}^{ab} = &\pounds_{\xi} \omega_{\mu}{}^{ab} + \nabla_{\mu} \theta^{ab}\,. 
\end{split}
\end{align}
Here $\pounds_{\xi}$ is a Lie derivative along $\xi$, and $\nabla_{\mu}$ denotes a covariant derivative,
\begin{equation}
	\nabla_{\mu}Q^a{}_{\nu} = \partial_{\mu}Q^{a}{}_{\nu} + \omega_{\mu}{}^{a}{}_{c} Q^{c}{}_{\nu} - \Gamma^{\alpha}{}_{\mu\nu} Q^a{}_{\alpha}\,. 
\end{equation}

If $\omega_{\mu}{}^{ab}$ and $e^{a}{}_{\mu}$ are independent fields, the transformations \eqref{E:deltaew} will lead to two non conservation laws: one for $T^{\mu}{}_{a}$ and one $S^{\mu}{}_{ab}$. Conversely,  if $\omega_{\mu}{}^{ab}$ is determined from $e^{a}{}_{\mu}$ then the only current is the stress tensor and \eqref{E:deltaew} will lead to two non conservation laws for it. The improvement term which allows for an ambiguity in the definition of the spin current can be traced to the latter and its absence to the former. Before showing this explicitly, let us introduce some notation. We define the torsion free spin connection $\mathring{\omega}{}_{\mu}{}^{ab}$ via
\begin{subequations}
\label{E:defring}
\begin{align}
	\mathring{\omega}{}_{\mu}{}^{ab} &= e_{\nu}{}^{a} \left(\partial_{\mu} e^{\nu b} + \mathring{\Gamma}^{\nu}{}_{\sigma\mu}e^{\sigma b} \right) \\
\intertext{where}
	\mathring{\Gamma}^{\alpha}{}_{\beta\gamma} &= \frac{1}{2} g^{\alpha\delta}\left(\partial_{\beta}g_{\gamma\delta} + \partial_{\gamma}g_{\beta\delta} - \partial_{\delta} g_{\beta\gamma} \right)\,.
\end{align}
\end{subequations}
The covariant derivative associated with $\mathring{\omega}_{\mu}{}^{ab}$ or $\mathring{\Gamma}^{\alpha}{}_{\beta\gamma}$ will be denoted by $\mathring{\nabla}{}_{\mu}$. For instance,
\begin{equation}
\label{E:ringnabla}
	\mathring{\nabla}{}_{\mu}Q^a{}_{\nu} =  \partial_{\mu}V^{a}{}_{\nu} + \mathring{\omega}_{\mu}{}^{a}{}_{c} V^{c}{}_{\nu} - \mathring{\Gamma}^{\alpha}{}_{\mu\nu} V^a{}_{\alpha}\,. 
\end{equation}
Likewise, $\mathring{R}^{\alpha}{}_{\beta\gamma\delta}$ and other ringed curvature tensors are associated with ringed connections in contrast to their non-ringed counterparts, e.g., $R^{\alpha}{}_{\beta\gamma\delta}$.
The difference between the spin connection and the ringed spin connection is the contorsion tensor
\begin{equation}
\label{E:contorsion}
	\omega_{\mu}{}^{ab} = \mathring{\omega}{}_{\mu}{}^{ab} + K_{\mu}{}^{ab}\,.
\end{equation}
The contorsion tensor, $K_{\mu}{}^{ab}$, is related to the torsion tensor, $T^{\alpha}{}_{\beta\gamma}$, defined via
\begin{equation}
\label{E:torsion}
	\Gamma^{\alpha}{}_{\beta\gamma} - \Gamma^{\alpha}{}_{\gamma\beta} =  T^{\alpha}{}_{\beta\gamma}\,,
\end{equation}
through
\begin{equation}
	T^{\alpha}{}_{\mu\nu} = K_{\mu}{}^{\alpha}{}_{\nu} - K_{\nu}{}^{\alpha}{}_{\mu}\,.
\end{equation}

Going back to \eqref{E:deltaew}, suppose that $\omega_{\mu}{}^{ab} = \mathring{\omega}{}_{\mu}{}^{ab}$. In this case \eqref{E:deltaew} will reduce to 
\begin{equation}
\label{E:deltae}
	\delta S[\phi,\,e^{a}{}_{\mu},\,{\omega}_{\mu}{}^{ab}(e)]\Big|_{\omega_{\mu}{}^{ab} = \mathring{\omega}{}_{\mu}{}^{ab}} = \int \left(T^{\mu}_{\hbox{c}}{}_{a} \delta e^{a}{}_{\mu} + E \cdot \phi \right)|e|d^dx
\end{equation}
where 
\begin{equation}
\label{E:defTc}
	T_{\hbox{c}}^{\mu}{}_{a}(x) = T^{\mu}{}_{a}(x) + \frac{1}{|e|}\int \frac{1}{2} S^{\nu}{}_{cb}(y) \frac{\delta \mathring{\omega}_{\nu}{}^{cb}(y)}{\delta e^{a}{}_{\mu}(x)} |e(y)|d^dy\,.
\end{equation}
The conservation equations associated with the spurionic symmetries \eqref{E:deltaew} read
\begin{equation}
\label{E:Tceqns}
	\mathring{\nabla}_{\mu}T^{\mu\nu}_{\hbox{c}} = 0\,,
	\qquad
	T^{\mu\nu}_{\hbox{c}} - T^{\nu\mu}_{\hbox{c}} = 0\,.
\end{equation}
Note that in our current formulation, the absence of an antisymmetric component of the energy momentum tensor is a dynamical statement. That is, it is satisifed only if the equations of motion, $E=0$, are satisfied. 

However, one can construct an improved stress tensor,
\begin{equation}
\label{E:defTI}
	T^{\mu\nu}_{\hbox{\tiny{I}}} = T^{\mu\nu}_{\hbox{c}} - \frac{1}{2} \left(T^{\mu\nu}_{\hbox{c}} - T^{\nu\mu}_{\hbox{c}}\right)\,.
\end{equation}
Then \eqref{E:Tceqns} read
\begin{equation}
	\mathring{\nabla}_{\mu}T^{\mu\nu}_{\hbox{\tiny{I}}} = 0 \,,
	\qquad
	T^{\mu\nu}_{\hbox{c}} - T^{\nu\mu}_{\hbox{c}} = 0\,,
\end{equation}
under the equations of motion,
but now the improved energy momentum tensor is symmetric, $T^{[\mu\nu]}_{\tiny{\hbox{I}}}=0$, independently of the equations of motion.

So far, we have set $\omega_{\mu}{}^{ab} = \mathring{\omega}{}_{\mu}{}^{ab}$ and then taken the variation \eqref{E:variation} in order to get the stress tensor. We could have carried out the same procedure in the reverse order, first take the variation \eqref{E:variation} and then set $\omega_{\mu}{}^{ab} = \mathring{\omega}{}_{\mu}{}^{ab}$. In this case we would have obtained the non conservation equations
\begin{equation}
\label{E:Teqns}
	\mathring{\nabla}{}_{\mu} T^{\mu\nu} = \frac{1}{2}\mathring{R}^{\nu}{}_{\alpha\beta\gamma}S^{\alpha\beta\gamma}\,,
	\qquad
	T^{\mu\nu}-T^{\nu\mu} = \mathring{\nabla}_{\alpha}S^{\alpha\mu\nu}\,.
\end{equation}
Note that the first equality in \eqref{E:Teqns} can be written in the equivalent form,
\begin{equation}
	\mathring{\nabla}{}_{\mu} T^{\mu\nu} = \frac{1}{2} \mathring{\nabla}_{\mu} \mathring{\nabla}_{\alpha} \left(S^{\alpha\mu\nu} - S^{\mu\alpha\nu} - S^{\nu\alpha\mu} \right)\,.
\end{equation}
The second equality in \eqref{E:Teqns} is equivalent to angular momentum conservation, c.f., \eqref{E:JandS}.

Using \eqref{E:defring} and \eqref{E:defTc} it is straightforward to show that $S^{\alpha\mu\nu}$ is related to $T_{\hbox{c}}^a{}_{\mu}$ via
\begin{equation}
\label{E:TcT}
	T_{\hbox{c}}^{\mu\nu} = T^{\mu\nu} - \frac{1}{2}\mathring{\nabla}_{\alpha} \left( S^{\alpha\mu\nu} - S^{\mu\alpha\nu} - S^{\nu\alpha\mu}\right) \,.
\end{equation}
Thus, equations \eqref{E:Tceqns} and \eqref{E:Teqns} are, obviously, equivalent, and no information is lost by solving one or the other. In fact, since $\mathring{\omega}{}_{\mu}{}^{ab}$ is a function of $e^a{}_{\mu}$ the dependence of the action, $S$, on the spin connection $\mathring{\omega}{}_{\mu}{}^{ab}$ is ambiguous: we may always shuffle a dependence of the action on $\mathring{\omega}{}_{\mu}{}^{ab}$ into a dependence on $e^a{}_{\mu}$ and its derivatives. Thus, in general, we find
\begin{equation}
\label{E:Tpeqns}
	\mathring{\nabla}_{\mu} T^{\prime \mu\nu} =  \frac{1}{2}\mathring{R}^{\nu}{}_{\alpha\beta\gamma}S^{\prime\,\alpha\beta\gamma}\,,
	\qquad
	T^{\prime \mu\nu} - T^{\prime \nu\mu} = \mathring{\nabla}_{\alpha} S^{\prime\,\alpha\mu\nu} \,,
\end{equation}
where
\begin{subequations}
\label{E:BRterms}
\begin{equation}
\label{E:BRTmn}
	T^{\prime\,\mu\nu} = T^{\mu\nu} + \frac{1}{2}\mathring{\nabla}_{\alpha} \left( B^{\alpha\mu\nu} - B^{\mu\alpha\nu} - B^{\nu\alpha\mu}\right) \,.
\end{equation}
with
\begin{equation}
\label{E:BRSabc}
	B^{\alpha\mu\nu} = S^{\prime\,\alpha\mu\nu} - S^{\alpha\mu\nu}\,.
\end{equation}
\end{subequations}
Equations \eqref{E:Tceqns} and \eqref{E:Teqns} are special cases of \eqref{E:Tpeqns}.

Let us emphasize once again that given an action, $S$, equations \eqref{E:Tceqns}, \eqref{E:Teqns} and \eqref{E:Tpeqns} are all equivalent and will take the same functional form. The difference between $T^{\mu\nu}$ and $T^{\prime\,\mu\nu}$ exhibited in equation \eqref{E:BRTmn}, is often referred to as as a Belinfante-Rosenfeld improvement term, and amounts to exchanging derivatives of the vielbein in the action between the spin connection and the stress tensor. In equations, it amounts to a decomposition of the stress tensor $T_{\hbox{c}}^{\mu\nu}$ of the form
\begin{equation}
		T_{\hbox{c}}^{\mu\nu}  = T^{\prime\,\mu\nu} - \frac{1}{2}\mathring{\nabla}_{\alpha} \left( S^{\prime\,\alpha\mu\nu} - S^{\prime\,\mu\alpha\nu} - S^{\prime\,\nu\alpha\mu}\right) \,.
\end{equation}
Colloquially, due to the similarity between \eqref{E:Tpeqns} and \eqref{E:Teqns}, one oftentimes refers to the decomposition in \eqref{E:TcT} 
as an ambiguity in the spin current. The construction presented in this section makes it clear that the modification of the spin current a' la \eqref{E:BRSabc} is compensated by a modification of the energy momentum tensor given by \eqref{E:BRTmn} so that the equations of motion are unchanged.
 
As we've seen in \eqref{E:defTI}, in addition to the Belinfante-Rosenfeld improvement term one can add to the stress tensor additional improvement terms to obtain a manifestly symmetric stress tensor, $T^{\mu\nu}_{\hbox{\tiny{I}}}$. As mentioned in the introduction, and as we will see later, in the absence of torsion the improved stress tensor $T^{\mu\nu}_{\hbox{\tiny{I}}}$ is indistinguishable from the standard stress tensor of hydrodynamic theory, in accordance with general expectations.
Nevertheless, the current $S^{\lambda\mu\nu}$ still exists and still satisfies the non-conservation equation \eqref{E:Teqns}. In a hydrodynamic theory, where there is no access to the equations of motion of the fundamental fields and only conservation equations play a role in determining the dynamics, we will use \eqref{E:Teqns} to determine the dynamics of the spin chemical potential.
\footnote{As is perhaps expected, and as we discuss in section \ref{S:discussion}, we will find that the resulting equations for the 
spin chemical potential are independent of the choice of improvement terms and, in addition, the dynamics of the temeprature and velocity field, as determined by the improved stress tensor $T^{\mu\nu}_{\hbox{\tiny{I}}}$, will be identical to those of a standard theory of hydrodynamics in the absence of the spin conservation equations and torsion.}

Let us now turn our attention to the somewhat simpler situation where $\omega_{\mu}{}^{ab}$ and $e^{a}{}_{\mu}$ are independent parameters. Coordinate invariance and local Lorentz invariance imply the non-conservation equations
\begin{align}
\begin{split}
\label{E:EOMs}
	\mathring{\nabla}_{\mu}T^{\mu\nu} & = \frac{1}{2} R^{\rho\sigma\nu\lambda}S_{\rho\lambda\sigma} - T_{\rho\sigma}K^{\nu a b}e^{\rho}{}_{a} e^{\sigma}{}_{b} \\
	&= \frac{1}{2} \mathring{\nabla}_{\mu} \left(  \mathring{\nabla}_{\lambda} \left( S^{\lambda\mu\nu}- S^{\mu\lambda\nu} - S^{\nu\lambda\mu}  \right) - S^{\mu}{}_{\rho\sigma} K^{\nu\rho\sigma} \right) + \frac{1}{2} S_{\lambda\rho\sigma}\mathring{\nabla}^{\nu} K^{\lambda\rho\sigma} \,,\\
	\mathring{\nabla}_{\lambda}S^{\lambda}{}_{\mu\nu} &= 2 T_{[\mu\nu]} + 2 S^{\lambda}{}_{\rho[\mu}e_{\nu]}{}^{a} e_{\rho}{}^{b} K_{\lambda}{}_{ ab}\,.
\end{split}
\end{align}
Since, in the presence of torsion, $\omega_{\mu}{}^{ab}$ and $e^{a}{}_{\mu}$ are independent sources the freedom leading to \eqref{E:BRterms} is absent and the stress tensor and spin current can not be transmuted into one another.  In the next section we will use this feature of torsionful backgrounds to compute the spin current and stress tensor in hydrostatic equilibrium. In the torsionless limit equations \eqref{E:EOMs} reduce to \eqref{E:Teqns} (or equivalently, \eqref{E:Tceqns} or \eqref{E:Tpeqns}) as expected.

\section{Hydrostatics}
\label{S:hydrostatic}
A fluid which is acted on by time independent external forces will settle down to a hydrostatically equilibrated configuration. Since a hydrostatically equilibrated configuration must be a solution to the hydrodynamic equations of motion, the explicit form of the hydrostatic configuration must be compatible with the constitutive relations of the fluid. Put differently, the existence of hydrostatic equilibrium poses constraints on the constitutive relations of the fluid. Therefore, if we can generate a valid hydrostatically equilibrated configuration we may use it to simplify the construction of the constitutive relations for the fluid.

To obtain a valid hydrostatically equilibrated configuration we use the methods developed in \cite{Jensen:2012jh,Banerjee:2012iz}: Consider a hydrostatically equilibrated state. In such a state  Euclidean correlation functions of generic operators at equal Euclidean time will decay exponentially at large distances (assuming we are not at a critical point or that there are no long range forces at play). That is, there exists a power series expansion of zero frequency correlation functions of generic operators around zero spatial momentum. Let us consider a generating function for such correlation functions valid up to, say, $m$ powers of the spatial momentum. In real space such a generating function will contain $m$ derivatives of its arguments, the sources for the operators in question. For instance, the hydrostatic generating function for the energy momentum tensor will be a local function of the background metric and its derivatives. 

There exists an abundant body of literature on the construction of the hydrostatic generating function, and the associated constraints on constitutive relations for a variety of operators and sources. See e.g., \cite{Jain:2012rh,Jensen:2012jy,Valle:2012em,Bhattacharyya:2012xi,Banerjee:2012cr,Jensen:2012kj,Bhattacharya:2012zx,Eling:2013bj,Armas:2013hsa,Jensen:2013vta,Chapman:2013qpa,Manes:2013kka,Jensen:2013kka,Jensen:2013rga,Bhattacharyya:2013lha,Armas:2013goa,Bhattacharyya:2014bha,Megias:2014mba,Jensen:2014ama,DiPietro:2014bca,Harder:2015nxa,Valle:2015hfa,Banerjee:2015uta,Armas:2016xxg,Hernandez:2017mch,Armas:2018zbe,Kovtun:2019wjz,Kovtun:2019hdm,Armas:2019gnb,Armas:2020mpr}. For the current work, we are interested in hydrostatics in the presence of a spin current, as defined in \eqref{E:variation}. To this end, we need to construct a Lorentz and coordinate invariant generating function associated with a time independent external vielbein, $e^{a}{}_{\mu}$ and spin connection, $\omega_{\mu}{}^{ab}$ (and their covariant derivatives). 

To make time independence of the sources manifest, we denote by $V^{\mu}$ the Killing vector along which the metric is time independent, and by $\theta_V{}^{a}{}_{b}$ the Lorentz transformation parameter under which $e^{a}{}_{\mu}$ and $\omega_{\mu}{}^{ab}$ are invariant. Then, time independence of $\omega_{\mu}{}^{ab}$ and $e^a{}_{\mu}$ amounts to
\begin{align}
\begin{split}
\label{E:timeindependence}
	0& = \pounds_{V} e^{a}{}_{\mu} - \theta_V{}^{a}{}_{b} e^{b}{}_{\mu} \\
	0& = \pounds_V \omega_{\mu}{}^{a}{}_b + \partial_{\mu}\theta_V{}^a{}_{b} + \omega_{\mu}{}^{a}{}_{c} \theta_V{}^{c}{}_{b} - \theta_V{}^{a}{}_{c}\omega_{\mu}{}^{c}{}_{b} \,.
\end{split}
\end{align}
While it is often convenient to choose a particular gauge where $V^{\mu} = (1,0,\ldots,0)$ and $\theta_V{}^a{}_b = 0$, expressions of the form \eqref{E:timeindependence} are useful in order to make the notation covariant.

In order for the generating function
\begin{equation}
\label{E:Wini}
	W = \int |e| d^dx \mathcal{W}[e^{a}{}_{\mu},\,\omega_{\mu}{}^{ab},\,V^{\mu},\,\theta_V{}^a{}_{b},\,\nabla_{\mu}]\,,
\end{equation}
to be invariant under coordinate transformations and local Lorentz transformations, the vielbein, spin connection, Killing vector and Lorentz parameter $\theta_V{}^{a}{}_{b}$, together with their covariant derivatives, must be contracted to form scalars. Coordinate transformations of the vielbein and spin connection are given by \eqref{E:deltaew} while those of $V^{\mu}$ and $\theta_V{}^a{}_{b}$ are given by
\begin{align}
\begin{split}
\label{E:deltaVtheta}
	\delta V^{\mu} &= \pounds_{\xi}V^{\mu} \\
	\delta \theta_V{}^a{}_b &= \pounds_{\xi} \theta_V{}^a{}_b - \pounds_V \theta^a{}_b + \theta_V{}^a{}_c \theta^c{}_b - \theta^a{}_c \theta_V{}^c{}_b
\end{split}
\end{align}
The last two equalities can be obtained by writing $\delta_2 \delta_1 Q^a{}_{\mu}$ in terms of the action of $\delta_2$ on the parameters $\xi_1^{\mu}$ and $\theta_1{}^a{}_b$ and comparing it to the same expression when treating $\delta_1 Q^{a}{}_{\mu}$ as a tensor on which $\delta_2$ acts. 

The transformation properties of the constituents of the generating function, \eqref{E:deltaew} and \eqref{E:deltaVtheta} imply that $e^{a}{}_{\mu}$, $V^{\mu}$ and $\nabla_{\mu}$ transform as tensors, but $\omega_{\mu}{}^{ab}$ and $\theta_V{}^{a}{}_{b}$ do not. Since our goal is to generate a scalar expression for the density $\mathcal{W}$, it is convenient to replace $\theta_V{}^{a}{}_{b}$ with the combination
\begin{subequations}
\label{E:defthermal}
\begin{equation}
\label{E:defmuab}
	\mu^{ab} = \frac{V^{\mu}\omega_{\mu}{}^{ab} + \theta_V{}^{ab}}{\sqrt{-V^2}}
\end{equation}
which transforms as a tensor. Altertnately, one may view $\mu^{ab}$ as the holonomy of the spin connection around the thermal circle. Defining,
\begin{equation}
	T^{-1} = \sqrt{-V^2}
	\qquad
	u^{\mu} = \frac{V^{\mu}}{\sqrt{-V^2}}\,,
\end{equation}
\end{subequations}
allows us to write $\mathcal{W}$ in the form
\begin{equation}
\label{E:FinalW}
	\mathcal{W} = \mathcal{W}\left(T,\,u^{\mu},\,\mu^{ab},\,e^{a}{}_{\mu},\,K_{\mu}{}^{ab},\,\mathring{\nabla}_{\mu}\right)\,.
\end{equation}
In writing \eqref{E:FinalW} we have replaced the spin connection $\omega_{\mu}{}^{ab}$ with the contorsion tensor, $K_{\mu}{}^{ab}$, see \eqref{E:contorsion}, and the covariant derivative with a ringed one, c.f., \eqref{E:ringnabla}. 

In order to obtain the most general generating function W, we need to construct all possible scalars out of the constituents of $\mathcal{W}$ specified in \eqref{E:FinalW}. Due to the presence of the covariant derivative, $\mathring{\nabla}_{\mu}$, in \eqref{E:FinalW} there are an infinite number of such scalars. Often, it is convenient to expand $\mathcal{W}$ in terms of a derivative expansion in which case there are a finite number of scalars at each order in the expansion. In all systems which have been studied so far, the expansion of $\mathcal{W}$ in derivatives paralleled the derivative expansion of the associated constitutive relations. That is, order $n$ terms in $\mathcal{W}$ corresponded to order $n$ terms in the constitutive relations derived from $\mathcal{W}$. In our current setup this is not quite the case:
in \eqref{E:FinalW} we have replaced the non tensorial spin connection with the contorsion tensor, $K_{\mu}{}^{ab} = \omega_{\mu}{}^{ab} - \mathring{\omega}{}_{\mu}{}^{ab}$. Since, 
\begin{align}
\begin{split}
\label{E:deltaK}
	\frac{\delta}{\delta e^{c}{}_{\nu}} \int |e| B^{\mu}{}_{ab} K_{\mu}{}^{ab} d^dx &= \left( \frac{\delta}{\delta e^{c}{}_{\nu}} \int |e| B^{\mu}{}_{ab}  d^dx \right) K_{\mu}{}^{ab} + |e| \mathring{\nabla}_{\lambda} \left(B^{\lambda\nu\mu} - B^{\nu\lambda\mu} - B^{\mu\lambda\nu}\right)e_{c\mu}\,, \\
	\frac{\delta}{\delta \omega_{\nu}{}^{cd} }\int |e| B^{\mu}{}_{ab} K_{\mu}{}^{ab} d^dx &= \left( \frac{\delta}{\delta \omega_{\nu}{}^{cd}} \int |e| B^{\mu}{}_{ab}  d^dx \right) K_{\mu}{}^{ab} + |e| B^{\nu}{}_{cd}\,, 
\end{split}
\end{align}
then the contribution of the contorsion to the constitutive relations of the stress tensor will be one order higher in derivatives than its contribution to the spin current. We will refer to terms which contribute to the spin current at order $m$ and the stress tensor at order $m+1$ as order $m+1$ terms.\footnote{The alert reader may have noticed that \eqref{E:deltaK} has a structure very similar to \eqref{E:BRTmn}. We will comment on this resemblance shortly.}

Keeping in mind the unwonted contributions of the contorsion tensor to the constitutive relations, we can now consistently expand the generating function $W$ in a derivative expansion. From it, we may construct the hydrostatic energy momentum tensor and spin current $T_h^{\mu\nu}$ and $S_{h}^{\lambda\mu\nu}$ via
\begin{equation}
\label{E:defTS}
	T_h^{\mu}{}_{a} = \frac{1}{|e|}\frac{\delta W}{\delta e^{a}{}_{\mu}} 
	\qquad
	S_h^{\mu}{}_{ab} = \frac{2}{|e|}\frac{\delta W}{\delta \omega_{\mu}{}^{ab}}\,.
\end{equation}
We will carry out this task in the remainder of this section.

\subsection{The ideal fluid}
\label{SS:ideal}
The generating function whose variation will give us constitutive relations with no explicit derivatives is given by 
\begin{equation}
\label{E:defideal}
	\mathcal{W} = P(T,\,u^{\mu},\,\mu^{ab},\,e^{a}{}_{\mu}) \,.
\end{equation}
A fluid whose constitutive relations are obtained by varying the expression on the right hand side of \eqref{E:defideal} is often referred to as an ideal fluid. Ideal fluids are fluids whose constitutive relations contain no explicit derivatives of the hydrodynamic variables. Note that we have phrased the previous sentence rather carefully. In most instances an equivalent definition of an ideal fluid is one whose constitutive relations are zeroth order in derivatives. We will see shortly that in our current scheme, these two definitions of an ideal fluid are not interchangeable. In the remainder of this work we will use the former definition exclusively.

To proceed, it is convenient to decompose $\mu^{ab}$ into components parallel and orthogonal to the velocity field, 
\begin{equation}
\label{E:mutoMm}
	\mu^{ab} = u^a m^b - u^b m^a + M^{ab}
\end{equation}
where $u_{a}M^{ab}=0$ and $u_a m^a=0$. With this notation the available scalars in a generic number of spacetime dimensions $d$ are given by
contractions of chains of $M^{ab}$,
\begin{equation}
	\mathcal{M}_{(n)}{}^{a}{}_{b} = M^{a}{}_{c_2}M^{c_2}{}_{c_3} \ldots M^{c_n}{}_{b}\,, \qquad n \geq 2\,,
\end{equation}
with a pair of $m_a$'s or with themselves:
\begin{align}
\begin{split}
\label{E:Mms}
	m_{(n)} &= m_c\, \mathcal{M}_{(2n)}{}^{cd} m_d \,,\\
	M_{(n)} &= \mathcal{M}_{(2n)}{}^c{}_{c}\,.\\
\end{split}
\end{align}
For notational convenience it is useful to define
\begin{align}
\begin{split}
	\mathcal{M}_{(0)}{}^{a}{}_{b} = \delta^a{}_{b}\,,
	\qquad
	\mathcal{M}_{(1)}{}^{a}{}_{b} = M^{a}{}_{b}\,,
\end{split}
\end{align}
so that $m_{(0)} = m_c m^c$ and $M_{(0)}=d$ are well defined. 
The number of possible independent scalars is bound by the dimensionality of $M^{ab}$ and $m^a$. In a generic number, $d$, of spacetime dimensions the independent scalars are given by
$m_{(0)},\,\ldots,\,m_{\left(\left\lfloor \frac{d-2}{2} \right \rfloor \right)}$ for $d\geq 2$ and $M_{(1)},\,\ldots,\,M_{\left(\left\lfloor \frac{d-1}{2} \right\rfloor\right)}$ for $d\geq 3$.
Of course, in addition to \eqref{E:Mms} one may construct various dimension dependent pseudo scalars using the Levi Civitta tensor. For example, in $3+1$ dimensions one may use,
\begin{equation}
	\widetilde{M} = \epsilon^{abcd}u_a m_b M_{cd}\,.
\end{equation}
Since $\widetilde{M}^2 = 4 m_{(1)} -2 M_{(1)} m_{(0)}$ keeping all four of $m_{(0)}$, $m_{(1)}$, $M_{(1)}$ and $\widetilde{M}$ is redundant and we may keep only $m_{(0)}$, $M_{(1)}$ and $\widetilde{M}$.
In what follows we will consider contributions to the constitutive relations for a generic $d$ dimensional spacetime theory.

The dependence of the ideal stress tensor and current on the external fields can now be computed by varying $P(m_{(n)},\,M_{(n)},\,T)$ with respect to the vielbein or spin connection respectively. 
Inserting \eqref{E:defideal} into \eqref{E:defTS} and denoting the resulting stress tensor and spin current by $T_{id}^{\mu\nu}$ and $S_{id}{}^{\mu\nu\rho}$, viz.,
\begin{equation}
	T_{id}{}^{\mu}{}_{a} = \frac{1}{|e|}\frac{\delta}{\delta e^{a}{}_{\mu}}  \int  d^dx \sqrt{g} P\,,
	\qquad
	S_{id}{}^{\mu}{}_{ab} = \frac{2}{|e|}\frac{\delta}{\delta \omega_{\mu}{}^{ab}}  \int  d^dx \sqrt{g} P\,,
\end{equation}
we find 
\begin{align}
\begin{split}
\label{E:idealconstitutive}
	T_{id}^{\mu\nu} & =  \epsilon u^{\mu}u^{\nu} + P \Delta^{\mu\nu} + u^{\mu}\Delta^{\nu\alpha}P_{\alpha} \\ 
	S_{id}{}^{\lambda}{}_{\mu\nu} & = u^{\lambda} \rho_{\mu\nu} 
\end{split}
\end{align}
where
\begin{equation}
	\Delta^{\mu\nu} = g^{\mu\nu} + u^{\mu}u^{\nu}
\end{equation}
is a projection orthogonal to the velocity field,
$\epsilon$ is given by
\begin{equation}
\label{E:GibbsDuhem}
	\epsilon = -P + s T + \frac{1}{2}\rho_{\alpha\beta}\mu^{\alpha\beta}\,,
\end{equation}
and $s$, $\rho_{ab}$ and $P_a$ are given by the variation of the pressure with respect to the variables $T$, $\mu^{ab}$ and $u^a$,
\begin{equation}
\label{E:Pressurederivatives}
	s = \frac{\partial P}{\partial T}\,,\qquad
	\frac{1}{2}\rho_{ab} = \frac{\partial P}{\partial \mu^{ab}}\,,\qquad
	P_a = \frac{\partial P}{\partial u^a}\,.
\end{equation}
Their explicit dependence on derivatives of the pressure with respect to our basis of scalars is given by
\begin{multline}
	\rho_{ab}  =
	4  \sum_{n=0}^{\left\lfloor\frac{d-2}{2}\right\rfloor} \frac{\partial P}{\partial m_{(n)}} \left(m_c \mathcal{M}_{(2n)}{}^{c}{}_{[a}u_{b]} + \sum_{k=0}^{\left\lfloor\frac{2n-1}{2}\right\rfloor} m_c \mathcal{M}_{(k)}{}^{c}{}_{[a} \mathcal{M}_{(2n-1-k)}{}_{ b]}{}^d m_d \right)
	\\
	- 4 \sum_{n=1}^{\left\lfloor\frac{d-1}{2}\right\rfloor} n \frac{\partial P}{\partial M_{(n)}} \mathcal{M}_{(2n-1)}{}_{ab} 
\end{multline}
and 
\begin{multline}
	P_{a} = 	
	  -2\sum_{n=0}^{\left\lfloor\frac{d-2}{2}\right\rfloor} \frac{\partial P}{\partial m_{(n)}} \left( m_{(n)} u_a - m_b \mathcal{M}_{(2n+1)}{}^{b}{}_{a} + \sum_{k=0}^{\left\lfloor \frac{2n-1}{2}\right\rfloor} m_{b} \mathcal{M}_{(2n - 1-2k)}{}^b{}_a m_{(k)} \right)
	\\
	- 4 \sum_{n=1}^{\left\lfloor\frac{d-1}{2}\right\rfloor} n \frac{\partial P}{\partial M_{(n)}} m_b \mathcal{M}_{(2n-1)}{}^b{}_{a} \,,
\end{multline}
where square brackets represent an antisymmetric combination, $A_{[\mu\nu]} = \frac{1}{2} \left(A_{\mu\nu}-A_{\nu\mu}\right)$. Note that the stress tensor is not symmetric due to the term proportional to (the transverse part of) $P^{\mu}$.

A few comments are in order. The relations given in \eqref{E:idealconstitutive} describe the dependence of the energy momentum tensor and spin current on the external sources which have been conveniently packaged into the variables, $T$, $\mu^{ab}$ and $u^{a}$ given in \eqref{E:defthermal}. Since these are hydrostatically equilibrated configurations the resulting stress tensor and current should also be expressible in terms of solutions to the hydrodynamic equations. This interpretation allows us to identify $T$ as the temperature, $u^{\mu}$ as the velocity field and $\mu^{ab}$ as the spin chemical potential. From this point of view, the relations \eqref{E:idealconstitutive} are the constitutive relations for an ideal fluid and \eqref{E:defthermal} give the dependence of the hydrodynamic variables on the external sources once hydrostatic equilibrium is achieved. 

Once the hydrodynamic variables $T$, $u^{\mu}$ and $\mu^{ab}$ have been ascertained, we may identify $P$ with the pressure, $\epsilon$ with the energy density, $s$ with the entropy density, $\rho_{\alpha\beta}$ with the spin charge density and $P_a$ with what we refer to as a momentum density. With the above interpretation of thermodynamic variables, equation \eqref{E:GibbsDuhem} can be interpreted as the Gibbs Duhem Relation relating energy density and pressure. This expression together with \eqref{E:Pressurederivatives} gives us the first law of thermodynamics
\begin{equation}
\label{E:firstlaw}
	d\epsilon = T ds + \frac{1}{2} \mu^{ab} d\rho_{ab} - P_a du^a\,.
\end{equation}

From \eqref{E:firstlaw} (or \eqref{E:Pressurederivatives}) we observe that the momentum density is the variable conjugate to the velocity field, $u^a$. Such a quantity is absent in relativistic thermodynamics as long as the velocity field is the only tensor which breaks Lorentz invariance. Thus, normal, charged and uncharged fluids, do not support such a variable. The two component model of superfluids includes a superfluid velocity field which further breaks Lorentz invariance but there, since superfluid flow is a gradient flow, thermodynamic quantities are packaged in a somewhat different way than here. The only instance we are aware of where the momentum density plays a thermodynamic role similar to the one in \eqref{E:firstlaw} is in non relativistic thermodynamics without boost invariance. See \cite{deBoer:2017ing}. 

\subsection{An aside on non hydrostatic terms}

By construction, the constitutive relations \eqref{E:idealconstitutive} do not contain explicit derivatives of the vielbein, $e^a{}_{\mu}$, spin connection, $\omega_{\mu}{}^{ab}$, and hydrostatic parameters, $V^{\mu}$, and $\theta_V{}^{ab}$. In the remainder of this section we will follow standard procedure and add to the generating function $W$ terms which contain derivatives of the vielbein and hydrostatic parameters, order by order in a derivative expansion.
By varying the generating function with respect to the sources we will obtain the stress tensor and spin current as a function of these sources, c.f., equation \eqref{E:defTS}. We can then use the identifications in \eqref{E:defthermal} to obtain the constitutive relations for a fluid in hydrostatic equilibrium. 

Having said that, let us pause the computation of higher order corrections to the hydrostatic constitutive relations, and comment on the possible, additional, constitutive relations which are not hydrostatic which we will compute later in this work. 
Recall that the constitutive relations for a fluid do not depend on the particular configuration the fluid is in, be it hydrostatic or not. While we use the hydrostatic partition function to obtain those constitutive relations which do not vanish in hydrostatic equilibrium, we must, later, supplement these by any additional expressions which vanish in hydrostatic equilibrium. Let us comment on such expressions.

In hydrostatic equilibirum, the vielbein and spin connection are time independent in the sense of \eqref{E:timeindependence}. Since the external sources are eventually identified with hydrodynamic variables, as in \eqref{E:defthermal}, hydrostatic equilibrium poses constraints on the hydrodynamic variables themselves. Indeed, the stationarity condition on the background metric $\pounds_V g_{\mu\nu}=0$ (which follows from \eqref{E:timeindependence}) leads to
\begin{equation}
\label{E:gconstraints}
	u^{\mu}\partial_{\mu}T = 0\,,
	\qquad
	a^{\mu} + \frac{\Delta^{\mu\nu}\mathring{\nabla}_{\nu}T}{T} = 0\,,
	\qquad
	\theta = 0\,,
	\qquad
	\sigma^{\mu\nu} = 0\,,
\end{equation}
where $a^{\mu}$, $\theta$ and $\sigma^{\mu\nu}$ are components of the derivative of the velocity field,
\begin{equation}
\label{E:dudecomposition}
	\mathring{\nabla}_\mu u_\nu = \frac{1}{d-1} \theta \Delta_{\mu \nu} -u_\mu a_\nu + \sigma_{\mu \nu} + \Omega^{\mu \nu} \, , 
\end{equation}
with $a^{\mu}u_{\mu} = \sigma^{\mu\nu}u_{\nu} = \Omega^{\mu\nu}u_{\nu} = \sigma^{\mu}{}_{\mu}=0$ and $\sigma^{\mu\nu} = \sigma^{\nu\mu}$ and $\Omega^{\mu\nu} = -\Omega^{\nu\mu}$.  A compendium of decompositions of various quantities appearing in this work relative to the velocity field, including \eqref{E:dudecomposition} can be found in appendix \ref{A:decompositions}.
Similarly, the constraint on $e^{a}{}_{\mu}$ and $\omega_{\mu}{}^{ab}$ given in \eqref{E:timeindependence} read
\begin{equation}
\label{E:cpconstraint}
	u^{\mu} K_{\mu}{}^{ab} = \mu^{ab} + e^{a}{}_{\mu} e^{b}{}_{\nu} \left(\Omega^{\mu\nu} -2 u^{[\mu}a^{\nu]}\right) 
\end{equation}
and 
\begin{subequations}
\begin{equation}
\label{E:dcpconstraint}
	T e^{\rho}{}_{a} e^{\sigma}{}_{b} \mathring{\nabla}_{\lambda} \frac{\mu^{ab}}{T} = R^{\rho\sigma}{}_{\lambda\alpha} u^{\alpha}  - 2 e^{\rho}{}_{a} e^{\sigma}{}_{b} K_{\lambda c}{}^{[a}\mu^{b]c} \,, 
\end{equation}
with
\begin{equation}
	R^{\rho\sigma}{}_{\lambda\alpha} = \mathring{R}^{\rho\sigma}{}_{\lambda\alpha} + 2 e^{\rho}{}_{a} e^{\sigma}{}_{b} \left( \mathring{\nabla}_{[\lambda} K_{\alpha}{}^{ab} - K_{\lambda}^{c[a}K_{\alpha c}{}^{b}\right)
\end{equation}
\end{subequations}
respectively.

Equations \eqref{E:gconstraints} maintain that the shear and expansion of the flow vanish and that the acceleration must be proportional to gradients of the temperature if the system is in hydrostatic equilibrium. 
Likewise, equation \eqref{E:cpconstraint} implies that in hydrostatic equilibrium
\begin{equation}
\label{E:hattedM}
	\hat{M}^{\mu\nu} = 0
	\qquad
	\hbox{and}
	\qquad
	\hat{m}^{\mu} = 0
\end{equation}
where
\begin{align}
\begin{split}
\label{E:defhMm}
	\hat{M}^{\mu\nu} &= M^{\mu\nu} - K^{\mu\nu} + \Omega^{\mu\nu} \\
	\hat{m}^{\mu} &= m^{\mu} - k^{\mu} - a^{\mu} 
\end{split}
\end{align}
and 
\begin{equation}
	k^{\alpha} = u^{\mu}u_{\beta}K_{\mu}{}^{\alpha\beta}
	\qquad
	K^{\alpha\beta} = \Delta^{\alpha}{}_{\gamma} \Delta^{\beta}{}_{\delta} u^{\mu}K_{\mu}{}^{\gamma\delta}\,.
\end{equation}
(A full decomposition of the contorsion tensor into components can be found in appendix \ref{A:decompositions}.) The last equation, \eqref{E:dcpconstraint} implies that covariant derivatives of the spin chemical potential may be written in terms of the spin chemical potential and the Riemann tensor. These relations, combined with \eqref{E:hattedM} will also constrain gradients of the acceleration and vorticity two form. Since the explicit form of these constraints is somewhat long and not very useful, we will not write them explicitly. 

Relation \eqref{E:cpconstraint} is key. Suppose we are working in a flat torsionless background where $g_{\mu\nu} = \eta_{\mu\nu}$ and $K_{\mu}{}^{ab}=0$. In such a background an equilibrated fluid will have vanishing acceleration and vanishing vorticity, $a^{\mu}=0$ and $\Omega^{\mu\nu}=0$, and therefore vanishing spin chemical potential, $\mu^{ab}=0$. In fact, from \eqref{E:cpconstraint} we find that (in the absence of torsion) a spin chemical potential can be sustained in equilibrium only if we have non zero acceleration (implying a temperature gradient due to the second equality in \eqref{E:gconstraints}) or non zero vorticity. 

The relation \eqref{E:cpconstraint} was obtained by different means in \cite{Becattini:2007sr}, where it took the form
\begin{equation}
\label{E:cpconstraint2}
	\mu^{ab} = -T e^{[a}{}_{\rho}e^{b]}{}_{\sigma} \mathring{\nabla}^{\rho} \frac{u^{\sigma}}{T}\,,
\end{equation}
valid in the absence of torsion, where the last term on the right, $-\frac{T}{2} \left(\mathring{\nabla}_{\rho} \frac{u_{\sigma}}{T} - \mathring{\nabla}_{\sigma} \frac{u_{\rho}}{T}\right)$ is referred to as the thermal vorticity. We emphasize, in light of the above discussion, that the relation \eqref{E:cpconstraint} or its equivalent form \eqref{E:cpconstraint2}, is, a priori, only valid in hydrostatic equilibrium where the velocity and temperature are time independent and are fixed in terms of the background metric. Of course, it is possible, and as we shall see shortly, practically relevant, that the relation \eqref{E:cpconstraint} will hold dynamically. Meaning, it will be enforced as an equation of motion. At this point however, we can not ascertain this.

To summarize, the relation \eqref{E:cpconstraint} implies that a non vanishing spin chemical potential can only be supported by acceleration, vorticity or torsion (in hydrostatic equilibrium). Therefore, in order for a consistent derivative expansion to exist the spin chemical potential, torsion, acceleration and vorticity should all be of the same order. The simplest way to implement this would be to require that the spin chemical potential and contorsion tensor are first order in derivatives. While a bit surprising at first, it is, perhaps, not so strange once the previous discussion is taken into account. A spin chemical potential will naturally vanish once first order in gradient quantities are small. Likewise, the contorsion tensor, in its role as a connection, is often associated with explicit derivatives. Therefore, in the remainder of this work we will treat the spin chemical potential and the contorsion tensor as a first order in derivative quantity. This implies, among other things, that the ideal stress tensor and current, while not containing explicit derivatives, may be expanded in a derivative expansion by counting $m^{a}$ and $M^{ab}$ as first order in derivative objects. Thus, while the ideal constitutive relations given in \eqref{E:idealconstitutive} do not contain any explicit derivatives, they may still be expanded in a derivative expansion once $m^a$ and $M^{ab}$ are counted as first order in derivative quantities.

Let us remark that there may be other possible derivative counting schemes compatible with \eqref{E:cpconstraint}. Perhaps the spatial component of the contorsion tensor may be counted as zeroth order in derivatives while its temporal component and the spin chemical potential count as first order in derivatives. Or, it might be possible to decompose the temporal component of the contorsion tensor and the spin chemical potential into an order zero and order one term, treating each component separately in some sense. While we haven't ruled out such possibilities so far, we seem to encounter difficulties when trying to implement them effectively. Another counting scheme implicitly implemented in e.g., \cite{Florkowski:2017ruc,Becattini:2018duy} is to consider a background vorticity. We will discuss some of these further in section \ref{S:discussion}.

\subsection{The hydrostatic expansion to leading order in derivatives}
Having set up the derivative expansion for fluids with spin, we are almost ready to compute higher order corrections to the ideal constitutive relations \eqref{E:idealconstitutive}. The last obstacle we need to deal with is a mismatch between derivative counting of the constitutive relations and the derivative counting of the equations of motion. Since the hydrodynamic equations are conservation laws it is often the case that the $m$'th order in derivative constitutive relations contribute to the $m+1$'th order terms in the equations of motion. As discussed in detail in section \ref{S:Canonicalspincurrent} the equations of motion in the presence of torsion are non conservation equations, \eqref{E:EOMs}. In terms of derivative counting, we observe that the antisymmetric components of the stress tensor enter the equations of motion at the same order as derivatives of the stress tensor and spin current. Therefore, in order to have a consistent perturbative scheme for the hydrodynamic equations of motion we must count the constitutive relations for the antisymmetric components of the stress tensor as one order higher than that of the symmetric components and of the spin current. In the remainder of this work we will construct the leading, non ideal, contribution to the constitutive relations for fluids with spin. This implies that we will expand the spin current and symmetric components of the stress tensor to first order in derivatives, and the antisymmetric components of the stress tensor to second order in derivatives. To avoid the latter lengthy articulation of our perturbative expansion, we will often state that we are computing the constitutive relations such that the equations of motion are second order in derivatives. 

Keeping in mind the derivative counting scheme described above and in the previous section, in order to compute the hydrostatic constitutive relations to the required order we must enumerate all possible hydrostatic scalar quantities up to first order in derivatives, and all hydrostatic scalar quantities which may contribute to the antisymmetric components of the stress tensor to second order in derivatives.

At zero order in derivatives, the temperature $T$ is the only available scalar and its contribution to the constitutive relations has been computed in \eqref{E:idealconstitutive}. At first order in derivatives, the only available scalar is the trace of the contorsion tensor, $\kappa$, defined in \eqref{E:kappas}. At second order in derivatives, we need to consider only those terms which contribute to the antisymmetric component of the stress tensor. The scalars $M_{(1)}$ and $m_{(0)}$  have been dealt with in the previous section which studied the all order in derivative ideal fluid. Second order scalars involving derivatives of $m^{\mu}$ and $M^{\mu\nu}$ can be traded with the Riemann tensor using \eqref{E:dcpconstraint}. Following \cite{Banerjee:2012iz,Jensen:2012jh} the latter contributes to the symmetric components of the energy momentum tensor and is less relevant to our current study. Similarly, \eqref{E:hattedM} may be used to remove scalars involving the vorticity two-form and the acceleration, $\Omega^{\mu\nu}$ and $a^{\mu}$, in place of the components of the contorsion, $K^{\mu\nu}$ and $k^{\mu}$ and the chemical potential $M^{\mu\nu}$ and $m^{\mu}$. Thus, at the end of the day, the only two contorsion independent scalars are $M_{(1)}$ and $m_{(0)}$ which we have treated earlier in our discussion of the ideal fluid. In addition, we find that there are four scalars linear in contorsion and ten additional scalars quadratic in contorsion.\footnote{
If we allow for terms which explicitly break parity or time reversal (involving the Levi-Civitta tensor), then more terms will be available, their explicit form will depend on the spacetime dimension. We leave a discussion of such terms to future work.
}
\begin{table}[hbt!]
\begin{center}
		\begin{tabular}{| c   c      c  c    | }
		\hline
		\multicolumn{4}{| l |}{Order zero hydrostatic scalars} \\
		\hline
		$T$ & & &   \\ 
		\hline
		\multicolumn{4}{| l |}{Order one hydrostatic scalars} \\
		\hline
		$\kappa$ & & & \\ 
		\hline
		\multicolumn{4}{| l |}{Order two, contorsion independant hydrostatic scalars} \\
		\hline
			$M_{(1)}$ & 
			$m_{(0)}$ & & \\
		\hline
		\multicolumn{4}{| l |}{Order two, hydrostatic scalars linear in contorsion} \\
		\hline
			$S_{(1)} = M_{\mu\nu} \kappa_A^{\mu\nu}$ & 
			$S_{(2)} = M_{\mu\nu}K^{\mu\nu}$ & 
			$S_{(3)} = m_{\mu}k^{\mu} $ &
			$S_{(4)} = m_{\mu}\mathcal{K}_V^{\mu} $  \\ 
		\hline
		\multicolumn{4}{| l |}{Order two hydrostatic scalars, quadratic in contorsion} \\
		\hline
			$S_{(5)} = \kappa^2$ &
			$S_{(6)} =k \cdot k$ &
			$S_{(7)} =K_{\mu\nu}K^{\mu\nu}$ &
			$S_{(8)} =k \cdot \mathcal{K}_V$ \\
			$S_{(9)} =\mathcal{K}_V \cdot \mathcal{K}_V$ &
			$S_{(10)} =K_{\mu\nu}\kappa_A{}^{\mu\nu}$ &
			$S_{(11)} =\kappa_{A\,\mu\nu}\kappa_A{}^{\mu\nu}$ &
			$S_{(12)} =\kappa_{S\,\mu\nu}\kappa_S{}^{\mu\nu}$ \\
			$S_{(13)} =\mathcal{K}_{A\,\mu\nu\rho} \mathcal{K}_A{}^{\mu\nu\rho}$ &
			$S_{(14)} =\mathcal{K}_{T\,\mu\nu\rho} \mathcal{K}_T{}^{\mu\nu\rho}$ &&  \\ 
		\hline
	\end{tabular}
	\caption{\label{T:allscalarsnoK} 
	Summary of zeroth, first and second order independent inequivalent scalars which may contribute to the equations of motion at second order in derivatives. The components of the spin chemical potential $M_{\mu\nu}$ and $m^{\mu}$ have been defined in \eqref{E:mutoMm}, the scalars $m_{(1)}$ and $M_{(1)}$ have been defined in \eqref{E:Mms} and the various $K$, $k$, $\kappa$ and $\mathcal{K}$ denote components of the contorsion tensor whose definition can be found in appendix \ref{AA:contorsion}.}
\end{center}
\end{table}

To obtain the contribution of the constitutive relations to equations of motion to second order in derivatives we may write the partition function, $W = \int \sqrt{-g} d^dx \mathcal{W}$, as
\begin{equation}
\label{E:hydrostatic2}
	\mathcal{W} = P + \mathcal{W}_{(1)}+ \mathcal{W}_{(2)}
\end{equation}
where $P$ is the pressure function given in \eqref{E:defideal} which determines the constitutive relations for the ideal fluid, and
\begin{align}
\begin{split}
\label{E:W1W2}
	\mathcal{W}_{(1)} &= \chi^{(1)}_1 \kappa\,, \\
	\mathcal{W}_{(2)} & = \sum_i \chi^{(2)}_i S_{(i)}\,.
\end{split}
\end{align}

In writing \eqref{E:hydrostatic2} as a derivative expansion, we may, for consistency, expand the pressure term, $P$, to quadratic order in the chemical potential
\begin{equation}
\label{E:Pdecomposition}
	P(m_{(n)},\,M_{(n)},\,T)  = P_0(T) + \rho_m(T) m_{(0)} + \rho_M(T) M_{(1)} + \mathcal{O}(\nabla^4).
\end{equation}
Put differently, we may expand the ideal constitutive relations, \eqref{E:idealconstitutive}, to quadratic order in the chemical potential (such that the equations of motion are second order in derivatives). Explicitly, one finds
\begin{align}
\begin{split}
\label{E:idealexpansion}
	T_{id}^{\mu\nu} =& \left(\epsilon_0+(\rho_m+T \rho_m') m_{\alpha}m^{\alpha} + (\rho_M+T \rho_M') M_{\alpha\beta}M^{\alpha\beta} \right) u^{\mu}u^{\nu} \\
	&+ (P_0+\rho_m m_{\alpha}m^{\alpha} + \rho_M M^{\alpha\beta}M_{\alpha\beta}) \Delta^{\mu\nu} \\
	&+ u^{\mu}m_{\alpha}M^{\alpha\nu}\left(2 \rho_m - 4 \rho_M\right)  +\mathcal{O}(\nabla^4)\\
	S_{id}^{\lambda\mu\nu} =& u^{\lambda} \left(4 \rho_m m^{[\mu}u^{\nu]} - 4 \rho_M M^{\mu\nu} \right)  +\mathcal{O}(\nabla^4) \,,
\end{split}
\end{align}
where
\begin{equation}
\label{E:gotepsilon0}
	\epsilon_0 = T \frac{\partial P_0}{\partial T} - P_0\,,
\end{equation}
and primes denote a derivative with respect to the temperature.

The constitutive relations associated with \eqref{E:hydrostatic2} can be computed using \eqref{E:defTS}.
Consider
\begin{equation}
\label{E:defTj}
	T_{h(j)}{}^{\mu}{}_{a} = \frac{1}{|e|} \frac{\delta W_{(j)}}{\delta e^a{}_{\mu}}\Bigg|_{K_{\mu}{}^{ab}=0}
	\qquad
	S_{h(j-1)}{}^{\mu}{}_{ab} = \frac{2}{|e|} \frac{\delta W_{(j)}}{\delta \omega_{\mu}{}^{ab}}\Bigg|_{K_{\mu}{}^{ab}=0}
\end{equation}
with $j=1,2$ and $W_{(j)} = \int |e| d^dx \mathcal{W}_{(j)}$. The subscript `$h$' specifies that we are considering hydrostatic components of the stress tensor or current and the subscript `$(j)$' specifies the order in derivatives of the resulting constitutive relations for the stress tensor and current.

Writing
\begin{align}
\begin{split}
\label{E:hdecomposition}
	S_{h(j)}{}^{\lambda\mu\nu} =&  S_{h(j)BR}^{\lambda\mu\nu} + S_{h(j)nBR}^{\lambda\mu\nu} \\
	T_{h(j)}{}^{\mu\nu} =& T_{h(j)BR}^{\mu\nu} + T_{h(j)nBR}^{\mu\nu}
\end{split}
\end{align}
with
\begin{equation}
\label{E:defTBR}
	T_{h(j)BR}^{\mu\nu} = {\frac{1}{2}} \mathring{\nabla}_{\lambda}   \left(
		S_{h(j-1)BR}{}^{\lambda\mu\nu} - S_{h(j-1)BR}{}^{\mu\lambda\nu} - S_{h(j-1)BR}{}^{\nu\lambda\mu}\right)\Bigg|_{\hbox{\tiny hydrostatic equilibrium}}
\end{equation}
we find
\begin{subequations}
\label{E:constitutive12}
\begin{align}
\begin{split}
\label{E:S01}
	S_{h(0)BR}{}^{\lambda\mu\nu} = & 2 \chi^{(1)}_1 \Delta^{\lambda[\mu}u^{\nu]} \\
	S_{h(0)nBR}{}^{\lambda\mu\nu} = & 0 \\
	S_{h(1)BR}{}^{\lambda\mu\nu} =&
		 - 2 \chi_1^{(2)} M^{\lambda[\mu}u^{\nu]} 
		 + 2  \chi_2^{(2)} u^{\lambda}  M^{\mu\nu} 
		 - 2 \chi_3^{(2)} u^{\lambda}  u^{[\mu}m^{\nu]}  
		 + 4 \chi_4^{(2)} \Delta^{\lambda[\mu}m^{\nu]} \\
		&+4 \chi_5^{(2)} \kappa \Delta^{\lambda[\mu}u^{\nu]} \\
		&+2u^\lambda \left( 
			2 \chi^{(2)}_6 k^{[\mu} u^{\nu]} 
			+ 2 \chi^{(2)}_7 K^{\mu \nu} 
			+ \chi^{(2)}_{10} \kappa_A{}^{\mu \nu}   \right) \\ 
		&+ 2u^\lambda   \chi^{(2)}_8 \mathcal{K}_V{}^{[\mu} u^{\nu]} 
		+ 4\Delta^{\lambda [\mu} \left( 
			\chi^{(2)}_8 k^{\nu]} 
			+2 \chi^{(2)}_9 \mathcal{K}_V{}^{\nu]} \right)\\ 
		&-  2\chi^{(2)}_{10} K^{\lambda [\mu} u^{\nu]} 
		- 4 \chi^{(2)}_{11} \kappa_A{}^{\lambda [\mu}u^{\nu]}
		+ 4 \chi^{(2)}_{12} \kappa_S{}^{\lambda [\mu} u^{\nu]} \\ 
		&+ 4 \chi^{(2)}_{13} \mathcal{K}_A{}^{\lambda \mu \nu} 
		+ 4 \chi^{(2)}_{14} \mathcal{K}_T{}^{\mu \nu \lambda} \\
	S_{h(1)nBR}{}^{\lambda\mu\nu} = 
		&+2 u^{\lambda} \left(\chi^{(2)}_1 \kappa_A{}^{\mu\nu} + \chi_2^{(2)} K^{\mu\nu} -  \chi^{(2)}_3 u^{[\mu} k^{\nu]}\right)
		- 2 \chi_4^{(2)} u^{\lambda} u^{[\mu}\mathcal{K}_V{}^{\nu]}  \\
\end{split}
\end{align}
and
\begin{align}
\begin{split}
\label{E:T12}
	T_{h(1)nBR}^{\mu\nu}  = &   \left( T \frac{\partial \chi^{(1)}_1}{\partial T}- \chi^{(1)}_1 \right)\kappa u^\mu u^\nu 
		   + \chi^{(1)}_1 \frac{d-2}{d-1} \kappa   \Delta^{\mu \nu} \\ 
		& + 2\chi^{(1)}_1 u^{(\mu} k^{\nu)} + \frac{1}{2} \chi^{(1)}_1 u^\mu \mathcal{K}_V{}^\nu   
		   -\chi^{(1)}_1 \kappa_S{}^{\mu \nu} - \chi^{(1)}_1 \kappa_A{}^{\mu \nu}   \\
	T_{h(2)nBR}^{\mu\nu} = &   \chi_1^{(2)} \left(\mathcal{K}_T{}^{ \alpha \beta[\mu} + \mathcal{K}_A{}^{\alpha \beta[\mu} \right)u^{\nu]}M_{\alpha \beta}
		   +\frac{\chi^{(2)}_1 \kappa M^{\mu \nu}}{d-1}  \\ 
		&+ \frac{\chi_1^{(2)}} { 2(d-2)} u^{[\mu} M^{\nu] \beta} \mathcal{K}_{V\,\beta} - 2\chi^{(2)}_1 u^{[\mu} \kappa_A{}^{\nu] \beta} m_\beta       
		  + \chi^{(2)}_1 M^{\alpha [\mu} \left(  \kappa_A{}^{\nu]}{}_\alpha -  \kappa_S{}^{\nu]}{}_\alpha \right) \\
		&- 2 \chi^{(2)}_2 u^{[\mu} \left( M^{\nu] \alpha} k_\alpha + K^{\nu]\alpha} m_\alpha \right) 
		  - \chi^{(2)}_3 u^{[\mu} \left( M^{\nu]\alpha} k_\alpha+ K^{\nu]\alpha} m_\alpha  \right)  \\
		& + 2 \chi^{(2)}_4  \mathcal{K}_A{}^{\mu \nu \alpha} m_\alpha  
		  + 2 \chi^{(2)}_4  \mathcal{K}_T{}^{[\mu \nu]\alpha } m_\alpha 
		  + \frac{\chi^{(2)}_4}{(d-2)} \mathcal{K}_V^{[\mu} m^{\nu]} \\
		&  - \chi^{(2)}_4 u^{[\mu} M^{\nu]\alpha} \mathcal{K}_{V\,\alpha}  
	 	+ 2 \chi^{(2)}_4 u^{[\mu} \left(\kappa_A{}^{\nu]\alpha} -  \kappa_S{}^{\nu]\alpha} \right) m_\alpha 
		   + \frac{2\chi^{(2)}_4 (d-2)}{(d-1)} \kappa u^{[\mu} m^{\nu]}  
\end{split}
\end{align}
\end{subequations}

A few comments are in order. First, we note that $T_{h(j)nBR}^{\mu\nu}=0$ and $S_{h(j)nBR}^{\lambda\mu\nu}=0$ when the contorsion tensor vanishes. Thus, in the absence of torsion, the structure of the constitutive relations \eqref{E:hdecomposition} is identical to the ambiguity in the definition of the stress tensor, c.f., \eqref{E:BRTmn}. Following the discussion leading to, and succeeding \eqref{E:BRTmn}, it is  not surprising that all hydrostatic contributions to the stress tensor coming from variations of the contorsion tensor are of this type. Further following the discussion, it should be clear from \eqref{E:Tpeqns} that the constitutive relations \eqref{E:constitutive12} will not contribute to the equations of motion in the absence of torsion. Be that as it may, the terms associated with the coefficients $\chi_1^{(1)}$ and $\chi_i^{(2)}$ do affect the expectation value (and correlation functions) of the stress tensor and spin current and should not be ignored or removed. In the remainder of this work we will refer to 
constitutive relations of the form \eqref{E:defTBR} as Belinfante Rosenfeld terms, or BR terms for short. 
The subscripts `$BR$' and `$nBR$' in \eqref{E:hdecomposition} specifies those constitutive relations which contribute to BR terms or non BR terms respectively.

We would also like to point out the importance of imposing the hydrostatic relations after taking the derivatives of the spin current on the right hand side of \eqref{E:defTBR}. A naive evaluation of the right hand side of \eqref{E:defTBR} without imposing the hydrostatic relations may result in non hydrostatic contributions to the constitutive relations. This point is probably best explained through an example. Consider $S_{h(0)BR}{}^{\lambda\mu\nu}$ and $T_{h(1)BR}^{\mu\nu}$. Inserting \eqref{E:S01} into \eqref{E:defTBR} yields
\begin{align}
\begin{split}
\label{E:T1}
	T_{h(1)BR}^{\mu\nu} =& \chi_1^{(1)} a^{\mu}u^{\nu}  
		+ \chi_1^{(1)} \Omega^{\mu\nu} 
		- \frac{\partial \chi_1^{(1)}}{\partial T} \Delta^{\mu\alpha}\partial_{\alpha}T u^{\mu} \\
		&+ \frac{\partial \chi_1^{(1)}}{\partial T} u^{\alpha}\partial_{\alpha}T \Delta^{\mu\nu} 
		+ \chi_1^{(1)}\theta \left(\eta^{\mu\nu}-\frac{\Delta^{\mu\nu}}{d-1}\right) 
		- \chi_1^{(1)}\sigma^{\mu\nu} \Bigg|_{\hbox{\tiny hydrostatic equilibrium}}\\
		=&
		 \chi_1^{(1)} a^{\mu}u^{\nu}  
		+ \chi_1^{(1)} \Omega^{\mu\nu} 
		+ \frac{\partial \chi_1^{(1)}}{\partial T} T u^{\mu}a^{\nu} \,,
\end{split}
\end{align}
where in going from the first equality to the second equality we have used \eqref{E:gconstraints}. A similar computation can be carried out for $T_{(2)}^{\mu\nu}$.

\section{Non hydrostatic terms}
\label{S:nonhydrostatic}
So far we have considered only those terms in the constitutive relations for the stress tensor and current which do not vanish in the hydrostatic limit, $T_h^{\mu\nu}$ and $S_h^{\lambda\mu\nu}$, c.f., equation \eqref{E:defTS}. The full stress tensor and spin current are given by a combination of hydrostatic terms and non hydrostatic terms,
\begin{equation}
	T^{\mu\nu} = T_h^{\mu\nu} + T_{nh}^{\mu\nu}
	\qquad
	S^{\lambda\mu\nu} = S_h^{\lambda\mu\nu} + S_{nh}^{\lambda\mu\nu}\,,
\end{equation}
where the non hydrostatic contributions include the most general constitutive relations which vanish in hydrostatic equilibrium.

Note that while the non hydrostatic terms are well defined, the hydrostatic terms are not. One may always add to the hydrostatic expressions terms which vanish in hydrostatic equilibrium. More formally, there exists an equivalence class of hydrostatic stress tensors and spin currents where elements of the same class differ by non hydrostatic terms (expressions which vanish in hydrostatic equilibrium). In the previous section we have chosen a particular representative for the hydrostatic components of the stress tensor and spin current. A somewhat different representative which will be useful in characterizing the full set of constitutive relations is to choose the BR type terms for the stress tensor such that they are BR terms also outside of hydrostatic equilibrium. We will refer to such a representative as $T_{h+}^{\mu\nu}$. 

For instance, in place of $T_{h(j)BR}^{\mu\nu}$ defined in \eqref{E:defTBR} we consider
\begin{equation}
\label{E:defTBR2}
	T_{h+(j)BR}^{\mu\nu} = {\frac{1}{2}} \mathring{\nabla}_{\lambda}   \left(
		S_{(j-1)}{}^{\lambda\mu\nu} - S_{(j-1)}{}^{\mu\lambda\nu} - S_{(j-1)}{}^{\nu\lambda\mu}\right)\,.
\end{equation}
The expression for $T_{h+(j)BR}^{\mu\nu}$ differs from $T_{h(j)BR}^{\mu\nu}$ by non hydrostatic terms, so that $T_{h+(j)BR}^{\mu\nu}=T_{h(j)BR}^{\mu\nu}$ in hydrostatic equilibrium. As an explicit demonstration of this fact consider
\begin{multline}
\label{E:ThpBR}
	T_{h+(1)BR}^{\mu\nu} = \chi_1^{(1)} a^{\mu}u^{\nu}  
		+ \chi_1^{(1)} \Omega^{\mu\nu} 
		- \frac{\partial \chi_1^{(1)}}{\partial T} \Delta^{\mu\alpha}\partial_{\alpha}T u^{\mu} \\
		+ \frac{\partial \chi_1^{(1)}}{\partial T} u^{\alpha}\partial_{\alpha}T \Delta^{\mu\nu} 
		+ \chi_1^{(1)}\theta \left(\eta^{\mu\nu}-\frac{\Delta^{\mu\nu}}{d-1}\right) 
		- \chi_1^{(1)}\sigma^{\mu\nu} 
\end{multline}
which coincides with $T_{h(1)}^{\mu\nu}$, c.f., equation \eqref{E:T1}, in hydrostatic equilibrium.

Following the above discussion, we will, without loss of generality, consider the decomposition
\begin{equation}
\label{E:ourdecomposition}
	T^{\mu\nu} = T_{h+}^{\mu\nu} + T_{nh}^{\mu\nu}
	\qquad
	S^{\lambda\mu\nu} = S_h^{\lambda\mu\nu} + S_{nh}^{\lambda\mu\nu}\,.
\end{equation}
Our strategy for writing down the full set of non hydrostatic constitutive relations is the standard one. We first tabulate all possible non hydrostatic constitutive relations at the order we are interested in, and then impose restrictions enforced by the second law of hydrodynamics, Onsager relations or unitarity. We can also use field redefinitions, often referred to as frames, to remove ambiguous terms in the constitutive relations and use the equations of motion at one order lower in derivatives to identify what would otherwise seem like distinct tensor structures. 

\subsection{The leading order constitutive relations}
There are no non hydrostatic zero order in derivative constitutive relations that we may add to the stress tensor and current. 
Recall, however, that the equation of motion for the spin current, \eqref{E:EOMs},  relates the divergence of the spin current to the antisymmetric component of the stress tensor. Therefore the antisymmetric component of the stress tensor at first order in derivatives contributes to the equations of motion at the same order as the zero order in derivative contribution to the spin current. Thus, a full analysis of the leading order in derivative equations of motion requires us to consider all possible non hydrostatic first order in derivative contributions to the antisymmetric components of the stress tensor. 

At first order in derivatives there are two antisymmetric tensor structures which are orthogonal to the velocity field, only one of which vanishes in hydrostatic equilibrium, and five antisymmetric tensor structures which have one leg parallel to the velocity field, two of which vanish in equilibrium. See table \ref{T:antisymmfirst}. Thus, the non hydrostatic constitutive relations contributing to the leading order equations of motion take the form
\begin{equation}
\label{E:leadingnh}
	T_{nh}^{[\mu\nu]} = \sigma_A(T) A^{[\mu} u^{\nu]} + \sigma_m(T) \hat{m}^{[\mu}u^{\nu]} + \sigma_M(T) \hat{M}^{\mu\nu} + \mathcal{O}(\nabla^2)\,,
	\qquad
	S_{nh}^{\lambda\mu\nu} = \mathcal{O}(\nabla)\,.
\end{equation}
\begin{table*}[hbt]
\begin{center}
	\begin{tabular}{| c | c | c |}
	\hline
	Tensor type & All data & Non hydrostatic data \\
	\hline
	Vectors & 
		$ A^{\mu} = \Delta^{\mu \nu} \mathring{\nabla}_\nu T + T a^\mu$, $\hat m^\mu$, $m^{\mu}$, $k^\mu$, $\mathcal{K}_V{}^\mu$ &  
		$ A^{\mu} = \Delta^{\mu \nu} \mathring{\nabla}_\nu T + T a^\mu$, $\hat m^\mu$ \\
	\hline
	Antisymmetric & $M^{\mu\nu}$, $\hat{M}^{\mu\nu}$, $\kappa_A{}^{\mu\nu}$, $K^{\mu\nu}$ & $\hat{M}^{\mu\nu}$ \\
	\hline
	\end{tabular}
	\caption{\label{T:antisymmfirst} A table of all first order tensors which may contribute to first order antisymmetric tensor structures, independent of the equations of motion. A vector $Y^{\mu}$ can be combined with a velocity field, $u^{\mu}$, to construct an antisymmetric tensor with one leg parallel to the velocity field, $Y^{[\mu}u^{\nu]}$. }
\end{center}
\end{table*} 

The resulting equations of motion at leading order in derivatives can be obtained by inserting \eqref{E:leadingnh}, \eqref{E:idealconstitutive} (appropriately expanded in derivatives as in \eqref{E:Pdecomposition}) and the expressions in \eqref{E:constitutive12} into \eqref{E:EOMs}. We find
\begin{align}
\begin{split}
\label{E:leadingEOM}
	(d-1) u^{\mu} \mathring{\nabla}_{\mu} T + T \mathring{\nabla}_{\mu} u^{\mu} &= \mathcal{O}(\nabla^2)
	\qquad
	A^{\mu} = \mathcal{O}(\nabla^2),
	\\
	\sigma_m \hat{m}^{\mu}+\sigma_A A^{\mu} &= \mathcal{O}(\nabla^2),
	\qquad
	\sigma_M \hat{M}^{\mu\nu} = \mathcal{O}(\nabla^2)\,.
\end{split}
\end{align}
The first two expressions coincide with the equations of motion for an ideal uncharged fluid without spin. The last two equations follow from the equations of motion for the spin current and are new. Before proceeding with the second order in derivative equations of motion, let us discuss them and their implications in some detail.

Unless $\sigma_m$ and $\sigma_M$ vanish, the leading order equations of motion for the spin current imply that
\begin{equation}
\label{E:dynamichatM}
	\hat{m}^{\mu}  = \mathcal{O}(\nabla^2)\,,
	\qquad
	\hbox{and}
	\qquad
	\hat{M}^{\mu\nu} = \mathcal{O}(\nabla^2)\,.
\end{equation}
These equations of motion are identical to the constraints coming from the requirement of hydrostatic equilibrium, see \eqref{E:hattedM} . Thus, even though the spin chemical potential is constrained to satisfy \eqref{E:hattedM} in hydrostatic equilibrium, we see from \eqref{E:dynamichatM} that it is dynamically constrained to take on the same value even outside of hydrostatic equilibrium. At least to leading order in derivatives. We remind the reader that there is an analagous situation for $A^{\mu}$. In hydrostatic equilibrium we have the constraint $A^{\mu}=0$ as in \eqref{E:gconstraints}, in addition, the dynamics ensure that $A^{\mu}=0$ also outside of equilibrium as in \eqref{E:leadingEOM}. 

The equalities in \eqref{E:dynamichatM} (and $A^{\mu}=0$) valid to leading order in derivatives has significant repercussions on the higher order in derivative structure of the constitutive relations. For instance, when going to next to leading order in derivatives one may remove tensor structures which are equivalent under the equations of motion. Thus, ${A}^{\mu}$, $\hat{M}^{\mu\nu}$ and $\hat{m}^{\mu}$ will not contribute to the stress tensor and spin current at that order in derivatives. 

Having $\hat{M}^{\mu\nu}=0$ and $\hat{m}^{\mu}=0$ essentially forces the dynamics of the spin current to take on its hydrostatic value at least to leading order in derivatives. 
This situation is consequentially modified if, for some reason $\sigma_M = \sigma_m=0$ (which was studied in \cite{Gallegos:2021bzp} and referred to as the dynamical spin limit) 
or if $\sigma_M$ and $\sigma_m$ are made perturbatively small in an appropriate sense (as discussed in \cite{Li:2020eon,Hongo:2021ona}). A priori there is no reason to consider either case, but it may be that in certain dynamical situations argued for in \cite{Li:2020eon} the latter may be physically relevant. Here, and in the remainder of this work, we will consider non infinitesimal $\sigma_M$ and $\sigma_m$. We expect that dynamical situations in which $\sigma_M$ and $\sigma_m$ are less relevant to the equations of motion may be worked out using the formalism developed in this letter. We will discuss this further in section \ref{S:discussion}.

\subsection{Fluid frames}
Gradient expansions allow for field redefinitions. In the framework of relativistic hydrodynamics without spin this implies that the definition of the velocity field, temperature and possible chemical potentials may be modified at every order in the gradient expansion. For instance, if we denote the velocity field by $u^{\mu}$ then one may equally define, say, $u^{\prime\,\mu} = u^{\mu} + u^{\alpha}\mathring{\nabla}_{\alpha} u^{\mu}$. The alternate velocity field, $u^{\prime\,\mu}$ will coincide with $u^{\mu}$ at zero order in gradients but will deviate from it at first order in a derivative expansion.  

Of course, physical observable will be independent of the particular choice of frame as long as the derivative expansion is used consistently. While the solutions to the equations of motion for $u^{\mu}$ and $u^{\prime\,\mu}$ will differ, the expectation value of the stress tensor under the equations of motion, appropriately truncated in a derivative expansion will be the same whichever frame is used. For this reason it is often convenient to choose a useful definition of the hydrodynamic variables (in terms of higher order derivative corrections) which will simplify computations. 

An often used frame in hydrodynamics without a spin current is the Landau frame where the velocity field and temperature are chosen such that $u_{\mu}T^{\mu\nu} = - \epsilon_0 u^{\nu}$ (with $\epsilon_0$ defined in \eqref{E:gotepsilon0}) is valid to all orders in a derivative expansion. To show that such a frame is possible one starts with the constitutive relations at zero order in derivatives, $T^{\mu\nu} = \epsilon_0 u^{\mu} u^{\nu} + P_0 \Delta^{\mu\nu}$ (with $P_0$ the pressure in equilibrium) and considers how shifts in $u^{\mu}$ and $T$, $u^{\prime\,\mu} = u^{\mu} + \delta u^{\mu}$, $T' = T + \delta T$ affect them. One may always choose $u^{\prime}$ and $T'$ such that the Landau frame condition holds, order by order in a derivative expansion. See, e.g., \cite{Bhattacharya:2011tra}, for a modern discussion.

Another frame which has been discussed in the literature is the hydrostatic frame. The hydrostatic frame is the frame in which the hydrostatic constitutive relations naturally appear after varying the generating function with respect to the sources \cite{Jensen:2012jh,Banerjee:2012iz}. 
To go from, say, the Landau frame, to the hydrostatic frame, one needs to redefine the velocity field and temperature field by appropriate higher derivative terms.

In the presence of a spin current one may carry out field redefinitions of the velocity field, temperature, and spin chemical potential,
\begin{equation}
	u^{\prime\,\mu} = u^{\mu} + \delta u^{\mu}\,,\qquad T' = T + \delta T\,,\qquad \mu^{\prime ab} = \mu^{ab} + \delta \mu^{ab}\,.
\end{equation}
Since the velocity and temperature are zeroth order in derivatives and the spin chemical potential is first order in derivatives then modifications in the velocity field and temperature, $\delta u^{\mu}$ and $\delta T$ can be of first or higher order in derivatives, and modifications to the spin chemical potential, $\delta \mu^{ab}$ can be of second or higher order in derivatives. 

Since the zeroth order in derivatives expression for the stress tensor, given in \eqref{E:idealexpansion}, is identical to that of a neutral, spinless, fluid, first order shifts in the temperature and velocity field can only affect the first order symmetric components of the stress tensor, much like a neutral spinless fluid. Modifications to the spin chemical potential will affect the second order and higher antisymmetric components of the energy momentum tensor (as well as second order components of the spin current). Given \eqref{E:leadingnh} (and assuming that $\sigma_m$ and $\sigma_M$ are non zero) we may always choose $\mu^{ab}$ such that, e.g., second order and higher contributions to the antisymmetric components of the energy momentum tensor vanish.

Taking the above considerations into account, 
we could define a Landau like frame as the frame where
\begin{align}
\begin{split}
\label{E:Landau}
	u_{\nu}T^{(\mu\nu)} =& -\epsilon_0 u^{\mu} \,, \\
	T^{[\mu\nu]} = &\sigma_A(T) A^{[\mu} u^{\nu]} + \sigma_m(T) \hat{m}^{[\mu}u^{\nu]} + \sigma_M(T) \hat{M}^{\mu\nu}\,.
\end{split}
\end{align}
The drawback of the Landau like frame in the current context is that useful features of the hydrostatic frame, such as the grouping of the contributions of certain expressions into BR terms, is somewhat obscure. 

Thus, let us consider a decomposition of the non hydrostatic ($nh$) terms into non hydrostatic BR terms ($nhBR$) and non hydrostatic non BR terms ($nhnBR$),
\begin{subequations}
\label{E:nhdecomposition}
\begin{align}
\begin{split}
	T_{nh}^{\mu\nu} =& T_{nhBR}^{\mu\nu} + T_{nhnBR}^{\mu\nu} \\
	S_{nh}^{\lambda\mu\nu} = & S_{nhBR}^{\lambda\mu\nu} + S_{nhnBR}^{\lambda\mu\nu}
\end{split}
\end{align}
where
\begin{equation}
	T_{nhBR}^{\mu\nu} =  {\frac{1}{2}} \mathring{\nabla}_{\lambda}   \left(
		S_{nhBR}{}^{\lambda\mu\nu} - S_{nhBR}{}^{\mu\lambda\nu} - S_{nhBR}{}^{\nu\lambda\mu}\right)\,,
\end{equation}
\end{subequations}
and the distinction between $S_{nhBR}$ and $S_{nhnBR}$ is somewhat superficial and can be chosen conveniently. As we will see shortly, up to the order we are working in, the choice $S_{nhnBR}^{\lambda\mu\nu}=0$ is possible, but other choices are also allowed. With this decomposition in mind, a slightly more useful frame which we will use in most of what follows is a hybrid frame which combines the Landau like frame above, the hydrostatic frame, and non hydrostatic BR terms, such that
\begin{align}
\begin{split}
\label{E:hybrid}
	u_{\mu}T^{(\mu\nu)} = &u_{\mu} (T_{h+}^{(\mu\nu)} + T_{nhBR}^{(\mu\nu)}) \,, \\
	T^{[\mu\nu]} = &\left(T_{h+}^{[\mu\nu]} + T_{nhBR}^{[\mu\nu]}\right) + \sigma_A(T) A^{[\mu} u^{\nu]} + \sigma_m(T) \hat{m}^{[\mu}u^{\nu]} + \sigma_M(T) \hat{M}^{\mu\nu}\, ,
\end{split}
\end{align}
where $T_{h+}$ was defined at the beginning of section \ref{S:nonhydrostatic}.
Of course, one can go from one frame to another using standard field redefinitions. In section \ref{S:conformal} we will see that, for a certain range of parameters, the hybrid frame above is incompatible with conformal invariance and homogenous scaling of the hydrodynamic variables. In that case it is more convenient to work with a hybrid conformal frame. We will discuss this frame further in section \ref{S:conformal} where it is more relevant.

\subsection{Tabulating the constitutive relations}
To obtain the most general constitutive relations in our hybrid frame we must tabulate all possible inequivalent non hydrostatic contributions to the stress tensor and current. While the operative procedure for such a classification has been discussed extensively in the literature, see for instance, \cite{Bhattacharyya:2012nq}, we will briefly outline it for completeness. 

To obtain, say, $m$'th order in derivatives constitutive relations for the stress tensor, we need to classify all possible $m$'th order rank 2 tensor structures, barring expressions which do not contribute in the Landau-like frame we are interested in. It is convenient to decompose these rank 2 tensors into traceless symmetric tensors transverse to the velocity field, antisymmetric tensors transverse to the velocity field, vectors transverse to the velocity field and scalars. To obtain the $m$'th order in derivative contributions to the spin current we need, in addition, $m$'th order transverse vectors, and $m$'th order rank three tensors whose last two indices are antisymmetric. We will refer to tensors of this type as tensors possessing spin symmetry.

The various scalars, vectors, and higher rank tensors which we need to construct can be further grouped into composite and non composite tensors. Composite tensors are those obtained from products of lower rank tensors, e.g., a composite rank 2 vector can be obtained by, say, taking a product of a rank 1 scalar and a rank 1 vector. Non composite tensors can not be reduced to a product of lower order tensors. Once we have all appropriate non composite tensors at a given order, it is a straightforward combinatorical task to generate the composite ones at the same order. (Note that one might need to classify higher rank non composite tensors of lower order for this purpose.)

Finally, since we are working in a derivative expansion, when enumerating tensor structures at order $m$ we may use the equations of motion up to that order to equate seemingly different tensor structures. Therefore, when constructing the constitutive relations we need to classify all appropriate composite and non composite, inequivalent (under the equations of motion) non hydrostatic tensor structures in the absence of torsion or curvature, to an appropriate order in derivatives. The main tensor structures required to do so can be found in tables \ref{T:allfirst} and \ref{T:allsfirst}.
\begin{table*}[hbt]
\begin{center}
		\begin{tabular}{| p{0.17\linewidth} |  c |  c |  p{0.17\linewidth} | p{0.2\linewidth}  | }
		\hline
		Tensor type & All data & EOM & Independent data & Independent non hydrostatic data \\
		\hline
		Scalars & $ u^\mu \mathring{\nabla}_\mu T , \theta, \kappa $ & $u_\nu D^\nu=0$ & $\theta,\kappa $ & $\theta$ \\
		\hline
		\multirow{2}{*}{Vectors} & $ \Delta^{\mu \nu} \mathring{\nabla}_\nu T + T a^\mu$, $\hat m^\mu$, & $\Delta^\mu_\nu D^\nu=0$ & \multirow{2}{*}{$m^\mu$, $k^\mu$, $\mathcal{K}_V{}^\mu$} &  \\ 
						    & $m^\mu$, $k^\mu$, $\mathcal{K}_V{}^\mu$ & $\Delta^{\mu}{}_{\alpha} u_{\beta} L^{\alpha\beta} = 0$ &  & \\
		\hline
		Symmetric & \multirow{2}{*}{$\sigma^{\mu\nu}$} & & \multirow{2}{*}{$\sigma^{\mu\nu}$} &\multirow{2}{*}{$\sigma^{\mu\nu}$} \\
		traceless & & & & \\
		\hline
		Antisymmetric & $M^{\mu\nu}$, $\hat{M}^{\mu\nu}$ & $\Delta^{\mu}{}_{\alpha} \Delta^{\nu}{}_{\beta} L^{\alpha\beta} = 0$ & $M^{\mu\nu}$  &  \\
		\hline
		Spin symmetry & $\mathcal{K}_A^{\mu\nu\rho}$, $\mathcal{K}_T^{\mu\nu\rho}$ & & $\mathcal{K}_A^{\mu\nu\rho}$, $\mathcal{K}_T^{\mu\nu\rho}$ & \\
		\hline
	\end{tabular}
	\caption{\label{T:allfirst} A list of all first order in derivative non composite data, the constraints generated by the equations of motion, and constraints from hydrostatic equilibrium. The expressions for $\hat{M}^{\mu\nu}$ and $\hat{m}^{\mu}$ are given in \eqref{E:defhMm} and $D^{\mu}$ and $L^{\mu\nu}$ denote energy conservation and angular momentum conservation respectively (e.g., $L^{\mu\nu} = \mathring{\nabla}_{\lambda}S^{\lambda}{}_{\mu\nu} - 2 T_{[\mu\nu]} - 2 S^{\lambda}{}_{\rho[\mu}e_{\nu]}{}^{a} e_{\rho}{}^{b} K_{\lambda}{}_{ ab}$, see \eqref{E:EOMs}.)}
\end{center}
\end{table*}

\begin{table*}[hbt]
\begin{center}
		\begin{tabular}{| c   |}
		\hline
		Composite 1st order spin symmetric data \\
		\hline
		$\sigma^{\mu[\rho}u^{\nu]}$, \, $\theta \Delta^{\mu[\rho}u^{\sigma]}$ \\
		\hline
	\end{tabular}
	\caption{\label{T:allsfirst} A list of all composite non hydrostatic first order data with spin symmetry.}
\end{center}
\end{table*}

From the above data we may now construct the most general $T_{nh}^{(\mu\nu)} = \frac{1}{2}\left( T^{\mu\nu}_{nh} + T^{\nu\mu}_{nh} \right)$ and $S_{nh}^{\lambda\mu\nu}$ to first order in derivatives, and $T_{nh}^{[\mu\nu]}$ to second order in derivatives.
We find, using the decomposition in \eqref{E:nhdecomposition},
\begin{align}
\begin{split}
\label{E:nhterms}
	S_{nhnBR}^{\lambda\mu\nu} &=  0 \,, \\
	S_{nhBR}^{\lambda\mu\nu} & = 2 \sigma_1 \sigma^{\lambda[\mu}u^{\nu]} + 2 \sigma_2 \theta \Delta^{\lambda[\mu}u^{\nu]} \,, \\
	T^{(\mu\nu)}_{nhnBR} &= - \zeta \theta \Delta^{\mu\nu} - \eta \sigma^{\mu\nu} \\
	T^{[\mu\nu]}_{nhnBR} & =   \sigma_A(T) A^{[\mu} u^{\nu]} + \sigma_m(T) \hat{m}^{[\mu}u^{\nu]} + \sigma_M(T) \hat{M}^{\mu\nu} 
\end{split}
\end{align}
As discussed earlier, the distinction between $S_{nhBR}^{\lambda\mu\nu}$ and $S_{nhnBR}^{\lambda\mu\nu}$ is somewhat arbitrary and we have chosen to set $S_{nhnBR}^{\lambda\mu\nu}=0$ for convenience. 

Putting together \eqref{E:idealexpansion}, \eqref{E:constitutive12} and  \eqref{E:nhterms}, and setting the torsion to zero, we find
\begin{align}
\begin{split}
\label{E:fullhybrid}
	S^{\lambda\mu\nu} & =  u^{\lambda} \left(4 \rho_m m^{[\mu}u^{\nu]} - 4 \rho_M M^{\mu\nu} \right) + 2 \chi^{(1)}_1 \Delta^{\lambda[\mu}u^{\nu]} \\
		& {}- 2 \chi_1^{(2)} M^{\lambda[\mu}u^{\nu]} 
		 + 2  \chi_2^{(2)} u^{\lambda}  M^{\mu\nu} 
		 - 2 \chi_3^{(2)} u^{\lambda}  u^{[\mu}m^{\nu]}  
		 + 4 \chi_4^{(2)} \Delta^{\lambda[\mu}m^{\nu]} \\  & + 2 \sigma_1 \sigma^{\lambda[\mu}u^{\nu]} + 2 \sigma_2 \theta \Delta^{\lambda[\mu}u^{\nu]} \,, \\
	T^{\mu\nu} & = \left(\epsilon_0+(\rho_m+T \rho_m') m_{\alpha}m^{\alpha} + (\rho_M+T \rho_M') M_{\alpha\beta}M^{\alpha\beta} \right) u^{\mu}u^{\nu} \\
	&+ (P_0+\rho_m m_{\alpha}m^{\alpha} + \rho_M M^{\alpha\beta}M_{\alpha\beta}) \Delta^{\mu\nu} + u^{\mu}m_{\alpha}M^{\alpha\nu}\left(2 \rho_m - 4 \rho_M\right) \\ & + \nabla_\lambda \bigg(\chi^{(1)}_1 \Delta^{\lambda[\mu}u^{\nu]}- 2 \chi_1^{(2)} M^{\lambda[\mu}u^{\nu]} 
		 + 2  \chi_2^{(2)} u^{\lambda}  M^{\mu\nu} 
		 - 2 \chi_3^{(2)} u^{\lambda}  u^{[\mu}m^{\nu]}  
		 + 4 \chi_4^{(2)} \Delta^{\lambda[\mu}m^{\nu]} \\
		 &+\sigma_1 \sigma^{\lambda[\mu}u^{\nu]} +  \sigma_2 \theta \Delta^{\lambda[\mu}u^{\nu]} - \lambda \leftrightarrow \mu  + \lambda \leftrightarrow \nu  \bigg) \\
 &- \zeta \theta \Delta^{\mu\nu} - \eta \sigma^{\mu\nu} +  \sigma_A(T) A^{[\mu} u^{\nu]} + \sigma_m(T) \hat{m}^{[\mu}u^{\nu]} + \sigma_M(T) \hat{M}^{\mu\nu} \,.
\end{split}
\end{align}
The full set of constitutive relations for the stress tensor and spin current, including torsion, can be found in appendix \ref{A:fullconstitutive}.

\subsection{The entropy current}
\label{S:entropycurrent}
One of the main constraints on the constitutive relations in hydrodynamic theory follows from a local version of the second law of thermodynamics. Following \cite{landau2013fluid} (see also \cite{Glorioso:2016gsa,Jensen:2018hhx,Haehl:2018uqv} for a modern treatment) we posit that there exists an entropy current $J_S^{\mu}$ satisfying 
\begin{equation}
\label{E:defJS1}
	J_S^{\mu} = s u^{\mu} + \mathcal{O}(\mathring{\nabla})
\end{equation}
with $s = \partial P / \partial T$ the entropy density, and
\begin{equation}
\label{E:defJS2}
	\mathring{\nabla}_{\mu}J_S^{\mu} \geq 0\,.
\end{equation}
Equation \eqref{E:defJS2} must be statisfied under the equation of motion and poses constraints on both the constitutive relations for the stress tensor and spin current and on the higher order corrections to the entropy current in \eqref{E:defJS1}. In all cases studied so far, \eqref{E:defJS1} and \eqref{E:defJS2} completely fix the entropy current to first order in a derivative expansion.

To implement \eqref{E:defJS2}, we first note that
\begin{equation}
\label{E:idealentropy}
	u_{\nu} \mathring{\nabla}_{\mu} T_{id}^{\mu\nu} + \frac{1}{2} \mu_{ab}\left( \mathring{\nabla}_{\lambda}S_{id}^{\lambda ab} - 2 T_{id}^{[ab]}\right) = -T \mathring{\nabla}_{\mu} \left( s u^{\mu} \right)
\end{equation}
implying that $s u^{\mu}$ is conserved for an ideal fluid. Put differently, the entropy current for an ideal fluid is given by the first term on the right hand side of \eqref{E:defJS1} and the inequality in \eqref{E:defJS2} is saturated. 

Motivated by \eqref{E:idealentropy} we define a canonical entropy current
\begin{equation}
	J_c^{\mu} = s u^{\mu} - \frac{u_{\nu}}{T} \left(T^{\mu\nu} - T_{id}^{\mu\nu}\right) - \frac{1}{2} \frac{\mu_{ab}}{T} \left(S^{\mu ab} - S_{id}^{\mu ab}\right)
\end{equation}
which satisfies
\begin{align}
\begin{split}
\label{E:divJc}
	\mathring{\nabla}_{\mu}J_c^{\mu} = &-\mathring{\nabla}_{\mu} \left(\frac{u_{\nu}}{T}\right) \left(T^{\mu\nu}-T_{id}^{\mu\nu}\right) 
	-\frac{1}{2} \mathring{\nabla}_{\lambda}\left(\frac{\mu_{ab}}{T} \right) \left(S^{\lambda ab}-S_{id}^{\lambda ab}\right)
	\\
	&- \frac{1}{2} \frac{u_{\nu}}{T} 
	\left( 
	\mathring{\nabla}_{\mu} \left(
		\mathring{\nabla}_{\lambda} \left( S^{\lambda\mu\nu}-S^{\mu\lambda\nu}-S^{\nu\lambda\mu}\right)
		-S^{\mu}{}_{\rho\sigma}K^{\nu\rho\sigma}\right) 
	+S_{\lambda\rho\sigma }\mathring{\nabla}^{\nu}K^{\lambda\rho\sigma}
	\right)
	\\
	& - \frac{\mu_{ab}}{T} \left( T^{ab}-T_{id}^{ab} + S^{\lambda}{}_{\rho}{}^{a} K_{\lambda}{}^{b\rho} \right)\,,
\end{split}
\end{align}
under the equations of motion. We define the full entropy current as
\begin{equation}
\label{E:entropydecomposition}
	J_S^{\mu} = J_c^{\mu} + J_{nc}^{\mu}
\end{equation}
where $J_{nc}^{\mu}$ is the most general current one can construct (within a derivative expansion) such that \eqref{E:defJS2} can be satisfied.

To proceed, it is convenient to insert the constitutive relations into \eqref{E:divJc} and expand them to second order in derivatives. We find that, under the equations of motion,
\begin{align}
\begin{split}
\label{E:divJc2}
	\mathring{\nabla}_{\mu} J_c^{\mu} = & 
		-\mathring{\nabla}_{\mu} \left(\frac{\chi_1^{(1)}}{T} \left(\theta u^{\mu} + \kappa u^{\mu} + \hat{m}^{\mu} \right) \right) \\
		&+  \frac{\zeta\theta^2}{T}  + \frac{\eta \sigma_{\mu\nu}\sigma^{\mu\nu}}{T} + \mathcal{O}(\nabla^3)\,.
\end{split}
\end{align}
In order for the full entropy current to be positive semidefinite we must take
\begin{align}
\begin{split}
\label{E:gotJnc}
	J_{nc}^{\mu} &= \frac{\chi_1^{(1)}}{T} \left(\theta u^{\mu} + \kappa u^{\mu} + \hat{m}^{\mu} \right)
\end{split}
\end{align}
and
\begin{equation}
\label{E:positivity}
	\zeta \geq 0,\qquad \eta \geq 0\,.
\end{equation}

The following comments are in order. First, we emphasize that the equality \eqref{E:divJc2} is satisfied only under the equations of motion. Had the equations of motion not been satisfied we would have obtained several additional terms on the right hand side of \eqref{E:divJc}, among them $\sigma_M \hat{M}_{\mu\nu} \hat{M}^{\mu\nu}$ and $\sigma_m \hat{m}_{\mu} \hat{m}^{\mu}$. Had the latter type of terms appeared we would have been forced to set $\sigma_M \geq 0$ and $\sigma_m \geq 0$. The fact that $\hat{M}_{\mu\nu}=0$ and $\hat{m}_{\mu}=0$ under the equations of motion implies that, at least to order $\mathcal{O}(\nabla^3)$ that we are working in, $\sigma_M$ and $\sigma_m$ are not constrained.

In fact, the structure of the canonical entropy current is such that the spin current of order $\mathcal{O}(\nabla)$ contributes to the entropy current at order $\mathcal{O}(\nabla^3)$ and therefore, can not be constrained by positivity of $\mathring{\nabla} \cdot J_S$ at the order we are working in.
To fully constrain all the transport coefficients associated with the terms in \eqref{E:nhterms} we need to compute $\mathring{\nabla}_{\mu}J_S^{\mu}$ to the next order in derivatives. While we will not carry out such an analysis here, we point out that, at least for fluids without spin, the constraints obtained from the hydrostatic partition function together with the first order constraints on positivity completely determine the entropy current \cite{Bhattacharyya:2013lha}. We believe that a similar analysis for fluids with spin will lead to the same result implying that the constraints obtained by the hydrostatic partition function together with \eqref{E:positivity} are the full set of constraints following from \eqref{E:defJS2}. We will return to this issue in a future publication.  

We also note that the entropy current obtained in \eqref{E:gotJnc} is unique. By requiring that \eqref{E:defJS1} and \eqref{E:defJS2} are valid in an arbitrary background geometry, we were forced to fix \eqref{E:gotJnc}. The non vanishing contribution of $J_{nc}^{\mu}$ to the entropy current is neccesary to  obtain \eqref{E:defJS1} and \eqref{E:defJS2}. Note also that it ensures that the total entropy current $J_S^{\mu}$ is independent of the choice of improvement terms once the torsion is set to zero.
This observation resolves some issues raised in previous work \cite{Fukushima:2020ucl} regarding the effect of improvement terms (sometimes referred to as pseudo-gauge transformations) on entropy production.

Apart from the constraints coming from \eqref{E:defJS2} and constraints coming from Onsager relations (which may be related to \eqref{E:defJS2}, see \cite{landau2013fluid}) it was shown in \cite{Jensen:2018hse} that there may exist extra constraints on transport coming from unitarity. To compute these additional constraints one would have to construct a Schwinger-Keldysh effective action for hydrodynamics with spin. This too, will be the topic of a future publication.

The full constitutive relations for a fluid with spin can be obtained by combining \eqref{E:idealconstitutive}, \eqref{E:S01}, \eqref{E:T12} and \eqref{E:nhterms}. For ease of reference, we have collected these terms together in equation (\ref{E:fullhybrid}) in the vanishing torsion limit, and in appendix \ref{A:fullconstitutive} in generality. But before ending this section let us comment on one particularly puzzling feature of the constitutive relations associated with $\chi_1^{(1)}$ in \eqref{E:T1}. A careful look at \eqref{E:T1} and \eqref{E:nhterms} reveals that the coefficient multiplying the shear tensor (usually referred to as the shear viscosity) is given by $-\chi_1^{(1)}-\eta$, and not just $-\eta$. Yet, it is only $\eta$ that contributes to entropy production and it is also only $\eta$ which is restricted to be positive (likewise, as we shall see in the next section, standard positivity constraints on correlation functions constrain $\eta$ and not $\chi_1^{(1)}$). Further, it is only $\eta$ which contributes to the equations of motion on account of terms associated with $\chi_1^{(1)}$ being of BR type. A similar observation can be made regarding $\chi^{(1)}_1$ and the bulk viscosity $\zeta$. At this point one might wonder whether $\chi_1^{(1)}$ contributes to any physical observable. In section \ref{S:linearized} we will show that it does.  In particular, it will contribute to the value of the stress tensor slightly out of equilibrium, and to stress tensor correlators in equilibrium.

\section{Linear response and Kubo formulas}
\label{S:linearized}
It is often convenient to relate the various coefficient functions of the constitutive relations to low momenta and low frequency correlation functions. Especially from an experimental standpoint. In what follows we will compute the Kubo formula for all coefficients which contribute to linear response theory. 

Recall that retarded correlation functions of the stress tensor in flat space can be obtained by varying the on-shell stress tensor with respect to a background metric. That is, the response of the on-shell stress tensor in the presence of a perturbative background metric relates to the retarded correlation function in the absence of such,
\begin{subequations}
\label{E:correlators}
\begin{equation}
	G^{\nu}{}_{b}{,}^{\mu}{}_{a} = \frac{\delta}{\delta e^{a}{}_{\mu}} |e| T^{\nu}{}_{b} \Big|_{e=\delta_0,\omega=\mathring{\omega}} \,,
\end{equation}
where by $e=\delta_0$ we mean that we evaluate the variation of the stress tensor on a flat background, and $\omega = \mathring{\omega}$ implies that we have set the torsion to zero.
Likewise,
\begin{align}
	G^{\nu}{}_{c}{,}^{\mu}{}_{ab}  &=  2 \frac{\delta}{\delta \omega_{\mu}{}^{ab}}  |e| T^{\nu}{}_{c}\Big|_{e=e_0,\omega=\mathring{\omega}}\,, \\
	G^{\lambda}{}_{ab}{,}^{\mu}{}_{c}  &=  \frac{\delta}{\delta e^{c}{}_{\mu}} |e| S^{\lambda}{}_{ab}\Big|_{e=e_0,\omega=\mathring{\omega}} \,, \\
	G^{\mu}{}_{cd}{ ,}^{\nu}{}_{ab}  & =2 \frac{\delta}{\delta \omega_{\nu}{}^{ab}} |e| S^{\mu}{}_{cd}\Big|_{e=e_0,\omega=\mathring{\omega}} \,.
\end{align}
\end{subequations}
Thus, contributions to the expectation value of the spin current and stress tensor which are linear in contorsion will contribute to correlation functions in a torsionless background and must therefore be included when classifying all possible constitutive relations for the stress tensor and current (which is the reason we have included them in our analysis). 

To evaluate the on shell stress tensor and spin current in the presence of a perturbed metric and perturbed contorsion tensor we must first solve the linearized hydrodynamic equations. The hydrostatic solution to the equations of motion \eqref{E:EOMs}, in a flat torsionless background is given by $u^{\mu} = u^{\mu}_0$, $T=T_0$ and $\mu^{ab} = 0$ where
\begin{equation}
\label{E:backgroundu0}
	u_0^{\mu} = (1,0)
\end{equation}
in Cartesian coordinates and with $T_0$ constant. To compute \eqref{E:correlators} we consider $u^{\mu} = u^{\mu}_0 + \delta u^{\mu}$, $T=T_0+\delta T$ and $\mu^{ab} = \delta \mu^{ab}$, and solve the linearized equations for $\delta u^{\mu}$, $\delta T$ and $\delta \mu^{ab}$ in a background geometry with a linearly perturbed metric $g_{\mu\nu} = \eta_{\mu\nu} + h_{\mu\nu}$ and linearly perturbed spin connection $\omega_{\mu}{}^{ab} = \mathring{\omega}_{\mu}{}^{ab} + o_{\mu}{}^{ab}$, where $\mathring{\omega}_{\mu}{}^{ab}$ is expanded to linear order in $h_{\mu\nu}$. These linearized equations take the schematic form
\begin{equation}
\label{E:linearizedEOM}
	B X = S
\end{equation}
where $B$ is a $d(d+1)/2 \times d(d+1)/2$ matrix, independent of the hydrodynamic sources, $\delta T$, $\delta u^{\mu}$ and $\delta\mu^{ab}$ or on the background perturbations $h_{\mu\nu}$ and $o_{\mu}{}^{ab}$, $X$ is a $d(d+1)/2$ dimensional vector of the hydrodynamic sources, and $S$ is a $d(d+1)/2$ dimensional vector composed of linear combinations of the background perturbations. By solving \eqref{E:linearizedEOM} and inserting the solution into the expressions for the stress tensor and current (equations \eqref{E:constitutivefull}, \eqref{E:genericid}, \eqref{E:generich}, and \eqref{E:genericnh}) we can compute their variations explicitly using \eqref{E:correlators} and obtain the associated correlation functions. Before doing so, and as an intermediate result, we can set $S=0$ on the right hand side of \eqref{E:linearizedEOM} and study propagating modes in the linearized hydrodynamic theory.

\subsection{An aside on sound modes}
Solving the linearized hydrodynamic equations often leads to the observation of sound modes or other propagating modes associated with an appropriate dispersion relation. Indeed, a non trivial solution to \eqref{E:linearizedEOM} with $S=0$ will exist only if $|B|=0$. Going to Fourier space with the conventions
\begin{equation}
\label{E:Fourierconventions}
	f(t,\vec{x}) = \int e^{-i \omega t + i \vec{k} \cdot \vec{x}} \hat{f}(\omega,\vec{k})d^3k d\omega\,,
\end{equation}
we find
\begin{equation}
\label{E:detB}
	|B| = \left(\sigma_M - 2 i \omega \rho_M\right) P_d(\omega\,,k)P_s(\omega\,,k)  \,,
\end{equation}
with $P_d$ and $P_s$ fourth order polynomials in $\omega$.
Solving $|B|=0$ reveals three gapped modes in addition to the more standard diffusion and sound modes present in fluids without spin. In what follows we will discuss these solutions in some detail. We comment that our analysis is restricted to configurations for which $\sigma_m \neq 0$ and $\sigma_M \neq 0$. If one of these coefficients vanishes the dynamical equations for the spin chemical potential will be modified, making this analysis moot. We will not discuss the dynamical spin limit in this work. 

At the derivative order that we are working in, the first term on the right hand side of \eqref{E:detB} is associated with gapped modes, $\omega=\omega_0$ with
\begin{equation}
\label{E:defw0}
	\omega_0  = i \frac{\sigma_M}{2  \rho_M}\,.
\end{equation} 
In this case the solution to the linearized equations is given by
\begin{equation}
\label{E:w0modes}
	\delta T =0,
	\qquad
	\delta u^i = 0,
	\qquad
	\delta m_i = 0,
	\qquad
	\delta M_{ij} = \epsilon_{i j \ell_1 \ldots \ell_{d-2}}  k^{\ell_1} \phi^{\ell_2 \ldots \ell_{d-2}}\,.
\end{equation}
In real space this is equivalent to $\delta M = * d \phi$ with $\phi$ a $d-3$ form and $*$ proportional to the hodge dual. Stability requires $\sigma_M \rho_M \leq 0$. 

The four solutions to $P_d=0$ are given by 
\begin{align}
\label{E:Pdmodes}
\omega &= \omega_0 - \frac{2i \sigma_M \rho_M \left(\rho_M \sigma_m + \rho_m \sigma_M \right) }{4 P'_0 T \rho_M^2 \sigma_m + \rho_m \sigma_M \left( \sigma_A \sigma_M T + 4 P'_0 \rho_M T - \sigma_m \sigma_M    \right)   } k^2 + \mathcal{O}(k^4) \\
	\omega_d &= -\frac{i \eta}{2 T_0 P_0'} k^2 + \mathcal{O}(k^3) \\
	\omega_{\pm} &=
	\frac{i T_0 P_0' }{T_0 \sigma_A -\sigma_m }
	\pm
	i\frac{\sqrt{T_0 \rho_m  P_0'  \left(T_0 \rho_m P_0' -T_0 \sigma_A  \sigma_m +\sigma_m^2\right)}}{T_0 \rho_m \sigma_A -\rho_m  \sigma_m }+ \mathcal{O}(k^2)\,,
\end{align}
where $\omega_0$ was defined in \eqref{E:defw0}. One can check that stability of $\omega_+$ and $\omega_-$ together requires $T_0 \sigma_A - \sigma_m <0$ and $\sigma_m \rho_m <0$.
The constraint \eqref{E:positivity} suggests that $P_0' \geq 0$ is a neccessary and sufficient condition for $\hbox{Im}(\omega_d) \leq 0$, and $\hbox{Im}(\omega_{\pm}) \leq 0$ poses stability constraints involving $\rho_m$, $\sigma_A$, $\sigma_m$ and derivatives of the pressure. 
The linearized solutions associated with $\omega=\omega_0 + \mathcal{O}(k^2)$ are the same as in \eqref{E:w0modes} up to $\mathcal{O}(k)$ corrections. 
The gapped modes associated $\omega_{\pm}$ are of the form
\begin{equation}
 	\delta T= \mathcal{O}\left(k^2 \right) \,,
 	\qquad
 	\delta m^i \propto \delta u^i \,,
 	\qquad
 	\delta M^{ij} \propto k^i \delta u^j - k^j \delta u^i \,,
\end{equation} 
and the modes associated with $\omega_d$ are the standard diffusion modes
\begin{equation}
	\delta T = \mathcal{O}(k^2)
	\qquad
	\delta u^{i} = \epsilon^{i j_1,\ldots j_{d-1}} k_{j_1} v_{j_2\ldots j_{d-1}} 
	\qquad
	\delta m^i = \mathcal{O}(k^3) 
	\qquad
	\delta M^{ij} = \mathcal{O}(k^2)\,.
\end{equation}
where $v_{j_2\ldots j_{d-1}}$ is a d-2 form. 

The four solutions to $P_s=0$ are given by $\omega_{s\,\pm}$ which satisfy 
\begin{equation}
	\omega_{s\,\pm} = \pm c_s k - i \frac{(d-1) \zeta+ (d-2) \eta}{2 (d-1) T_0 P_0'} k^2 + \mathcal{O}(k^3)\,,
\end{equation}
and $\omega_{\pm}$ which have been defined in \eqref{E:Pdmodes}. 
Stability of these modes does not provide additional constraints on the transport coefficients of the linearized theory. The modes associated with $\omega_{\pm}$ are identical to those associated with the $\omega = \omega_{\pm}$ solution to $P_d=0$, while the sound modes (associated with $\omega_{s\,\pm}$) are given by
\begin{align}
	\delta T = \left( \pm k c_s +  i k^2 \frac{(d-1) \zeta+ (d-2)\eta}{2 (d-1) T_0 P_0'}\right) T_0  \phi + \mathcal{O}(k^3)
	\qquad
	\delta u^{i} = k^i \phi + \mathcal{O}(k^2)
 \\  \nonumber
	\delta m^i = \mathcal{O}(k^2)
	\qquad
	\delta M^{ij} = \mathcal{O}(k^3)
\end{align}
where $c_s$ is the speed of sound
\begin{equation}
	c_s^2 = \frac{P_0'}{T_0 P_0''}\,.
\end{equation}

Thus, as advertised, we find that in spin hydrodynamics there exist gapped modes $\omega_0$ and $\omega_{\pm}$ in addition to the standard sound and diffusion modes, $\omega_{s\,\pm}$ and $\omega_d$ present in hydrodynamics without spin. The modes presented here would be in agreement with \cite{Hongo:2021ona} were $\sigma_A$ and $\sigma_m$ to vanish.

\subsection{Kubo formula}
Once we go to Fourier space, equations \eqref{E:linearizedEOM} become algebraic and can be solved analytically. It is then straightforward, though somewhat tedious to insert the solution back into the stress tensor and current and carry out the variation in \eqref{E:correlators} to obtain the appropriate Greens functions. Since the original equations of motion were valid to second order in a derivative expansion we can use the Greens function to obtain Kubo formulae for the transport coefficients which play a role in linearized hydrodynamics. Our results are summarized below.

After a somewhat long computation we find that the Kubo formula for the bulk and shear viscosity remain unchanged from those of a normal fluid,
\begin{align}
\begin{split}
\label{E:OldKubo}
		\zeta &= \lim_{\omega \rightarrow 0} \lim_{k \rightarrow 0} \frac{1}{\omega} \frac{ \text{Im}\left( G^{\mu \nu, \rho \sigma } \Delta_{(0)\mu \nu} \Delta_{(0)\rho \sigma} \right) } {\left(d-1 \right)^2 } \, , \\ 
	\eta &= \frac{2}{\left(d-2\right)\left(d+1 \right)} \lim_{\omega \rightarrow 0} \lim_{k \rightarrow 0} \frac{1}{\omega} \text{Im} \left( G^{\mu \nu ,\rho \sigma} \Delta_{(0)\mu \nu \rho \sigma} \right)\,.
\end{split}
\end{align}
Here the subscript $(0)$ specifies that we are evaluating expressions in the background solution \eqref{E:backgroundu0}. More explicitly,
\begin{align}
\begin{split}
	u_{(0)}^{\mu} &= (1,\vec{0})\,,\\
	\Delta_{(0)}^{\mu\nu} & = \eta^{\mu\nu} + u_{(0)}^{\mu} u_{(0)}^{\nu} \\
	\Delta^{\mu \nu \rho \sigma}_{(0)} &= \Delta^{\mu (\rho}_{(0)} \Delta^{\sigma) \nu}_{(0)} - \frac{1}{d-1} \Delta^{\mu \nu}_{(0)} \Delta^{\rho \sigma}_{(0)}\,.
\end{split}
\end{align}

A naive glance at the contribution of $\chi_1^{(1)}$ to the constitutive relations in \eqref{E:ThpBR} and \eqref{E:nhterms} might lead one to argue that the shear viscosity should be given by $\eta+\chi^{(1)}_1$ and that the bulk viscosity should also be modified in a similar manner. However the results in \eqref{E:OldKubo}, which are consistent with the results of the entropy current analysis \eqref{E:positivity} suggest that it is $\eta$ and $\zeta$ which specify the shear and bulk viscosity and not the modified expression mentioned above.
The Kubo formula for $\chi^{(1)}_1$ (defined in \eqref{E:W1W2}) can be obtained from the zero momenta correlator,
\begin{equation}
	4 \chi^{(2)}_9 - \frac{\left(\chi^{(1)}_1 \right)^2 }{2 T_{0} P'_0} = - \frac{1}{\left(d-1\right)\left(d-2\right)^2 } \lim_{\omega \rightarrow 0 } \lim_{k \rightarrow 0 } \text{Re}\left( G^{\lambda \rho \sigma, \kappa \alpha \beta } \right) \Delta_{\lambda \rho} \Delta_{\sigma \alpha} \Delta_{\kappa \beta}   \,,
\end{equation}
where $\chi^{(2)}_9$ is determined via \eqref{E:chi2list1} below.
The Kubo formula for $\chi^{(2)}_i$ defined in \eqref{E:W1W2} can be determined from the following zero frequency correlation functions:
\begin{align}
 \begin{split}
 	\chi^{(2)}_1 + \chi^{(2)}_{10} &= - \frac{2}{\left(d-2 \right)\left(d-3\right)}  \lim_{k \rightarrow 0} \lim_{\omega \rightarrow 0} \text{Re}\left( G^{\lambda \rho \sigma, \kappa \alpha \beta} \Pi_{(0)\lambda \alpha} \Pi_{(0)\rho \beta} u_{(0)\sigma} u_{(0)\kappa} \right) \, , \\ 
 		\chi^{(2)}_2 + 2\chi^{(2)}_7 &= - \lim_{k \rightarrow 0} \lim_{\omega \rightarrow 0} \frac{1}{\left(d-2 \right)k^2}  \text{Im}\left( G^{\lambda \rho \sigma, \mu \nu} \Pi_{(0)\rho \nu} u_{(0)\lambda} k_{(0)\sigma} u_{(0)\mu} \right)  \,, \\
 	\chi^{(2)}_3 + 2 \chi^{(2)}_6 &= - \lim_{k \rightarrow 0} \lim_{\omega \rightarrow 0} \frac{1}{k^2} \text{Im}\left( G^{\lambda \rho \sigma,\mu \nu} u_{(0)\lambda} u_{(0)\rho} k_\sigma u_{(0)\mu} u_{(0)\nu}  \right) \, , \\
 	\chi^{(2)}_4 + \chi^{(2)}_8 &= \lim_{k \rightarrow 0} \lim_{\omega \rightarrow 0} \frac{1}{2 \left(d-2 \right)k^2} \text{Im} \left( G^{\lambda \rho \sigma, \mu \nu} \Pi_{(0)\mu \nu} u_{(0)\lambda} u_{(0)\rho} k_{(0)\sigma} \right) \, , \\
 	\chi^{(2)}_6 + \chi^{(2)}_3 - \rho_m &= \lim_{k \rightarrow 0} \lim_{\omega \rightarrow 0} \frac{1}{k^2} \text{Re} \left( G^{\lambda \rho \sigma, \kappa \alpha \beta} u_{(0)\lambda} u_{(0)\kappa} u_{(0)\rho} k_\sigma u_{(0)\alpha} k_\beta \right) \, , \\
 	\chi^{(2)}_7 + \chi^{(2)}_2 + \rho_M &= \lim_{k \rightarrow 0} \lim_{\omega \rightarrow 0} \frac{1}{2 \left(d-2 \right) k^2} \text{Re} \left( G^{\lambda \rho \sigma, \kappa \alpha \beta} \Pi_{(0)\rho \alpha} u_{(0)\lambda} k_{(0)\sigma} u_{(0)\kappa} k_\beta \right)
\end{split}
\end{align}
and
\begin{align}
\begin{split}
\label{E:chi2list1}
  	\chi^{(2)}_8 &= \lim_{k \rightarrow 0} \lim_{\omega \rightarrow 0} \frac{1}{2 \left(d-2\right) k^2} \text{Im}\left( G^{\lambda \rho \sigma, \mu \nu} \Pi_{(0)\lambda \rho} k_\sigma u_\mu u_\nu \right) \, , \\
  	\chi^{(2)}_9 &= \lim_{k \rightarrow 0} \lim_{\omega \rightarrow 0} \frac{1}{8 \left(d-2\right)^2 k^2} \text{Im}\left( G^{\lambda \rho \sigma, \mu \nu} k_\rho \Pi_{(0)\lambda \sigma} \Pi_{(0)\mu \nu} \right)  \, , \\
 	\chi^{(2)}_5 + \frac{ \left(d-2\right) \chi^{(2)}_{12}}{d -1} 
		&=  -\lim_{k \rightarrow 0} \lim_{\omega \rightarrow 0} \frac{1}{k^4} \text{Im} \left( G^{\lambda \rho \sigma, \mu \nu} k_\lambda u_{(0)\rho} k_\sigma u_{(0)\mu} k_\nu \right)  \, , \\ 
 	 	\chi^{(2)}_5 - \frac{\chi^{(2)}_{12}}{d-1} &= \lim_{k \rightarrow 0} \lim_{\omega \rightarrow 0} \frac{\left(d-1\right)}{2\left(d-2\right)k^2} \text{Im}\left( G^{\lambda \rho \sigma, \mu \nu} \Pi_{(0)\lambda \rho} u_{(0)\sigma} u_{(0)\mu} k_\nu  \right) \, , \\
 	\chi^{(2)}_{12} - \frac{\chi^{(2)}_{10}}{2} &= \lim_{k \rightarrow 0} \lim_{\omega \rightarrow 0} \frac{1}{k^2}\left[ \frac{2 \text{Re}\left( G^{\mu \nu, [\rho \sigma]} u_{(0)\mu} u_{(0)\sigma} \Pi_{(0)\nu \rho}  \right) }{d-2} - \epsilon \right] \, , \\
 	\chi^{(2)}_{11} + \frac{\chi^{(2)}_{10}}{2} &=  \lim_{k \rightarrow 0} \lim_{\omega \rightarrow 0} \frac{1}{k^2} \left[ \frac{2 \text{Re}\left( G^{\mu \nu, \rho \sigma} u_{(0)\nu} u_{(0)\sigma} \Pi_{(0)\mu \rho} \right) }{d-2}  + P \right] \, , \\
 	\chi^{(2)}_9 + \frac{d-3}{d-2} \frac{\chi^{(2)}_{14}}{2} &= \lim_{k \rightarrow 0} \lim_{\omega \rightarrow 0} \frac{1}{8k^2} \left[ \frac{ 2 \text{Re} \left(G^{\mu \nu, [\rho \sigma]} k_\nu k_\sigma \Pi_{(0)\mu \rho} \right) }{d-2} - P \right] \, , \\
 	\chi^{(2)}_{14} + \frac{\chi^{(2)}_{13}}{2} &=  \lim_{k \rightarrow 0} \lim_{\omega \rightarrow 0} \frac{1}{k^2} \left[ \frac{3 \text{Re}\left( G^{\mu \nu, [\rho \sigma]} \Pi_{(0)\mu \rho} \Pi_{(0)\nu \sigma} \right) }{2 \left(d-2\right)\left(d-3\right)}  - \frac{3}{4} P \right] \, , \\
\end{split}
\end{align}
and $\rho_m$ and $\rho_M$ are given by
\begin{align}
\begin{split}
	2 \left(d-2\right)\left(d-1 \right) \left( \rho_m + \chi_7^{(2)} + \chi^{(2)}_2 \right) &= \lim_{\omega \rightarrow 0} \lim_{k \rightarrow 0} \text{Re} \left( G^{\lambda \rho \sigma, \kappa \alpha \beta} u_{(0)\lambda} u_{(0)\kappa} \Delta_{(0)\rho \alpha} \Delta_{(0)\sigma \beta} \right)  \, , \\
	 \left(d-1\right)\left[\frac{ \left(\chi^{(1)}_1\right)^2}{2 \frac{\partial P_0}{\partial T} T} +  \rho_M - \chi_6^{(2)} - \chi^{(2)}_3    \right] &= \lim_{\omega \rightarrow 0} \lim_{k \rightarrow 0} \text{Re} \left( G^{\lambda \rho \sigma, \kappa \alpha \beta} u_{(0)\lambda} u_{(0)\kappa} u_{(0)\rho} u_{(0)\alpha} \Delta_{(0)\sigma \beta} \right) \, . \\ 
\end{split}
\end{align}
It may seem somewhat surprising that we need to consider zero momentum correlators to get a handle over transport coefficients generated by the hydrostatic generating function. We believe that had we had control over the second order symmetric components of the stress tensor, we would have been able to obtain zero frequency Kubo formula for $\rho_m$ and $\rho_M$.

Finally, the Kubo formula for the non hydrostatic terms $\sigma_1$ and $\sigma_2$ are given by
\begin{align}
\begin{split}
		\sigma_1 - 2 \chi^{(2)}_{12} &= -\frac{2}{ \left(d-2 \right) \left( d+1 \right) } \lim_{\omega \rightarrow 0} \lim_{k \rightarrow 0} \frac{1}{\omega}  \text{Im} \left( G^{\lambda \rho \sigma, \mu} u_{(0)\rho} \Delta_{(0)\lambda \sigma \mu \nu} \right) \\
		\sigma_2 - 2 \chi^{(2)}_5 + \left( \frac{\partial \chi^{(1)}_1 }{\partial s} \right) \left(\frac{\partial \chi^{(1)}_1 }{\partial T}  \right) &=- \lim_{\omega \rightarrow 0} \lim_{k \rightarrow 0}  \frac{1}{\omega} \text{Im} \left( G^{\lambda \rho \sigma, \mu \nu} \Delta_{(0)\rho \lambda}u_{(0)\sigma} \Delta_{(0)\mu \nu} \right) \, . 
\end{split}
\end{align}

At the order we are working in, there are no Kubo formula for $\sigma_m$, $\sigma_M$ and $\sigma_A$. This is a result of the algebraic equations of motion, \eqref{E:dynamichatM} which imply that the stress tensor is insensitive to these coefficients under the equations of motion. The same reasoning doesn't allow us to obtain constraints on $\sigma_m$ and $\sigma_M$ as we now show.

Kubo formulas can be used to constrain transport coefficients. As far as we are aware these constraints always match the constraints obtained from requiring positivity of the entropy current. One of the main observations used in this context is that for any Hermitian operator $O$ the imaginary part of the retarded function $G_{OO}(\omega,k)$ should be positive for $\omega>0$
\begin{align}\label{HermitianCond}
   Im G_{OO}(\omega,k) \geq 0\,.
\end{align}
From \eqref{HermitianCond} it follows that
\begin{align}
\begin{split}
	\hbox{Im}G^{ii,ii} & \geq 0 \quad \hbox{no sum on } i \,, \\
	\hbox{Im}G^{12,12} &\geq 0 \quad i\neq j \hbox{ and no sum on }i,j\,,
\end{split}
\end{align}
which implies that
\begin{equation}
	\eta \geq 0
	\qquad
	\zeta \geq 0\,,
\end{equation}
matching the entropy current analysis \eqref{E:positivity}. 

It is not possible to extract similar information from the spin current without access to higher order terms in the derivative expansion. For instance, the non hydrostatic second order contributions to the spin current, parameterized by $\lambda_1$ and $\lambda_3$,
\begin{align}
    S^{\lambda \mu \nu}_{(2)}&= \ldots + u^\lambda \left[ u^{[\mu} \Delta^{\nu]}_\rho \mathcal{L}_{\frac{u}{T}} \lambda_1   k^\rho + \Delta^{[\mu}_\rho \Delta^{\nu]}_\sigma \mathcal{L}_{\frac{u}{T}} \lambda_3 K^{\rho \sigma}  \right] + ...
\end{align}
will lead to
\begin{equation}\label{KuboSpin1}
    -\lim_{\omega\to0} \lim_{k\to 0} \frac{1}{\omega}  \text{Im} \left( G^{\lambda \mu \nu, \kappa \rho \sigma}(\omega,0) u_{(0)\lambda} u_{(0)\kappa} \Delta_{(0) \mu[\rho}\Delta_{(0)\sigma]\nu} \right) =- \frac{1}{(d-1)^2} \left( \frac{\left(\chi^{(2)}_2 - 2 \rho_M\right)^2}{\sigma_M}+\lambda_3\right) 
\end{equation}
and
\begin{multline}
    -\lim_{\omega\to0} \lim_{k\to 0} \frac{1}{\omega}  \text{Im} \left( G^{\lambda \mu \nu, \kappa \rho \sigma}(\omega,0) u_{(0)\lambda} u_{(0)\kappa} u_{(0)[\mu}\Delta_{(0) \nu][\rho}u_{(0)\sigma]} \right)
    \\
     =\frac{1}{d-1}\left(  \frac{\left(2 \rho_m + \chi^{(2)}_3 \right) \left(2 T \left(\chi^{(1)} \sigma_A + s \left(2 \rho_m + \chi^{(2)}_3 \right) \right)  - \sigma_m \chi^{(1)}   \right) }{s \sigma_m T}-\lambda_1 \right)\,.
\end{multline}

Thus, without including the additional factors of $\lambda$ we might have erroneously concluded that $\sigma_M >0$, or if $\sigma_A=0$ then $\sigma_m>0$. That the sign of $\sigma_m$ and $\sigma_M$ is undetermined, is compatible with the entropy current analysis. We note that equation \eqref{KuboSpin1} with $\chi^{(2)}_2 =0$, $\rho_M =0$, and $\lambda_3=0$ was obtained in \cite{Hongo:2021ona}. 

It can be shown that a generic combination of the stress tensor, $T^{\mu\nu}$, and spin current, $S^{\lambda\mu\nu}$, which we schematically write out as $O=a_1 T + a_2 S$ will not give any  additional constraints on transport, again, compatible with our entropy current analysis.

\section{Charged fluids}
\label{S:charge}

Often one is interested in fluids which conserve charge in addition to energy and momentum. In this case, the chemical potential, $\mu$, conjugate to the charge serves as an additional (scalar) hydrodynamic degree of freedom whose equation of motion is determined by charge current conservation. In this section we will extend our analysis to include an Abelian conserved current. Our analysis will not be as comprehensive as that carried out for uncharged fluids. We will mainly be interested in the constitutive relations for the stress tensor, spin current and charge current in the absence of torsion. We leave a full analysis of the Kubo formula associated with a charge current to future work.

Let us start with the conservation equations for energy, angular momentum and current. In place of \eqref{E:variation} we now have
\begin{equation}
\label{E:variation2}
	\delta S = \int d^dx |e| \left(T^{\mu}{}_{a} \delta e^{a}{}_{\mu} + \frac{1}{2} S^{\mu}{}_{ab} \omega_{\mu}{}^{ab} + J^{\mu} \delta A_{\mu}+ E \cdot \delta \phi\right)
\end{equation}
with $A_{\mu}$ an external gauge field associated with the conserved charge generated by $J^{\mu}$. (We will be using $A_{\mu}$ to denote the gauge field and also the particular combination of acceleration and temperature derivatives c.f., table \ref{T:antisymmfirst}. We hope that the reader will be able to distinguish the two from context.) Following the analysis in section \ref{S:Canonicalspincurrent} the resulting equations of motion are given by 
\begin{align}
\begin{split}
\label{E:fullcharged}
	\mathring{\nabla}_\mu T^{\mu \nu} &=F^{\nu \rho} J_\rho   +\frac{1}{2} R^{\rho \sigma \nu \lambda}   S_{\lambda \rho \sigma}  - T_{\rho \sigma} K^{\rho \sigma \nu} \, , \\
 	\mathring{\nabla}_\lambda S^\lambda_{\hphantom{\lambda}\mu \nu} &= 2 T_{[\mu \nu]} - 2 S^\lambda_{\hphantom{\lambda}\rho [\mu}K^\rho_{\hphantom{\rho} \nu] \lambda}\, , \\
 	\mathring{\nabla}_\rho J^\rho &= 0
\end{split}
\end{align}
where $F_{\mu\nu} = \partial_{\mu}A_{\nu} - \partial_{\nu} A_{\mu}$ 
(in the absence of anomalies). Equations \eqref{E:fullcharged} are a combination of \eqref{E:Jouleheating} and \eqref{E:EOMs}. 

As before, we separate the constitutive relations into two categories. Those constitutive relations that come from hydrostatics, and non hydrostatic constitutive relations. That is, in addition to \eqref{E:ourdecomposition} we have
\begin{equation}
	J^{\mu} = J_h^{\mu} + J_{nh}^{\mu}\,,
\end{equation}
where
\begin{equation}
\label{E:dWdmu}
	J_{h}^{\mu} = \frac{1}{|e|} \frac{\partial W}{\partial A_{\mu}}\,.
\end{equation}

Consider the hydrostatic sector. In addition to the vielbein and spin connection, we have added an external gauge field $A_{\mu}$ which couples to the charge current which, in hydrostatic equilibrium, is invariant under the timelike Killing vector, $V^{\mu}$, the Lorentz parameter $\theta_V{}^{a}{}_{b}$, and, a gauge parameter $\Lambda_V$ such that
\begin{equation}
\label{E:timeindependencemu}
	0 = \pounds_V A_{\mu} + \partial_{\mu} \Lambda_V\,.
\end{equation}
(Compare with \eqref{E:timeindependence}.) As in \eqref{E:Wini} we wish to construct a generating function $W$ from $A_{\mu}$ (and the vielbein and spin connection) and from $\Lambda_V$ (and $V^{\mu}$ and $\theta_V{}^a{}_b$) and their derivatives, which is coordinate, Lorentz, and gauge invariant. The relations \eqref{E:deltaVtheta} remain unchanged since gauge transformations do not affect the vielbein and spin connections. In addition to it, we have
\begin{equation}
	\delta \Lambda_V =  \pounds_{\xi} \Lambda_V - \pounds_V \Lambda \,.
\end{equation}
Since
\begin{equation}
	\mu = \frac{V^{\mu}A_{\mu} + \Lambda_V}{\sqrt{-V^2}}
\end{equation}
is a Lorentz and gauge invariant scalar ($\delta A_{\mu} = \pounds_{\xi} A_{\mu} + \partial_{\mu} \Lambda$), it is convenient to use it in the generating function in place of $\Lambda_V$. Similar to the situation with $u^{\mu}$, $T$ and $\mu^{ab}$ which were defined in \eqref{E:defthermal}, $\mu$ will be interpreted as the chemical potential in hydrostatic equilibrium. 

Thus, we should extend $\mathcal{W}$ in \eqref{E:hydrostatic2} to include gauge invariant contributions coming from $A_{\mu}$ and $\mu$. At zero order in derivatives this implies that the pressure term $P$ in \eqref{E:hydrostatic2} will also depend on $\mu$. At first order in derivatives we find that, for a generic number of dimensions, there are no further first order in derivative contributions to $\mathcal{W}$, but $\chi^{(1)}_1$ may depend on $\mu$ in addition to its dependance on $T$. At second order in derivatives we find that there are six additional scalars that can contribute to $\mathcal{W}$ which will affect the antisymmetric components of the stress tensor at second order in derivatives. These scalars can be found in table \ref{T:U1scalars}. 
\begin{table}[hbt]
\begin{center}
		\begin{tabular}{| c   c    c    | }
		\hline
		\multicolumn{3}{| l |}{Zeroth order hydrostatic scalars in the presence of charge} \\
		\hline
			$\mu$ &  &  \\
		\hline
		\multicolumn{3}{| l |}{Order two hydrostatic scalars in the presence of charge} \\
		\hline
			$\tilde{S}_{(1)} = E_\mu m^\mu$ &
			$\tilde{S}_{(2)} = E_\mu k^\mu$ &
			$\tilde{S}_{(3)} = E_\mu \left( \mathcal{K}_V \right)^\mu$ 
			 \\
			$\tilde{S}_{(4)} = B_{\mu \nu} M^{\mu \nu}  $ &
			$\tilde{S}_{(5)} =B_{\mu\nu}K^{\mu\nu}$ &
			$\tilde{S}_{(6)} =B_{\mu \nu}\kappa_A{}^{\mu\nu}$ \\
		\hline
	\end{tabular}
	\caption{\label{T:U1scalars} Second order hydrostatic scalars which are associated with a $U(1)$ charge and will contribute to the stress tensor and current at zero and first order in derivatives and to the antisymmetric components of the stress tensor at second order in derivatives. Here $E^{\mu} = F^{\mu\nu}u_{\nu}$ and $B_{\mu\nu} = \Delta_{\mu\rho}\Delta_{\nu\sigma} F^{\rho\sigma}$.}
\end{center}
\end{table} 
Using the scalars in table \ref{T:U1scalars} and \ref{T:allscalarsnoK}, and omitting those contributions which vanish in the absence of torsion, we have
\begin{equation}
\label{E:chargedW}
	\mathcal{W} = P + \chi_1^{(1)} \kappa + \sum_{i=1}^4 \chi_i^{(2)} S_{(i)} + \sum_{i=1}^6 \tilde{\chi}_i^{(2)} \tilde{S}_{(i)}\,.
\end{equation}
Where now $P=P(T,\,\mu,\,m_{(0)},\,M_{(1)})$, and $\chi^{(i)}_j = \chi^{(i)}_j(T,\mu)$.
With the generating function $\mathcal{W}$ at hand we can use \eqref{E:defTS} and \eqref{E:dWdmu}
to obtain the constitutive relations in hydrostatic equilibrium. The explicit form of the constitutive relations generated by the generating function in hydrostatic equilibrium have been collected in appendix \ref{A:fullconstitutive}.

Out of equilibrium we can add those scalars, vectors and tensors which vanish in hydrostatics due to the stationarity conditions \eqref{E:timeindependence} and \eqref{E:timeindependencemu}. We find that, in addition to the first order scalars, vectors, and tensors in table \ref{T:allfirst}, there is an additional scalar $u^{\alpha} \mathring{\nabla}_{\alpha} \mu$ which is not an independent scalar under the equations of motion, and one additional independent vector, $T\mathring{\nabla}_{\mu} \left(\frac{\mu}{T} \right) - E_{\rho}$ with $E^{\mu} = F^{\mu\nu}u_{\nu}$. Thus, out of equilibrium, we have one additional contribution to the second order antisymmetric stress tensor, one tensor with spin current symmetry and one vector contribution to the charge current, see table \ref{T:chargecontribution}. The full constitutive relations for the stress tensor, spin current and charge current can be found in appendix \ref{A:fullconstitutive}.
\begin{table}[hbt]
\begin{center}
		\begin{tabular}{| c  |  c  |   c    | }
		\hline
		2nd order antisymmetric tensor & 1st order antisymmetric tensor & 1st order vector \\
		\hline
			$u^{[\mu} \Delta^{\nu \rho}\left( T \mathring{\nabla}_\rho \left(\frac{\mu}{T} \right) -E_\rho  \right)$ & $u^\lambda u^{[\mu} \Delta^{\nu] \rho} \left( T \mathring{\nabla}_\rho \left(\frac{\mu}{T} \right) -E_\rho  \right)$   & $ \Delta^{\mu \rho}\left(T  \mathring{\nabla}_\rho \left(\frac{\mu}{T} \right) -E_\rho  \right)$ \\
		\hline
	\end{tabular}
	\caption{  \label{T:chargecontribution} All contributions to the constitutive relations at subleading order in derivatives which vanish in the absence of a $U(1)$ charge current. }
\end{center}
 \end{table} 

An interesting feature of hydrodynamics with spin and charge is that the hydrostatic condition  $\hat{m}^{\mu}=0$ (c.f., \eqref{E:hattedM}) is no longer satisfied under the full equations of motion. Instead, using \eqref{E:nhcharged}, we find 
\begin{equation}
\label{E:hatmwithcharge}
	\hat{m}^{\mu} = \left(  \frac{\tilde{\sigma}_A}{\frac{\frac{\partial P_0}{\partial T}}{\frac{\partial P_0}{\partial \mu}} + \frac{\mu}{T}}  + \tilde{\sigma}_e \right)\Delta^{\nu\rho} \left(T\mathring{\nabla}_{\rho} \left(\frac{\mu}{T}\right) - E_{\rho}\right) \,,
\end{equation}
(compare with \eqref{E:dynamichatM}). 
The dynamical equation $\hat{M}^{\mu\nu}=0$ is unchanged in the presence of charge.

\section{Conformal fluids }
\label{S:conformal}
In many instances the structure of the stress tensor and spin current is further constrained by symmetries. Since conformal invariance is often associated with fixed points of RG flow, it is sometimes particularly useful to consider constraints on transport which follow from conformal symmetry.

Conformal transformations includes all those coordinate transformations which scale the Minkowski metric by an overall multiplicative function. It is always possible to place a conformally invariant theory on a curved manifold such that the resulting dynamics are Weyl invariant \cite{Iorio:1996ad}. Infinitesimal Weyl transformations scale the vielbein by an overall mutliplicative function which we denote by $\phi$,
\begin{equation}
\label{E:deltaWvielbein}
	\delta_W e^{a}{}_{\mu} = \phi e^a{}_{\mu}\,.
\end{equation}
In what follows infinitesimal Weyl transformations of fields which result in an overall multiplicative constant will be referred to as homogenous. Tensors which transform homogenously under Weyl transformations will be referred to as Weyl covariant tensors. Note that the transformation \eqref{E:deltaWvielbein} implies that the ringed connection transforms inhomogenously,
\begin{equation}
\label{E:deltaWringomega}
	\delta_W \mathring{\omega}_{\mu}{}^{ab} = 2 e^{[a}{}_{\mu} e^{b]\nu} \partial_{\nu}\phi\,.
\end{equation}

If we take a conformally invariant theory and place it in a background with non trivial curvature and with non trivial torsion, the resulting dynamics will still be Weyl invariant as long as the spin connection transforms as the ringed spin connection, up to an overall multiplicative constant which we parameterize by $q$ \cite{Buchbinder:1985ux,Shapiro:2001rz},
\begin{equation}
\label{E:deltaWomega}
	\delta_W {\omega}_{\mu}{}^{ab} =2  (1-q) e^{[a}{}_{\mu} e^{b]\nu} \partial_{\nu}\phi\,.
\end{equation}
Equation \eqref{E:deltaWomega} implies, via \eqref{E:contorsion}, that the contorsion tensor satisfies
\begin{equation}
\label{E:deltaWK}
	\delta_W K_{\mu}{}^{ab} = - q 2 e^{[a}{}_{\mu} e^{b]\nu} \partial_{\nu}\phi \,.
\end{equation}

Thus, if our action is invariant under \eqref{E:deltaWvielbein} and \eqref{E:deltaWK} it will be invariant under Weyl transformations. This suggests that in a flat torsionless background the theory will be conformal invariant for $q=0$. If $q \neq 0$ it seems that $K_{\mu}{}^{ab}$ will transform inhomogenously under conformal transformations implying that a torsionless background will transform into a torsionfull background under conformal transformations. In what follows we will entertain the possibility of arbitrary $q$, but keep in mind that $q=0$ is more relevant for the physical setting we have in mind. 

If we manage to construct a Weyl invariant action then the stress tensor, derived from varying the action with respect to the vielbein, must satisfy
\begin{equation}
\label{E:tracelessness}
	T^{\mu}{}_{\mu}  = (1-q) \mathring{\nabla}_{\mu} S_{\lambda}{}^{\lambda\mu}\,,
\end{equation}
and Weyl transformations of the stress tensor and spin current read
\begin{align}
\begin{split}
\label{E:deltaTS}
	\delta_W T^{\mu\nu} &= -(d+2) \phi T^{\mu\nu} + (q-1) S^{\mu\nu\rho} \partial_{\rho}\phi - (q-1) S_{\lambda}{}^{\lambda\mu} \mathring{\nabla}^{\nu} \phi \\
	\delta_W S^{\lambda\mu\nu} &= -(d+2) \phi S^{\lambda\mu\nu}\,.
\end{split}
\end{align}
Thus, our strategy for constructing the conformally invariant constitutive relations will be to ensure that \eqref{E:tracelessness} and \eqref{E:deltaTS} are satisfied.
We note in passing that in a theory with $q=0$ we may recover the standard tracelessness condition and Weyl rescaling relations upon an appropriate BR transformation.

Let us start by constructing constitutive relations which transform homogenously under Weyl rescalings. Once such constitutive relations are available it is straightforward to construct Weyl covariant tensors, such as the spin current, from them. Constructing the energy momentum tensor, which is not Weyl covariant, requires some more work, as we shall see shortly.

To construct constitutive relations which transform homogenously under Weyl rescalings, we generalize the techniques of \cite{Loganayagam:2008is}. Consider the zeroth order in derivative hydrodynamic variables, $T$ and $u^{\mu}$.
Using the hydrostatic relations \eqref{E:defthermal} and the scaling of the vielbein and spin connection \eqref{E:deltaWvielbein} and \eqref{E:deltaWomega} we find that 
\begin{equation}
\label{E:deltaWhydrozero}
	\delta_W T = -\phi T\,,
	\qquad
	\delta_W u^{\mu} = - \phi u^{\mu}\,.
\end{equation}
Thus, zero order in derivative terms will always scale homogenously under Weyl transformations. 

At first order in derivatives we can construct tensors which contain explicit derivatives of the zero'th order hydrodyhamic variables, and, in addition, the various components of the contorsion tensor and the spin connection.  We have collected all relevant first order tensors and their transformation laws under Weyl rescalings in table \ref{T:Weyl}.
\begin{table}[hbt!]
\begin{center}
		\begin{tabular}{| l |  l | }
		\hline
		Hydrodynamic variables & Background torsion \\
		\hline
		$\delta_W m_{\alpha} = (1-q) \Delta_{\alpha}{}^{\beta} \partial_{\beta}\phi$ & $\delta_W k_{\mu} = -q \Delta_{\mu}{}^{\nu} \partial_{\nu} \phi$	\\
		$\delta_W M^{\alpha\beta} = -3 \phi M^{\alpha\beta}$ &				    $\delta_W K^{\mu\nu} = -3\phi K^{\mu\nu}$	\\
		$\delta_W \theta = -\phi \theta +(d-1) u^{\alpha}\partial_{\alpha}\phi$ &	    $\delta_W \kappa = - \phi \kappa - q(d-1)u^{\alpha}\partial_{\alpha}\phi$	\\
		$\delta_W a_{\mu} = \Delta_{\mu}{}^{\nu} \partial_{\nu} {\phi}$ &			    $\delta_W \kappa_S{}^{\mu\nu} = -3 \phi \kappa_S{}^{\mu\nu}$	\\
		$\delta_W \Omega^{\mu\nu} = -3 \phi \Omega^{\mu\nu}$ &			    $\delta_W \kappa_A{}^{\mu\nu} = -3 \phi \kappa_A{}^{\mu\nu}$ 	\\
		$\delta_W \sigma^{\mu\nu} = -3 \phi \sigma^{\mu\nu}$ &				    $\delta_W \mathcal{K}_{V\,\mu} = -2 q (d-2) \Delta_{\mu}{}^{\nu} \partial_{\nu} \phi$ \\
		$\delta_W u^{\mu}\partial_{\mu}T = -2 \phi u^{\mu}\partial_{\mu}T - T u^{\mu}\partial_{\mu}\phi$	& $\delta_W \mathcal{K}_T{}^{\alpha\beta\mu} = -4 \phi \mathcal{K}_T{}^{\alpha\beta\mu}$ \\
		$\delta_W \Delta_{\mu}{}^{\nu}\partial_{\nu}T = -\phi \Delta_{\mu}{}^{\nu}\partial_{\nu}T - T \Delta_{\mu}{}^{\nu}\partial_{\nu}\phi$	& $\delta_W \mathcal{K}_A{}^{\alpha\beta\mu} = -4 \phi \mathcal{K}_A{}^{\alpha\beta\mu}$ \\
		\hline
	\end{tabular}
	\caption{\label{T:Weyl} A listing of relevant Weyl transformations of first order quantities, derived from the transformation rules \eqref{E:deltaWvielbein}, \eqref{E:deltaWK} and \eqref{E:deltaWhydrozero}. }
\end{center}
\end{table}
The tensors $m_{\alpha}$, $a_{\mu}$, $\Delta_{\mu}{}^{\nu} \partial_{\nu}T$, $k_{\mu}$, $\mathcal{K}_{V}^{\mu}$, $\theta$, $\kappa$ and $u^{\alpha}\partial_{\alpha}T$ do not transform homogenously under Weyl rescalings (for generic $q$) but we may construct linear combinations of them which do transform homogenously. We have collected the independent homogenously transforming first order scalars and vectors in table \ref{T:W1hom}.
\begin{table}[hbt!]
\begin{center}
	\begin{tabular}{| l | l |}
	\hline
	Weyl covariant vectors & Weyl covariant scalars \\
	\hline
	$A_{\alpha} = T a_{\alpha} + \Delta_{\alpha}{}^{\mu} \mathring{\nabla}_{\mu}T$ & $\theta T + (d-1) u^{\alpha}\partial_{\alpha}T$ \\
	$(1-q) a_{\alpha} - m_{\alpha}$					     & $q \theta + \kappa$ \\
	$q a_{\alpha} + k_{\alpha}$ 					     & \\
	$2 q(d-2) a_{\alpha} + \mathcal{K}_{V\,\alpha}$		     & \\
	\hline
	\end{tabular}
	\caption{\label{T:W1hom} Composite, homogenously transforming (under Weyl rescalings) scalars and vectors whose components do not transform homogenously under Weyl rescalings. Note that $\kappa$ transforms homogenously under Weyl rescalings for $q=0$.}
\end{center}
\end{table}

To proceed, it is convenient to define a Weyl connection and a Weyl covariant derivative, constructed in such a way that derivatives of Weyl covariant tensors will also be Weyl covariant \cite{Loganayagam:2008is}. Indeed, suppose we have at our disposal a connection $\mathcal{A}_{\mu}$ satisfying
\begin{equation}
\label{E:deltaWA}
	\delta_W \mathcal{A}_{\mu} = \partial_{\mu}\phi\,,
\end{equation}
from which we construct $A^{\lambda}_{\mu\nu}$,
\begin{equation}
	\mathcal{A}^\lambda_{\mu \nu} = g_{\mu \nu} \mathcal{A}^\lambda - \delta^\lambda_\mu \mathcal{A}_\nu - \delta^\lambda_\nu \mathcal{A}_\mu\,.
\end{equation}
From $\mathcal{A}_{\mu}$ and $\mathcal{A}^{\lambda}_{\mu\nu}$ we can construct a Weyl covariant derivative,
\begin{equation}
	\mathring{\mathcal{D}}_{\mu} Q^{\alpha_1 \ldots}{}_{\beta_1 \ldots} = 
	\mathring{\nabla}_\mu Q^{\alpha_1 ...}{}_{\beta_1 ...} - \omega \mathcal{A}_\mu Q^{\alpha_1 ...}{}_{\beta_1 ...} - \mathcal{A}^{\rho_1}_{\mu \beta_1} Q^{\alpha_1 ...}{}_{\rho_1 ...} +... + \mathcal{A}^{\alpha_1}_{\mu \sigma_1} Q^{\sigma_1 ...}{}_{\beta_1 ...} + ...\,,
\end{equation}
where $\omega$ is the Weyl weight of $Q^{\alpha_1 \ldots}{}_{\beta_1 \ldots}$,
\begin{equation}
\label{E:deltaQ}
	\delta_W Q^{\alpha_1 \ldots}{}_{\beta_1 \ldots} = \omega Q^{\alpha_1 \ldots}{}_{\beta_1 \ldots} \,.
\end{equation}
The Weyl covariant derivative satisfies
\begin{equation}
	\delta_W \left(\mathring{\mathcal{D}}_{\mu} Q^{\alpha_1 \ldots}{}_{\beta_1 \ldots} \right) = \omega {\phi} \mathring{\mathcal{D}}_{\mu} Q^{\alpha_1 \ldots}{}_{\beta_1 \ldots} \,.
\end{equation}

It remains to construct the connection $\mathcal{A}_{\mu}$. One possibility for a connection is
\begin{equation}
\label{E:theA}
	\mathcal{A}_{\mu} = a_{\mu} - \frac{\theta}{d-1} u_{\mu}\,.
\end{equation}
There are other connections which we may construct out of the first order quantities in table \ref{T:Weyl} which will have the property \eqref{E:deltaWA}. All of these connections will differ from \eqref{E:theA} by the vectors and scalars in table \ref{T:W1hom}; differences of connections are proper tensors. Thus, we may, without loss of generality, use the connection \eqref{E:theA}. It is now straightforward to construct non composite, second order in derivative quantities by acting with $\mathring{\mathcal{D}}_{\mu}$ on first order in derivative quantities.

As mentioned earlier, now that we have a full classification of the relevant Weyl covariant tensors constructed from the hydrodynamical variables and sources, it is straightforward to construct Weyl covariant tensors, such as the spin current, from them. As is clear from  \eqref{E:deltaTS} the stress tensor is not Weyl covariant for $q=0$ (or any $q \neq 1$ for that matter). In order to construct constitutive relations for the stress tensor such that \eqref{E:tracelessness} and \eqref{E:deltaTS} are satisfied we first construct a Weyl covariant tensor $\mathcal{T}^{\mu\nu}$ such that
\begin{equation}	
	\delta_W \mathcal{T}^{\mu\nu} = -(d+2) \phi \mathcal{T}^{\mu\nu}
\end{equation}
and
\begin{equation}
\label{E:tracecalT}
	\mathcal{T}^{\mu}{}_{\mu} = (1-q) \mathring{\nabla}_{\mu} S_{\lambda}{}^{\lambda\mu}\,.
\end{equation}
Given $\mathcal{T}^{\mu\nu}$ we may construct $T^{\mu\nu}$ via the relation
\begin{equation}
\label{E:TfromcalT}
	T^{\mu\nu} = \mathcal{T}^{\mu\nu} + (q-1) S^{\mu\nu\rho}\mathcal{A}_{\rho} - (q-1) S_{\rho}{}^{\rho\mu}\mathcal{A}^{\nu}\,.
\end{equation}
The definition \eqref{E:TfromcalT} and the property \eqref{E:tracecalT} ensure that \eqref{E:tracelessness} and \eqref{E:deltaTS} are satisfied.

We now have all the necessary ingredients in order to construct the constitutive relations for a conformally invariant theory. In the remainder of this section we will follow the same path taken in previous sections. We first construct the conformally invariant hydrostatic partition function and then consider non hydrostatic contributions to the resulting constitutive relations.

A conformally invariant hydrostatic partition function, $W$, can be constructed by requiring that $W$ is invariant under Weyl rescalings, or $\delta_W \mathcal{W} = - d \phi \mathcal{W}$. In order to obtain all possible first and second order Weyl covariant scalars we take linear combinations of the scalars in table \ref{T:allscalarsnoK} using the data in table \ref{T:W1hom}. The resulting list of Weyl covariant scalars can be found in table \ref{T:Wscalars}.
\begin{table}[hbt!]
\begin{center}
	\begin{tabular}{| l |}
	\hline
	First order Weyl covariant scalars \\
	\hline
	$\delta_{q\,0} \kappa$ \\
	\hline
	Second order Weyl covariant scalars \\
	\hline
	$S_{(1)}^c = S_{(1)}$,
	$S_{(2)}^c = S_{(2)}$,
	$S_{(3)}^c = (1-q)((1-q)k + q m)^2$, 
	$S_{(4)}^{c} = ((1-q)k+q m) \cdot (\mathcal{K}_V - 2(d-2)k)$ \\
	\hline
	$S_{(5)}^c =  \kappa^2$,
	$S_{(6)}^c = S_{(7)}$,
	$S_{(7)}^c =  (\mathcal{K}_V - 2(d-2)k)^2 $,
	$S_{(8)}^c = S_{(10)} $, \\
	\hline
	$S_{(9)}^c = S_{(11)} $,
	$S_{(10)}^c = S_{(12)} $, 
	$S_{(11)}^c = S_{(13)}$,
	$S_{(12)}^c = S_{(14)}$ \\
	\hline
	\end{tabular}
	\caption{\label{T:Wscalars} 
	First and second order Weyl covariant scalars which may contribute to the parition funciton at the order we are interested in. 
	When $q=1$ the covariant vector $(1-q)k_{\mu} + q m_{\mu}$ reduces to $m_{\mu}$ in which case the contribution of $((1-q)k + q m)^2$ can be identified with that of $m^2$ which has been taken into account by considering the ideal constitutive relations. For this reason we have multiplied the expression associated with $S_{(3)}^c$ with an overall factor of $1-q$.}
\end{center}
\end{table}
Using the data in table \ref{T:Wscalars}, we find that the hydrostatic partition function reads,
\begin{equation}
	\mathcal{W} = P + \mathcal{W}_{(1)} + \mathcal{W}_{(2)}
\end{equation}
where
\begin{align}
\begin{split}
\label{E:conformalhydrostatic}
	P =& p_0 T^d + r_M T^{d-2} {M}_{(1)} + r_m T^{d-2} \delta_{q 1} m_{(0)} + \mathcal{O}(\mathring{\nabla}^4) \\
	\mathcal{W}_{(1)} =& x_{1}^{(1)}T^{d-1}    \,\kappa   \\
	\mathcal{W}_{(2)} =& \sum_{i=1}^{12} x_i^{(2)}T^{d-2} S_{(i)}^c \,.
\end{split}
\end{align}
Here, $p_0$, $r_m$, $r_M$ and the $x_i^{(j)}$'s are real numbers. The Kronecker delta 
$\delta_{q\,1}$ ensures that the hydrostatic partition function includes all possible scalars for generic values of $q$ and for $q=1$ 
Note that $\kappa$ is conformal invariant in the hydrostatic limit for any value of $q$, but is conformal invariant in general only for $q=0$. Therefore, later, when including non hydrostatic corrections, we will need to modify the expressions associated with $x_1^{(1)}$ and $x^{(2)}_5$ so that they are conformal invariant for all values of $q$.

The stress tensor $T_{h+}^{\mu\nu}$ and spin current $S_h^{\lambda\mu\nu}$ associated with \eqref{E:conformalhydrostatic} can be obtained by varying the generating function $W$ with respect to the vielbein and spin connection respectively, and adding non hydrostatic terms to expressions derived from terms linear in $K_{\mu}{}^{ab}$ so that the resulting stress tensor is of the BR type. (Recall the discussion around \eqref{E:T1}.) Instead of carrying out this variation explicitly, we can read the result off of \eqref{E:idealexpansion}--\eqref{E:gotepsilon0} and \eqref{E:hdecomposition}--\eqref{E:constitutive12} by making the substitutions
\begin{align}
\begin{split}
\label{E:conformalsubstiutions}
	P_0 &= p_0T^{d}
	\qquad
	\epsilon_0 = (d-1) p_0 T^d 
	\qquad
	\chi^{(1)}_1 = x^{(1)}_1 T^{d-1}
	\\
	\chi^{(2)}_1 & = x^{(2)}_1 T^{d-2} 
	\qquad
	\chi^{(2)}_2  = x^{(2)}_2 T^{d-2} 
	\qquad
	\chi^{(2)}_3  = \left( -2(1-q)^2q x^{(2)}_3 -2 (d-2) q x^{(2)}_4 \right) T^{d-2} 
	\\
	\chi^{(2)}_4 &= q x^{(2)}_4 T^{d-2}
	\qquad
	\chi^{(2)}_5 = x^{(2)}_5 T^{d-2} 
	\quad
	\chi^{(2)}_6 = \left( (1-q)^3 x^{(2)}_3 + 2 (d-2)(q-1) x^{(2)}_4 + 4 (d-2)^2 x^{(2)}_7 \right)T^{d-2}  
	\\
	\chi^{(2)}_7 &= x^{(2)}_6 T^{d-2} 
	\qquad
	\chi^{(2)}_8 = (1-q) x^{(2)}_4 T^{d-2}
	\qquad
	\chi^{(2)}_9 = x^{(2)}_7 T^{d-2} 
	\\
	\chi^{(2)}_{10} &= x^{(2)}_8 T^{d-2}
	\qquad
	\chi^{(2)}_{11} = x^{(2)}_{9} T^{d-2}
	\qquad
	\chi^{(2)}_{12} = x^{(2)}_{10} T^{d-2}
	 \\
	\chi^{(2)}_{13} &= x^{(2)}_{11} T^{d-2}
	\qquad
	\chi^{(2)}_{14} = x^{(2)}_{12} T^{d-2}
	\qquad
	\rho_m = (q^2 (1-q) x^{(2)}_3 + r_m \delta_{q\,1})T^{d-2}
	\qquad
	\rho_M = r_M T^{d-2}\,.
\end{split}
\end{align}

The explicit form of the resulting constitutive relations can be found in appendix \ref{A:fullconstitutive}.

Weyl invariance of the generating function $W$ guarantees that, in the hydrostatic limit,  the stress tensor and spin current will satisfy \eqref{E:tracelessness} and \eqref{E:deltaTS}. 
Yet, a naive computation of the Weyl transformation of $T_{h+}^{\mu\nu}$ (or $T_{h}^{\mu\nu}$) resulting from the constitutive relations given in appendix \ref{A:fullconstitutive} do not reproduce \eqref{E:deltaTS}. This is due to the inhomogenous transformation properties of $k_{\mu}$ and $\mathcal{K}_{V\,\mu}$ under Weyl rescalings 
together with the fact that we have dropped terms quadratic in contorsion. Once we include the full hydrostatic constitutive relations derived from $W$, quadratic in contorsion, into the stress tensor, then \eqref{E:deltaTS} will be satisfied. Since the constitutive relations for the stress tensor at second order in the contorsion tensor are somewhat long we will have chosen not to list those terms here. The interested reader can find the terms relevant for this computation in appendix \ref{A:quadratic}. We have also checked that $T_{h+}^{\mu\nu}$ and $S_{h}^{\lambda\mu\nu}$ satisfy \eqref{E:tracelessness} and \eqref{E:deltaTS} outside of  hydrostatic equilibrium.

In order to construct a Weyl covariant non hydrostatic tensor $\mathcal{T}_{nh}^{\mu\nu}$ (and from it the stress tensor $T_{nh}^{\mu\nu}$ using the prescription of \eqref{E:TfromcalT}) and the Weyl covariant non hydrostatic spin current $S_{nh}^{\lambda\mu\nu}$, we consider the non covariant constitutive relations \eqref{E:nhterms} and use table \ref{T:Weyl} to remove from them all non Weyl covariant expressions. We find that in order for the non hydrostatic spin current to transform covariantly under Weyl rescalings, we need to set $\sigma_1 = T^{d-2} s_1$ and $\sigma_2=0$ in \eqref{E:nhterms}, leading to
\begin{equation}
\label{E:s1term}
	S_{nhBR}^{\lambda\mu\nu} = 2 T^{d-2} s_1 \sigma^{\lambda[\mu}u^{\nu]}\,.
\end{equation}

To construct the Weyl covariant tensor $\mathcal{T}_{nh}^{\mu\nu}$ from the non Weyl covariant expression, $T_{nh}^{\mu\nu}$, of \eqref{E:nhterms} we must not only set $\zeta=0$ in \eqref{E:nhterms} and scale the remaining transport coefficients by powers of $T$, we must also replace the ringed derivatives in \eqref{E:nhterms} with Weyl covariant ones:
\begin{align}
\begin{split}
\label{E:calTterm}
	\mathcal{T}_{nh}^{(\mu\nu)} &= -h T^{d-1} \sigma^{\mu\nu} - \mathring{\mathcal{D}}_{\lambda} \left(S_{nh BR}^{\mu\lambda\nu} - S_{nh BR}^{\lambda\nu\mu} \right)\,\\
	\mathcal{T}_{nh}^{[\mu\nu]} &= \mathring{\mathcal{D}}_{\lambda}S^{\lambda\mu\nu}{}_{nhBR} + s_A T^{d-2} A^{[\mu}u^{\nu]} + s_m T^{d-2} \hat{m}^{[\mu}u^{\nu]} + s_M T^{d-2} \hat{M}^{\mu\nu}\,.
\end{split}
\end{align}
Note that \eqref{E:tracecalT} is trivially satisfied since $S_{nhBR\,\lambda}{}^{\lambda\mu}=0$. 

We can now construct the non hydrostatic stress tensor using the prescription of \eqref{E:TfromcalT} with $\mathcal{T}^{\mu\nu}$ and $S^{\mu\nu\rho}$ replaced by $\mathcal{T}_{nh}^{\mu\nu}$ and $S_{nh}^{\mu\nu\rho}$. The only missing terms we need to include are corrections to the $x_1^{(1)}$ and $x^{(2)}_5$ terms once $q\neq 0$ and we are outside of hydrostatic equilibrium. Some trial and error reveals that we need to add to the spin current a term of the form $4 x_5^{(2)} q \theta \Delta^{\lambda[\mu } u^{\nu]}  $ and to the stress tensor contributions of the form
$ T^{d-1} q x^{(1)}_1 \theta ((d-1) u^{\mu}u^{\nu} + \Delta^{\mu\nu})$ and $\frac{1}{2} x^{(2)}_5 T^{d-2} q (\mathcal{K}_V^{\nu} u^{\mu}-\mathcal{K}_V^{\mu}u^{\nu} - 4 \kappa_A^{\mu\nu}) \theta$. The resulting constitutive relations are presented in appendix \ref{A:fullconstitutive}. It is given in a non Landau like frame since the latter will not have manifest conformal symmetry.

The constitutive relations for the conformal fluid described in this section will be modified in the presence of a charge current. The chemical potential and external gauge field associated with the charge current allow for additional tensor structures both in the non conformally invariant case and the conformally invariant one. Since we have refrained from classifying the constitutive relations for the charge current in the presence of an external torsion field, we will only discuss the $q=0$ torsionless, conformal, constitutive relations for a charged fluid with spin. (Recall that since torsion does not transform homogenously under Weyl rescalings, a $q \neq 0$, torsionless background can be transformed into a torsionfull one.) 

Consider the hydrostatic generating function for a charged fluid, \eqref{E:chargedW}. When $q=0$ the vectors $k^{\mu}$ and $\mathcal{K}_V^{\mu}$ transform homogenously under Weyl rescalings, (see table \ref{T:Weyl}) so all the scalars in table \ref{T:U1scalars} will contribute to $\mathcal{W}$, except for $\tilde{S}_{(1)}$. Thus, the generating function for a conformal charged fluid will take the form
\begin{equation}
\label{E:Wconformalcharged}
	\mathcal{W} = p_0 T^{d} + T^{d-1} x_1^{(1)} \kappa  +\sum_{i=1}^2 T^{d-2}{x}_i^{(2)}{S}_i^{(c)}   +\sum_{i=2}^6 T^{d-2}\tilde{x}_i^{(2)}\tilde{S}_i \,,
\end{equation}
where $p_0$ the $x^{(2)}_i$, and the $\tilde{x}^{(2)}_i$ all depend on the chemical potential to temperature ratio $\mu/T$. The explicit form of the hydrostatic constitutive relations for a conformal charged fluid, obtained by the variation of \eqref{E:Wconformalcharged} can be found in appendix \ref{A:fullconstitutive}. 

For the non hydrostatic terms we note that $\Delta^{\mu\nu} \left(\mathring{\nabla}_{\nu} \left(T\frac{\mu}{T}\right) - E_{\nu}\right)$ transforms homogenously under Weyl rescalings. Thus, the constitutive relations for the stress tensor and current are similar to those for the uncharged fluid, with the addition of the new terms associated with $\Delta^{\mu\nu} \left(\mathring{\nabla}_{\nu} \left(T\frac{\mu}{T}\right) - E_{\nu}\right)$ (and the dependence of the transport coefficients on $\mu/T$). The full constitutive relations for a conformally invariant charged fluid with spin can be found in appendix \ref{A:fullconstitutive}.

\section{Summary and discussion}
\label{S:discussion}

In this work we have constructed the most general constitutive relations for an uncharged fluid with spin in the presence of external torsion and a charged fluid with spin in the absence of torsion. As emphasized in the introduction, one does not expect background torsion to play a role in, say, the dynamics of heavy ion collisions. The reason we have included torsion in the constitutive relations is that, without it, one would obtain incorrect Kubo formula for the various transport coefficients. (This is similar to the need for including a non trivial background metric in order to compute Kubo formula in flat space.) Another reason we have included background torsion is that it might be relevant as an effective theory for certain condensed matter systems with dislocations; see e.g. \cite{deJuan:2009ldt,Mesaros:2009az}. We have also carried out a brief analysis of the effect of charge and of conformal invariance. Our main results have been collected in appendix \ref{A:fullconstitutive} for ease of reference.

To summarize the essential features of our analysis, consider the equations of motion for an uncharged fluid, given by \eqref{E:EOMs}.
Using the constitutive relations in appendix \ref{A:fullconstitutive}, the leading order terms for these equations (in the absence of torsion) take the form
\begin{align}
\begin{split}
\label{E:zeroorderEOM}
	\frac{\partial s}{\partial T} u^{\mu}\partial_{\mu}T + \frac{\partial P}{\partial T} \theta & = 0 \\
	A^{\mu} &= 0 \\
	\sigma_m \hat{m}^{\mu} &= 0 \\
	\sigma_M \hat{M}^{\mu\nu} & = 0\,,
\end{split}
\end{align}
Where $s = \frac{\partial P}{\partial T}$ is the entropy density, $\theta = \mathring{\nabla}_{\mu} u^{\mu}$, $\hat{m}^{\mu} = \mu^{\mu\alpha}u_{\alpha}  - a^{\mu}$, $\hat{M}^{\mu} = \Delta^{\mu}{}_{\alpha} \Delta^{\nu}{}_{\beta} \mu^{\alpha\beta}+ \Omega^{\mu\nu}$ and $A^{\mu} = \Delta^{\mu\nu} \mathring{\nabla}_{\nu} T+T u^{\alpha} \mathring{\nabla}_{\alpha} u^{\mu}$ (not to be confused with the gauge potential $A_{\mu}$ which sources the charge current).

The first two equations in \eqref{E:zeroorderEOM} are the standard equations for energy and momentum conservation. The last two equations follow from spin current conservation and, as long as $\sigma_m$ and $\sigma_M$ are non zero imply
\begin{equation}
\label{E:mM0}
	\hat{m}^{\mu} = 0
	\qquad
	\hat{M}^{\mu\nu} = 0\,.
\end{equation}
The relations \eqref{E:mM0} were derived previously in \cite{Becattini:2013fla} by appealing to the Boltzmann equation. (See \cite{Becattini:2020sww,Becattini:2020ngo} for reviews on this subject.)
In our work these relations were obtained as dynamical equations in the absence of charge. In the presence of charge the relations \eqref{E:zeroorderEOM} are modified and as a result, the first equality in \eqref{E:mM0} is replaced by \eqref{E:hatmwithcharge}. We do find that, charged or not, \eqref{E:mM0} are always valid in hydrostatic equilibrium.

One of the unusual features of \eqref{E:mM0} is that the spin chemical potential is determined algebraically from the acceleration and vorticity. This means, among other things, that one can not impose initial conditions for the spin chemical potential. Rather, it is determined algebraically from derivatives of the velocity. In a previous paper, \cite{Gallegos:2021bzp}, we have worked out the constitutive relations for a conformal fluid with spin, in the absence of torsion, in what we referred to as the dynamical spin limit. In the notation of the current paper, this corresponds to setting $\sigma_m=0$ and $\sigma_M=0$. If $\sigma_m$ and $\sigma_M$ vanish, then the leading equations of motion for the spin current become second order equations (recall that the spin chemical potential is counted as first order in derivatives) involving derivatives of $\hat{m}^{\mu}$ and $\hat{M}^{\mu\nu}$. This implies that the equations of motion for the spin chemical potential become first order differential equations. For this reason we have referred to the $\sigma_m = 0$ and $\sigma_M=0$ limits as a dynamical spin limit. Note that, at least in our current formalism, one can not simply take the limit $\sigma_m \to 0$ or $\sigma_M \to 0$ of \eqref{E:zeroorderEOM} in order to obtain the dynamical spin limit since we are working in a frame where higher order corrections to the antisymmetric components of the stress tensor vanish. As discussed in section \ref{S:entropycurrent} this frame choice is only allowed for $\sigma_m \neq 0$ and $\sigma_M \neq 0$. Thus, in a sense, the work in \cite{Gallegos:2021bzp} treats a specialized limit of the constitutive relations for hydrodynamics with spin which can not be captured by our current formulation. As of the time of writing of this note, we could not find symmetry arguments which would enforce $\sigma_m=0$ or $\sigma_M=0$. 

The authors of \cite{Li:2020eon,Hongo:2021ona} have set $\sigma_m=0$ and treated $\sigma_M$ as perturbatively small by using dynamical considerations; see also \cite{Hongo:2022izs} for a perturbative computation of the latter in an effective model for QCD.
Thus, we expect their analysis to be valid only for flows for which this approximation is valid. We have found that their resulting constitutive relations and correlation functions match ours where a comparison can be made. An earlier derivation of the constitutive relations for the stress tensor and spin current can be found in \cite{Hattori:2019lfp}. One of the differences between the current work and that of \cite{Hattori:2019lfp} is that the authors of \cite{Hattori:2019lfp} used the spin current itself as a degree of freedom (replacing our $m^{\mu}$ and $M^{\mu\nu}$). If we only consider the constitutive relations for the stress tensor then our expressions match theirs once we remove the BR contributions.

One particular transport coefficient which is absent from the constitutive relations described both in \cite{Li:2020eon,Hongo:2021ona} and in \cite{Hattori:2019lfp} is the first order term $\chi_1^{(1)}$ which contributes to the spin current in the form
\begin{equation}
	S^{\lambda\mu\nu} = \ldots + 2 \chi_1^{(1)} \Delta^{\lambda[\mu}u^{\nu]}
\end{equation}
and contributes to the stress tensor via a BR term, \eqref{E:BRterms}. One argument that has been made for the vanishing of $\chi_1^{(1)}$ in \cite{Li:2020eon,Hongo:2021ona} is that such terms do not contribute to the fully antisymmetric components of the spin current and that minimal coupling of fermions to torsion generate a fully antisymmetric spin current. While true, it doesn't seem to us a necessary condition for setting $\chi_1^{(1)}=0$. Once the theory is coupled to torsion, even perturbatively, all terms compatible with the symmetry should be allowed and therefore $\chi_1^{(1)}$ may appear in the constitutive relations. An extended discussion of the behavior of $\chi_1^{(1)}$ under CPT transformations can be found towards the end of this section.

The hydrostatic equations \eqref{E:hattedM} (derived from \eqref{E:cpconstraint}) imply that the order in derivatives of the contorsion tensor is tied to the acceleration and vorticity terms which are usually counted as first order in derivatives. Indeed, in this work we have set the contorsion tensor (and therefore the spin chemical potential) to be first order in derivatives. But other possibilities are allowed. For instance, one could allow for $u^{\mu}K_{\mu}{}^{ab}$ to be first order in derivatives, but other components of the contorsion tensor to be zeorth order in derivatives. This would be similar to the derivative counting of the electromagnetic field strength in magnetohydrodynamics \cite{Hernandez:2017mch} where the magnetic field is counted as one order lower than the electric field. In both the former, and the latter cases, the spin chemical potential is counted as first order in derivatives. 

In order to have a zeroth order in derivatives spin chemical potential one would need the acceleration and vorticity to be zeroth order in derivatives quantities. We point out that it is impossible to achieve hydrostatic equilibrium (with real temperature and velocity field) in Minkowski space with a non zero acceleration and vorticity. Indeed, in hydrostatic equilibrium the acceleration and vorticity are given by appropriate derivatives of the velocity field which in Minkowski space must take the form $(1,\vec{0})$. Thus, in Minkwoski space, (and in the absence of torsion), the spin chemical potential must vanish in equilbirium. Since unforced fluid dynamics will tend to reach an equilibrium configuration, the spin chemcial potential will tend to vanish over time. For this reason it is also physically sensible to count the spin chemical potential as a first order in derivative quantity. This observation may be contrasted with an analysis of the spin chemical potential in some applications of heavy ion collisions (see, .e.g, \cite{DeGroot:1980dk,Becattini:2012tc,Becattini:2013fla,Florkowski:2017ruc} ) where a background velocity (in four dimensions) of the form
\begin{equation}
\label{E:rotating}
	u_{\mu}dx^{\mu}  =  \gamma \left(-\Omega y dx + \Omega x dy \right)
\end{equation}
with
$
	\gamma^{-1} = \sqrt{1 - \Omega^2 (x^2+y^2)}
$,
is used, and leads to
\begin{equation}
	a_{\mu}dx^{\mu} = -\gamma^2 \Omega^2 (x dx + y d y )
	\qquad
	\Omega_{\mu\nu}dx^{\mu} \wedge dx^{\nu}  = -\gamma^3 \Omega dx \wedge dy\,.
\end{equation}
This solution is not a hydrostatic solution since, in order to avoid complex solutions, one would either need to insert a boundary or glue this solution to a stationary one. Either way, \eqref{E:rotating} can, at best, be metastable and will eventually dissipate. 

To properly deal with solutions for which vorticity and acceleration can be properly counted as zero order in derivative quantities one would need to develop a formalism where arbitrary powers of vorticity and acceleration appear at the ``ideal'' fluid level. (Linearly perturbing normal fluid dynamics around the solution \eqref{E:rotating} is equivalent to an uncontrolled truncation of the latter.) One approach to constructing a hydrostatic theory which incorporates vorticity and acceleration as zero order quantities would be to place the theory on a curved manifold where the magnetic components of the Riemann curvature, c.f., \cite{Jensen:2013kka}, are counted as zero order in derivative quantities, similar to the procedure for magnetohydrodynamics mentioned above. On top of that, one should then be able to add contorsion and then obtain a zero order in derivative spin chemical potential. 
In \cite{Li:2020eon} elements of such an analysis have been carried out.

\begin{table}[hbt]
\vspace{5pt}
\begin{center}
\begin{subtable}[t]{.3\textwidth}
\begin{tabular}{ | l || c | c | c |  }
 \hline
 	& $T$ &  $PT$ & $CPT$ \\
\hline
	$T$ & + & + & + \\
	$\mu$ & + & + & - \\
	$u^{0}$ & + & + & + \\
	$u^{i}$ & - & - & + \\
	$\mu^{0i}$ & + & - & - \\
	$\mu^{ij}$ & - & - & - \\
	$m^{0}$ & - & - & - \\
	$m^{i}$ & + & - & - \\
	$M^{0i}$ & + & - & - \\
	$M^{ij}$ & - & - & - \\
	$\theta$ & - & - & - \\
	$a^{0}$ & - & - & - \\
	$a^{i}$ & + & - & - \\
	$\sigma^{00}$ & - & - & - \\
	$\sigma^{0i}$ & + & - & - \\
	$\sigma^{ij}$ & - & - & - \\	
	$\Omega^{0i}$ & + & - & - \\
	$\Omega^{ij}$ & - & - & - \\	
\hline
\end{tabular}
\caption{CPT transformation properties of the hydrodynamic variables and some of their derivatives.}
\end{subtable}
\begin{subtable}[t]{.3\textwidth}
\begin{tabular}{ | l || c | c | c |  }
\hline
 	& $T$ &  $PT$ & $CPT$ \\
\hline
	$g^{00}$ & + & + & + \\
	$g^{0i}$ & - & + & + \\	
	$g^{ij}$ & + & + & + \\
	$K_{0}^{i0}$ & + & - & - \\
	$K_{0}^{ij}$ & - & - & - \\
	$K_{i}^{j0}$ & - & - & - \\	
	$K_{i}^{jk}$ & + & - & - \\	
	$A^{0}$ & + & + & - \\
	$A^{i}$ & - & + & - \\
	$\kappa$ & - & - & - \\
	$K^{0i}$ & + & - & - \\
	$K^{ij}$ & - & - & - \\	
	$\kappa_A^{0i}$ & + & - & - \\
	$\kappa_A^{ij}$ & - & - & - \\	
\hline
\end{tabular}
\caption{Spurionic CPT transformation properties of sources and some sources contracted with hydrodynamic variables.}
\end{subtable}
\caption{\label{T:CPT} CPT transformation properties of the hydrodynamic variables, some of their derivatives, and spurionic transformation properties of the sources.}
\end{center}
\end{table}

As we have emphasized time and again, the background torsion we turn on is simply a means to identify a canonical spin current and stress tensor which are useful in describing the dynamics of fluids with spin. At the end of the day, if we are to describe a flat torsionless background we may set torsion to zero. Once we do so, we expect to be able to use the standard hydrodynamic description of the fluid at least for the temperature and velocity fields. Satisfyingly, this is indeed the case. If one uses the improved stress tensor $T_{\hbox{\tiny I}}^{\mu\nu}$ defined in \eqref{E:defTI} then energy momentum conservation will reduce to the usual relativistic hydrodynamic equations of motion for $T$ and $u^{\mu}$. In addition, one will find that the spin chemical potential follows the acceleration and vorticity as in \eqref{E:mM0} (or its modified version \eqref{E:hatmwithcharge} in the presence of charge). Thus, in the absence of torsion the standard rules of hydrodynamics apply and they are supplemented by an additional algebraic constraint relating the spin chemical potential to the other hydrodynamic variables. Similar conclusions were reported in \cite{Li:2020eon}.

Our result is also inline with the general perception that in the absence of torsion, improvement terms (sometimes referred to as pseduo-gauge transformations) should not affect physical observables. Indeed, in our formalism, carrying out a BR type of transformation (in the absence of torsion) will not modify the equations of motion of the fluid. 

We end this section with a discussion of the CPT transformation properties of the hydrodynamic variables $u^{\mu}$, $T$, $\mu^{ab}$ and $\mu$ (allowing for a chemical potential) and the spurionic version of these transformations on the sources $e^{a}{}_{\mu}$, $\omega_{\mu}^{ab}$ and $A_{\mu}$ (if a $U(1)$ charge current is present). In what follows, to keep our notation compact, we will consider even dimensional theories where parity flips the sign of all spatial directions. An analysis of CPT for odd dimensions is straightforward. 

The vielbein is even under CPT (and so is the metric), the spin connection is odd under CPT and the vector potential is also odd under it. To determine the transformation properties of the hydrodynamic variables we may use the hydrostatic relations \eqref{E:defthermal}. Since the timelike Killing vector $V^{\mu}$ is invariant under CPT, we find that $T$ is even and $\mu^{ab}$ and $\mu$ are odd under CPT. We have collected the CPT transformation rules of various quantities under CPT in table \ref{T:CPT}.

One can use the CPT transformation properties in table \ref{T:CPT} to constrain properties of transport coefficients. For instance, a CPT even partition function would imply that $\chi^{(1)}_1(\mu,T)$ depend on odd powers of $\mu$. A full analysis of constraints on CPT transformation properties of transport coefficients would require a Schwinger-Keldysh effective action type of analysis, similar to the one carried out in, e.g., \cite{Haehl:2014zda,Crossley:2015evo,Jensen:2018hse}. We leave such a study for future work.

\section*{Acknowledgements}

We thank Francesco Becattini, Saso Grozdanov, Zohar Komargodski, Giorgio Torrieri, Enrico Speranza, Toby Wiseman and Ho-Ung Yee. DG and UG are partially supported by the Delta-Institute for Theoretical Physics (D-ITP) funded by the Dutch Ministry of Education, Culture and Science (OCW). In addition, DG is supported in part by CONACyT through the program Fomento, Desarrollo y Vinculacion de Recursos Humanos de Alto Nivel. AY is supported in part by an Israeli Science Foundation excellence center grant 2289/18 and a Binational Science Foundation grant 2016324.

\begin{appendix}
\section{A compendium of decompositions}
\label{A:decompositions}
In this work we have often decomposed various external sources and hydrodynamic quantities with respect to the residual $SO(d-1)$ symmetry associated with spatial rotations orthogonal to the velocity field. In this appendix we have collected the decomposition of various quantities into representations of the $SO(d-1)$ symmetry for ease of access. Our main tool for constructing a decomposition is the projection
\begin{equation}
	\Delta^{\mu\nu} = g^{\mu\nu} + u^{\mu}u^{\nu}\,.
\end{equation}
We will also often use a symmetrized and antisymmetrized indices using round or square brackets respectively,
\begin{equation}
	A_{[\mu\nu]} = \frac{1}{2} \left(A_{\mu\nu} - A_{\nu\mu}\right) 
	\qquad
	S_{(\mu\nu)} = \frac{1}{2} \left( S_{\mu\nu} + S_{\nu\mu}\right)\,.
\end{equation}

\subsection{Hydrodynamic quantities}
We often need the decomposition of the gradient of the velocity field
\begin{equation}
	\mathring{\nabla}_\mu u_\nu = \frac{1}{d-1} \theta \Delta_{\mu \nu} -u_\mu a_\nu + \sigma_{\mu \nu} + \Omega^{\mu \nu} \, , \\
\end{equation}
where
\begin{align}
\begin{split}
	\theta &= \mathring{\nabla}_\mu u^\mu \, , \\
	a^\nu &= u^\mu \mathring{\nabla}_\mu u^\nu \, , \\
	\Omega_{\mu \nu} &= \Delta^\rho_\mu \Delta^\sigma_\nu \mathring{\nabla}_{[\rho} u_{\sigma]} \, , \\
	\sigma^{\mu \nu} &= \Delta^{\mu \nu \rho \sigma} \mathring{\nabla}_\rho u_\sigma\,
\end{split}
\end{align}
with 
\begin{equation}
	\Delta^{\mu \nu \rho \sigma} = \Delta^{\mu (\rho} \Delta^{\sigma) \nu} - \frac{1}{d-1} \Delta^{\mu \nu} \Delta^{\rho \sigma}\,.
\end{equation}

Similarly, the decomposition of the chemical potential is given by
\begin{equation}
	\mu^{ab} = u^a m^b - u^b m^a + M^{ab}
\end{equation}
so that
\begin{align}
\begin{split}
\label{E:mMdecomposition}
	m^a &= \mu^{ab}u_b \\
	M^{ab} & = \Delta^{a}{}_{c}\Delta^{b}{}_{d} \mu^{cd}\,.
\end{split}
\end{align}

\subsection{The contorsion tensor}
\label{AA:contorsion}
Similar to the hydrodynamic fields we can decompose the contorsion tensor into components,
\begin{align}
	K_\mu{}^{ab} = &- u_\mu \left( u^a k^b - u^b k^a + K^{ab} \right) + \frac{ 2\kappa}{d-1} u^{[a} \Delta^{b]}{}_\mu  +2 u^{[a}  \kappa_S{}^{b]}{}_\mu + 2 u^{[a}  \kappa_A{}^{b]}{}_\mu \\ 
	& + \frac{1}{d-2} \Delta_{\mu}{}^{[a} \mathcal{K}_V{}^{b]}  +  \mathcal{K}_{T\,\mu}{}^{ab}  + \mathcal{K}_A{}_{\mu}{}^{ab}\, , 
\end{align}
where
\begin{align}
\begin{split}
	k^a &= u^\mu u_b  K_{\mu}{}^{ab}  \, ,\\
	K^{ab} &= \Delta^a{}_c \Delta^b{}_d u^\mu K_{\mu}{}^{cd} \, ,
\end{split}
\end{align}
denote the decomposition of $u_{\mu}K^{\mu ab}$ into a vector and antisymmetric tensor,
\begin{align}
\begin{split}
\label{E:kappas}
	\kappa &= e^\mu{}_c  K_{\mu}{}^{cd}u_d \, , \\
	\kappa_S{}^{\mu \nu} &= \Delta^{\mu \nu\rho}_{\hphantom{\mu \nu \rho} c} K_{\rho}{}^{cd} u_d  \, , \\
	\kappa_A{}^{\mu \nu} &= \left(\frac{\Delta^\mu{}_c \Delta^{\rho \nu} - \Delta^\nu{}_c \Delta^{\mu \rho}}{2} \right)  K_{\rho}{}^{cd} u_d  \, , \\
\end{split}
\end{align}
denote the decomposition of $\Delta^{\nu\mu} K_{\mu}{}^{ab}u_{b}$ into a scalar, symmetric traceless tensor and antisymmetric tensor, and
\begin{align}
\begin{split}
   	\mathcal{K}_V{}^\mu &= 2 \Delta^\mu{}_d \Delta^\rho{}_c K_{\rho}{}^{cd} \, ,  \\
   	\mathcal{K}_A{}^{\lambda\rho \sigma} &=\Delta^{[\rho}{}_c \Delta^\sigma{}_d \Delta^{\lambda] \alpha}  K_{\alpha}{}^{cd}  \\
   	\mathcal{K}_T{}_{\mu}{}^{\alpha \beta} &=  \Delta^\alpha{}_c \Delta^d{}_\beta \Delta^\nu{}_\mu K_{\nu}{}^{cd} -\frac{1}{d-2} \Delta^{[\alpha}_\mu  \mathcal{K}_V{}^{\beta]} - \mathcal{K}_A{}_{\mu}{}^{\alpha \beta} 
\end{split}
\end{align}
denote the decomposition of $\Delta^{\nu\mu}\Delta_a{}^{\alpha}\Delta_b{}^{\beta} K_{\mu}{}^{ab}$ into a vector, an antisymmetric tensor, and a mixed traceless tensor,
\begin{equation}
	\mathcal{K}_T{}_{\mu}{}^{\mu \sigma}= 0 \, , \qquad 
	\mathcal{K}_T{}^{[\rho \sigma \lambda]} =0 \, , \qquad
	\mathcal{K}_A{}^{\rho \sigma \lambda} = \mathcal{K}_A{}^{[\rho \sigma \lambda]} \,.
\end{equation}

\subsection{Miscellaneous quantities and definitions}
\label{AA:misc}
Apart from the above decompositions, we often use the following quantities which vanish in hydrostatic equilibrium,
\begin{align}
\begin{split}
	A^{\mu} &= \Delta^{\mu\nu} \mathring{\nabla}_{\nu} T + T a^{\mu} \\
	\hat{m}^{\mu} &= m^{\mu} - k^{\mu} - a^{\mu} \\
	\hat{M}^{\mu\nu} &= M^{\mu\nu} - K^{\mu\nu}+ \Omega^{\mu\nu} 
\end{split}
\end{align}
The first entry was taken from table \ref{T:antisymmfirst} and the last two entries were taken from \eqref{E:defhMm}.

Also, we decompose the external field strength, $F^{\mu\nu}$, into a magnetic component, $B_{\mu\nu}$ and an electric component, $E_{\mu}$,
\begin{equation}
	E^{\mu} = F^{\mu\nu}u_{\nu}
	\qquad
	 B_{\mu\nu} = \Delta_{\mu\rho} \Delta_{\nu\sigma}F^{\rho\sigma}\,.
\end{equation}
Finally, when considering conformal theories, we use a Weyl covariant connection similar to that discussed in \cite{Loganayagam:2008is},
\begin{equation}
	\mathcal{A}_{\mu} = a_{\mu} - \frac{\theta}{d-1} u_{\mu}\,.
\end{equation}
(See \eqref{E:theA}.)

\subsection{Scalars, vectors and tensors}
\label{AA:collections}
Often, we need to classify various scalar, vector and tensor structures. We have collected these in the tables below.

\begin{table}[hbt!]
\begin{center}
		\begin{tabular}{| c   c   c   c  c   | }
		\hline
		\multicolumn{5}{| l |}{Order two hydrostatic scalars} \\
		\hline
			$S_{(1)} = M_{\mu\nu} \kappa_A^{\mu\nu}$ & 
			$S_{(2)} = M_{\mu\nu}K^{\mu\nu}$ & 
			$S_{(3)} = m_{\mu}k^{\mu} $ &
			$S_{(4)} = m_{\mu}\mathcal{K}_V^{\mu} $ &  
			$S_{(5)} = \kappa^2$ \\
			$S_{(6)} =k \cdot k$ &
			$S_{(7)} =K_{\mu\nu}K^{\mu\nu}$ &
			$S_{(8)} =k \cdot \mathcal{K}_V$ &
			$S_{(9)} =\mathcal{K}_V \cdot \mathcal{K}_V$ &
			$S_{(10)} =K_{\mu\nu}\kappa_A{}^{\mu\nu}$ \\
			$S_{(11)} =\kappa_{A\,\mu\nu}\kappa_A{}^{\mu\nu}$ &
			$S_{(12)} =\kappa_{S\,\mu\nu}\kappa_S{}^{\mu\nu}$ &
			$S_{(13)} =\mathcal{K}_{A\,\mu\nu\rho} \mathcal{K}_A{}^{\mu\nu\rho}$ &
			$S_{(14)} =\mathcal{K}_{T\,\mu\nu\rho} \mathcal{K}_T{}^{\mu\nu\rho}$  & \\ 
		\hline
	\end{tabular}
	\caption{Summary of second order independent inequivalent hydrostatic scalars which may contribute to the equations of motion up to second order in derivatives. This information can also be found in table \ref{T:allscalarsnoK}.}
\end{center}
\end{table}

\begin{table}[hbt!]
\begin{center}
	\begin{tabular}{| l |}
	\hline
	Second order Weyl covariant scalars \\
	\hline
	$S_{(1)}^c = S_{(1)}$,
	$S_{(2)}^c = S_{(2)}$,
	$S_{(3)}^c = (1-q)((1-q)k + q m)^2$, 
	$S_{(4)}^{c} = ((1-q)k+q m) \cdot (\mathcal{K}_V - 2(d-2)k)$ \\
	\hline
	$S_{(5)}^c = \kappa^2$,
	$S_{(6)}^c = S_{(7)}$,
	$S_{(7)}^c =  (\mathcal{K}_V - 2(d-2)k)^2 $,
	$S_{(8)}^c = S_{(10)} $, \\
	\hline
	$S_{(9)}^c = S_{(11)} $,
	$S_{(10)}^c = S_{(12)} $, 
	$S_{(11)}^c = S_{(13)}$,
	$S_{(12)}^c = S_{(14)}$ \\
	\hline
	\end{tabular}
	\caption{Summary of second order Weyl covariant hydrostatic scalars which may contribute to the equations of motion up to second order in derivatives. This information can also be found in table \ref{T:Wscalars}. }
\end{center}
\end{table}

\begin{table}[hbt]
\begin{center}
		\begin{tabular}{| c   c    c    | }
		\hline
		\multicolumn{3}{| l |}{Order two hydrostatic scalars in the presence of charge} \\
		\hline
			$\tilde{S}_{(1)} = E_\mu m^\mu$ &
			$\tilde{S}_{(2)} = E_\mu k^\mu$ &
			$\tilde{S}_{(3)} = E_\mu \left( \mathcal{K}_V \right)^\mu$ 
			 \\
			$\tilde{S}_{(4)} = B_{\mu \nu} M^{\mu \nu}  $ &
			$\tilde{S}_{(5)} =B_{\mu\nu}K^{\mu\nu}$ &
			$\tilde{S}_{(6)} =B_{\mu \nu}\kappa_A{}^{\mu\nu}$ \\
		\hline
	\end{tabular}
	\caption{Second order hydrostatic scalars which are associated with a $U(1)$ charge which may contribute to the equations of motion up to second order in derivatives. This information can also be found in table \ref{T:U1scalars}. }
\end{center}
\end{table} 

\begin{table}[hbt]
\begin{center}
		\begin{tabular}{| c   c    c    | }
		\hline
		\multicolumn{3}{| l |}{First order non hydrostatic independent scalars} \\
		\hline
			$\theta$ & & \\
		\hline
		\multicolumn{3}{| l |}{First order non hydrostatic independent vectors} \\
		\hline
			$T \mathring{\nabla}_{\mu} \left(\frac{\mu}{T}\right) - E_{\mu} $ & & \\
		\hline
		\multicolumn{3}{| l |}{First order non hydrostatic independent symmetric tensors} \\
		\hline
			$\sigma^{\mu\nu}$ & & \\
		\hline
	\end{tabular}
	\caption{Independent non hydrostatic data. The entries of this table are collected from table \ref{T:allfirst} and the discussion around table \ref{T:chargecontribution}. }
\end{center}
\end{table}

\section{The constitutive relations}
\label{A:fullconstitutive}

For ease of reference, we have collected the full set of constitutive relations for a fluid with a spin current. We will first present our most general result for the constitutive relations of the stress tensor and spin current in a curved torsionfull background relevant when expanding the equations of motion to second order in derivatives and keeping terms linear in torsion. We will then discuss various modifications of this result when we go to a flat spacetime and set the torsion to zero, take the conformal limit or consider the effect of charge. At appropriate places we refer to the main text where the computations have been carried out.

The constitutive relations on a curved torsionfull background are given by
\begin{align}
\begin{split}
\label{E:constitutivefull}
	T^{\mu\nu} &= T_{id}^{\mu\nu} +T_{h+}^{\mu\nu} + T_{nh}^{\mu\nu} \\
	S^{\lambda\mu\nu} & = S_{id}^{\lambda\mu\nu} + S_{h}^{\lambda\mu\nu} + S_{nh}^{\lambda\mu\nu}\,.
\end{split}
\end{align}
Here, the subscript $id$ refers to the ideal components of the fluid, generated from the pressure term in the generating function. The subscripts $h$ and $h+$ refer to components coming from higher order contributions to the hydrostatic generating function, with the $+$ denoting the BR completion of the hydrostatic terms as discussed at length at the beginning of section \ref{S:nonhydrostatic}. The subscript $nh$ refers to non hydrostatic terms. We further decompose these two terms into BR terms and non BR terms,
\begin{align}
\begin{split}
	T_{h+}^{\mu\nu} &= T_{h+BR}^{\mu\nu} + T_{h\,nBR}^{\mu\nu} \\
	T_{nh}^{\mu\nu} & = T_{nh\,BR}^{\mu\nu} + T_{nh\,nBR}^{\mu\nu} \\
	S_{h}^{\lambda\mu\nu} & = S_{h\,BR}^{\lambda\mu\nu} + S_{h\,nBR}^{\lambda\mu\nu} \\
	S_{nh}^{\lambda\mu\nu} & = S_{nh\,BR}^{\lambda\mu\nu} + S_{nh\,nBR}^{\lambda\mu\nu}\,,
\end{split}
\end{align}
with
\begin{align}
\begin{split}
\label{E:BRterms}
	T_{h+BR}^{\mu\nu} &=  \frac{1}{2} \mathring{\nabla}_{\lambda}   \left(
		S_{h\,BR}^{\lambda\mu\nu} - S_{h\,BR}^{\mu\lambda\nu} - S_{h\,BR}^{\nu\lambda\mu}\right)  \\
	T_{nh\,BR}^{\mu\nu} & =\frac{1}{2} \mathring{\nabla}_{\lambda}   \left(
		S_{nh\,BR}^{\lambda\mu\nu} - S_{nh\,BR}^{\mu\lambda\nu} - S_{nh\,BR}^{\nu\lambda\mu}\right) \,. 
\end{split}
\end{align}

\subsection{Generic, uncharged fluid in a torsionfull, curved background.}
The explicit form of the constitutive relations for the ideal components is given by
\begin{align}
\begin{split}
\label{E:genericid}
	T_{id}^{\mu\nu} =& \left(\epsilon_0+(\rho_m+T \rho_m') m_{\alpha}m^{\alpha} + (\rho_M+T \rho_M') M_{\alpha\beta}M^{\alpha\beta} \right) u^{\mu}u^{\nu} \\
	&+ (P_0+\rho_m m_{\alpha}m^{\alpha} + \rho_M M^{\alpha\beta}M_{\alpha\beta}) \Delta^{\mu\nu} \\
	&+ u^{\mu}m_{\alpha}M^{\alpha\nu}\left(2 \rho_m - 4 \rho_M\right)  +\mathcal{O}(\nabla^4)\\
	S_{id}^{\lambda\mu\nu} =& u^{\lambda} \left(4 \rho_m m^{[\mu}u^{\nu]} - 4 \rho_M M^{\mu\nu} \right)  +\mathcal{O}(\nabla^4) \,,
\end{split}
\end{align}
where
\begin{equation}
	\epsilon_0 = T \frac{\partial P_0}{\partial T} - P_0\,.
\end{equation}
(Taken from \eqref{E:idealexpansion} and \eqref{E:gotepsilon0}.) Here $\mathcal{O}(\nabla^4)$ denotes terms which are fourth order in derivatives but do not contain any explicit derivatives of the hydrodynamic fields.

The explicit form of the hydrostatic components of the stress tensor and current are given by
\begin{subequations}
\label{E:generich}
\begin{align}
\begin{split}
	S_{h\,BR}{}^{\lambda\mu\nu} = & 2 \chi^{(1)}_1 \Delta^{\lambda[\mu}u^{\nu]} \\
		 &- 2 \chi_1^{(2)} M^{\lambda[\mu}u^{\nu]} 
		 + 2  \chi_2^{(2)} u^{\lambda}  M^{\mu\nu} 
		 - 2 \chi_3^{(2)} u^{\lambda}  u^{[\mu}m^{\nu]}  
		 + 4 \chi_4^{(2)} \Delta^{\lambda[\mu}m^{\nu]}  \\
		&+4 \chi_5^{(2)} \kappa \Delta^{\lambda[\mu}u^{\nu]} 
		+2u^\lambda \left( 
			2 \chi^{(2)}_6 k^{[\mu} u^{\nu]} 
			+ 2 \chi^{(2)}_7 K^{\mu \nu} 
			+ \chi^{(2)}_{10} \kappa_A{}^{\mu \nu}   \right) \\ 
		&+ 2u^\lambda   \chi^{(2)}_8 \mathcal{K}_V{}^{[\mu} u^{\nu]} 
		+ 4\Delta^{\lambda [\mu} \left( 
			\chi^{(2)}_8 k^{\nu]} 
			+2 \chi^{(2)}_9 \mathcal{K}_V{}^{\nu]} \right)
		-  2\chi^{(2)}_{10} K^{\lambda [\mu} u^{\nu]}  \\
		&- 4 \chi^{(2)}_{11} \kappa_A{}^{\lambda [\mu}u^{\nu]}
		+ 4 \chi^{(2)}_{12} \kappa_S{}^{\lambda [\mu} u^{\nu]} \
		+ 4 \chi^{(2)}_{13} \mathcal{K}_A{}^{\lambda \mu \nu} 
		+ 4 \chi^{(2)}_{14} \mathcal{K}_T{}^{\mu \nu \lambda} \\
	S_{h\,nBR}{}^{\lambda\mu\nu} = 
		&+2 u^{\lambda} \left(\chi^{(2)}_1 \kappa_A{}^{\mu\nu} + \chi_2^{(2)} K^{\mu\nu} -  \chi^{(2)}_3 u^{[\mu} k^{\nu]}\right)
		- 2 \chi_4^{(2)} u^{\lambda} u^{[\mu}\mathcal{K}_V{}^{\nu]} + \mathcal{O}(\nabla^2) \\
\end{split}
\end{align}
and
\begin{align}
\begin{split}
	T_{h\,nBR}^{\mu\nu}  = &   \left( T \frac{\partial \chi^{(1)}_1}{\partial T}- \chi^{(1)}_1 \right)\kappa u^\mu u^\nu 
		   + \chi^{(1)}_1 \frac{d-2}{d-1} \kappa   \Delta^{\mu \nu} \\ 
		& + 2\chi^{(1)}_1 u^{(\mu} k^{\nu)} + \frac{1}{2} \chi^{(1)}_1 u^\mu \mathcal{K}_V{}^\nu   
		   -\chi^{(1)}_1 \kappa_S{}^{\mu \nu} - \chi^{(1)}_1 \kappa_A{}^{\mu \nu}   \\
		&+  \chi_1^{(2)} \left(\mathcal{K}_T{}^{ \alpha \beta[\mu} + \mathcal{K}_A{}^{\alpha \beta[\mu} \right)u^{\nu]}M_{\alpha \beta} +\frac{\chi^{(2)}_1 \kappa M^{\mu \nu}}{d-1}  \\ 
		&+ \frac{\chi_1^{(2)}} { 2(d-2)} u^{[\mu} M^{\nu] \beta} \mathcal{K}_{V\,\beta} - 2\chi^{(2)}_1 u^{[\mu} \kappa_A{}^{\nu] \beta} m_\beta       
		  + \chi^{(2)}_1 M^{\alpha [\mu} \left(  \kappa_A{}^{\nu]}{}_\alpha -  \kappa_S{}^{\nu]}{}_\alpha \right) \\
		&- 2 \chi^{(2)}_2 u^{[\mu} \left( M^{\nu] \alpha} k_\alpha + K^{\nu]\alpha} m_\alpha \right) 
		  - \chi^{(2)}_3 u^{[\mu} \left( M^{\nu]\alpha} k_\alpha+ K^{\nu]\alpha} m_\alpha  \right)  \\
		& + 2 \chi^{(2)}_4  \mathcal{K}_A{}^{\mu \nu \alpha} m_\alpha  
		  + 2 \chi^{(2)}_4  \mathcal{K}_T{}^{[\mu \nu]\alpha } m_\alpha 
		  + \frac{\chi^{(2)}_4}{(d-2)} \mathcal{K}_V^{[\mu} m^{\nu]} \\
		&  - \chi^{(2)}_4 u^{[\mu} M^{\nu]\alpha} \mathcal{K}_{V\,\alpha}  
	 	+ 2 \chi^{(2)}_4 u^{[\mu} \left(\kappa_A{}^{\nu]\alpha} -  \kappa_S{}^{\nu]\alpha} \right) m_\alpha 
		   + \frac{2\chi^{(2)}_4 (d-2)}{(d-1)} \kappa u^{[\mu} m^{\nu]}  \,.
\end{split}
\end{align}
\end{subequations}
The non hydrostatic components of the stress tensor and current are given by
\begin{align}
\begin{split}
\label{E:genericnh}
	S_{nh\,nBR}^{\lambda\mu\nu} &=  0 \,, \\
	S_{nh\,BR}^{\lambda\mu\nu} & = 2 \sigma_1 \sigma^{\lambda[\mu}u^{\nu]} + 2 \sigma_2 \theta \Delta^{\lambda[\mu}u^{\nu]} \,, \\
	T_{nh\,nBR}^{\mu\nu} &= - \zeta \theta \Delta^{\mu\nu} - \eta \sigma^{\mu\nu} + \sigma_A A^{[\mu} u^{\nu]} + \sigma_m \hat{m}^{[\mu}u^{\nu]} + \sigma_M \hat{M}^{\mu\nu}\,.
\end{split}
\end{align}
(Taken from \eqref{E:nhterms}.)

\subsection{Conformal uncharged fluid in a torsionfull, curved background.}
The explicit form of the constitutive relations for the ideal components of an uncharged conformal fluid is given by
\begin{align}
\begin{split}
	T_{id}^{\mu\nu} =& \left(p_0 (d-1) T^d +\delta_{q\,1} r_m (d-1) T^{d-2} m_{\alpha}m^{\alpha} + r_M (d-1) T^{d-2} M_{\alpha\beta}M^{\alpha\beta} \right) u^{\mu}u^{\nu} \\
	&+ (p_0 T^d +\delta_{q\,1} r_m T^{d-2} m_{\alpha}m^{\alpha} + r_M T^{d-2} M^{\alpha\beta}M_{\alpha\beta}) \Delta^{\mu\nu} \\
	&+ u^{\mu}m_{\alpha}M^{\alpha\nu}T^{d-2} \left(2 \delta_{q\,1} r_m - 4 r_M\right)  +\mathcal{O}(\nabla^4)\\
	S_{id}^{\lambda\mu\nu} =& u^{\lambda} T^{d-2} \left(4 \delta_{q\,1}  r_m m^{[\mu}u^{\nu]} - 4 r_M M^{\mu\nu} \right)  +\mathcal{O}(\nabla^4) \,,
\end{split}
\end{align}
where $p_0$, $r_m$ and $r_M$ are real constants.
(Taken by inserting \eqref{E:conformalsubstiutions} into \eqref{E:idealexpansion}.) 

The explicit form of the hydrostatic components of the stress tensor and current are given by
\begin{align}
\begin{split}
	T^{-(d-2)} S_{h\,BR}{}^{\lambda\mu\nu} = & 2 x^{(1)}_1 T \delta_{q\,0} \Delta^{\lambda[\mu}u^{\nu]} \\
		 &- 2 x_1^{(2)} M^{\lambda[\mu}u^{\nu]} 
		 + 2  x_2^{(2)} u^{\lambda}  M^{\mu\nu} 
		 - 4(1-q)^2 x_3^{(2)} u^{\lambda} (q u^{[\mu}u^{\nu]} + (q-1) k^{[\mu}u^{\nu]})  \\
		&+ x_4^{(2)} \Bigg(
		 	u^{\lambda} \left(4(d-2)q u^{[\mu}m^{\nu]} + 8 (d-2) (q-1) k^{[\mu}u^{\nu]} - 2(q-1) K_V^{[\mu}u^{\nu]}\right) 
			\\
			&+4 (1-q) \Delta^{\lambda[\mu}k^{\nu]} +4 q\Delta^{\lambda[\mu}m^{\nu]}  
		 \Bigg) \\
		&+4 x_5^{(2)}  \kappa \Delta^{\lambda[\mu}u^{\nu]} 
		+4 x_6^{(2)} u^\lambda K^{\mu \nu} \\
		&+ x_7^{(2)} \left(u^{\lambda}
			\left( 16 (d-2)^2 k^{[\mu}u^{\nu]}
				-8 (d-2) K_V^{\mu}u^{\nu]} 
			\right)
			+8 \Delta^{\lambda[\mu} K_V^{\nu]} - 16 (d-2) \Delta^{\lambda[\mu}  k^{\nu]}
		\right)
		\\ 
		&+2 x^{(2)}_8 \left(
			u^{\lambda} \kappa_A^{\mu\nu}
			-K^{\lambda[\mu}u^{\nu]}
		\right) 
		-4 x^{(2)}_9 \kappa_A^{\lambda[\mu}u^{\nu]}
		+4 x^{(2)}_{10} \kappa_S^{\lambda[\mu}u^{\nu]} 
		+4 x^{(2)}_{11} K_A^{\lambda\mu\nu} 
		+4 x^{(2)}_{12} K_T^{\mu\nu\lambda}
		\\
	T^{-(d-2)} S_{h\,nBR}{}^{\lambda\mu\nu} = 
		&2 x^{(2)}_1 u^{\lambda}  \kappa_A{}^{\mu\nu} 
		+ 2 x_2^{(2)} u^{\lambda}  K^{\mu\nu} 
		+4 q (1-q) x^{(2)}_3 u^{\lambda} \left((q-1)u^{[\mu}k^{\nu]} + q m^{[\mu}u^{\nu]}\right)
		\\
		&+2 q x^{(2)}_4 u^{\lambda} \left(2(d-2) u^{[\mu}k^{\nu]} - u^{[\mu}K_V^{\nu]}\right) 
\end{split}
\end{align}
and
\begin{align}
\begin{split}
	T^{-(d-2)} T_{h\,nBR}^{\mu\nu}  = & 
		(d-2) T x^{(1)}_1 \kappa u^\mu u^\nu 
		+ T x^{(1)}_1  \frac{d-2}{d-1} \kappa   \Delta^{\mu \nu} 
		+ 2 T x^{(1)}_1  u^{(\mu} k^{\nu)} 
		  \\
		  &+ \frac{1}{2} T x^{(1)}_1  u^\mu \mathcal{K}_V{}^\nu   
		   - T x^{(1)}_1  \kappa_S{}^{\mu \nu}
		   - T x^{(1)}_1  \kappa_A{}^{\mu \nu}   \\
	&+x^{(2)}_1 \Bigg(
		\frac{\kappa M^{\mu\nu}}{d-1}  
		+ \frac{1}{2(d-2)} u^{[\mu} M^{\nu]}{}_{\beta} K_{V}^{\beta}
		+\left(K_A^{\alpha\beta[\mu}u^{\nu]}+ K_T^{\alpha\beta[\mu}u^{\nu]}\right)M_{\alpha\beta} 
		\\
		&- u^{[\mu}\kappa_A^{\nu]\beta}m_{\beta}+M_{\alpha}{}^{[\mu}\kappa_A^{\nu]}{}_{\alpha} - M_{\alpha}{}^{[\mu}\kappa_S^{\nu]}{}_{\alpha} 
		\Bigg)
	-2 x^{(2)}_2 \left(
	u^{[\mu}K^{\nu]\alpha}m_{\alpha} + u^{[\mu}M^{\nu]\alpha}k_{\alpha}
	\right) \\
	&+(1-q) x^{(2)}_3 \Bigg(
	2(q-1) u^{[\mu} K^{\nu]\alpha}m_{\alpha} 
	- 2 q u^{[\mu} M^{\nu]}{}_{\alpha}m^{\alpha} 
	+2 (q-1) u^{[\mu}M^{\nu]}{}_{\alpha}k^{\alpha} 
	\Big) \\
	&+q x^{(2)}_4 \Bigg(
	2 K_A^{[\mu\nu]\alpha}m_{\alpha} 
	+ 2 K_T^{[\mu\nu]\alpha}m_{\alpha}
	+\frac{K_V^{[\mu}m^{\nu]}}{d-2}
	+2(d-2) u^{[\mu}K^{\nu]\alpha}m_{\alpha}
	\\
	&+\frac{2(d-2)}{d-1} \kappa u^{[\mu}m^{\nu]}
	+2(d-2) u^{[\mu}M^{\nu]\alpha}k_{\alpha}
	-u^{[\mu}M^{\nu]\alpha}K_{V\,\alpha}
	+u^{[\mu}\kappa_A^{\nu]\alpha}m_{\alpha}
	-u^{[\mu}\kappa_S^{\nu]\alpha}m_{\alpha}
	\Bigg)
	\,,
\end{split}
\end{align}
where the $\chi^{(i)}_j$'s are real constants. (Taken by inserting \eqref{E:conformalsubstiutions} into \eqref{E:constitutive12}.)
The non hydrostatic components of the stress tensor and spin current are given by
\begin{align}
\begin{split}
	S_{nh\,nBR}^{\lambda\mu\nu} =&  0 \,, \\
	S_{nh\,BR}^{\lambda\mu\nu} =& 4 x_5^{(2)} q \theta \Delta^{\lambda[\mu } u^{\nu]}  + 2 T^{d-2} s_1 \sigma^{\lambda[\mu}u^{\nu]} \,, \\
	T_{nh\,nBR}^{\mu\nu} =&  -h T^{d-1} \sigma^{\mu\nu} + s_A T^{d-2} A^{[\mu}u^{\nu]} + s_m T^{d-2} \hat{m}^{[\mu}u^{\nu]} + s_M T^{d-2} \hat{M}^{\mu\nu} 
	 \\
	&+ q T^{d-2} s_1 \left(\frac{\theta\sigma^{\mu\nu}}{d-1} - \sigma^{\mu\rho} a_{\rho} u^{\nu} \right)
	  +T^{d-1} q x^{(1)}_1 \theta ((d-1) u^{\mu}u^{\nu} + \Delta^{\mu\nu}) \\
	&+\frac{1}{2} x^{(2)}_5 T^{d-2} q (\mathcal{K}_V^{\nu} u^{\mu}-\mathcal{K}_V^{\mu}u^{\nu} - 4 \kappa_A^{\mu\nu}) \theta
\end{split}
\end{align}
where $s_1$, $s_A$, $s_m$ and $s_M$ are real constants.
(Taken from \eqref{E:s1term}, \eqref{E:calTterm} and by using 
\begin{equation}
	\frac{1}{2} \mathring{\mathcal{D}}_{\lambda}   \left(
		S^{\lambda\mu\nu} - S^{\mu\lambda\nu} - S^{\nu\lambda\mu}\right)
	=
		\frac{1}{2} \mathring{\nabla}_{\lambda}   \left(
		S^{\lambda\mu\nu} - S^{\mu\lambda\nu} - S^{\nu\lambda\mu}\right)
		+
		S^{\mu\nu\rho}\mathcal{A}_{\rho}
		-
		S_{\rho}{}^{\rho\mu}\mathcal{A}^{\nu}
\end{equation}
in implementing the prescription of \eqref{E:TfromcalT} as discussed below \eqref{E:calTterm} .) Note that the last term on the right hand side of $T^{\mu\nu}_{nh\,nBR}$ does not satisfy the Landau frame condition and may be removed by a frame transformation. Such a frame transformation will induce a non standard scaling relation for the hydrodynamic variables which will ensure conformal invariance. 

\subsection{Generic charged fluid in a torsionless background.}
In a torsionless background many of the previous constitutive relations simplify. In addition, there are additional contributions to the ideal and non ideal constitutive relations coming from the background electromagnetic field and the charge chemical potential. The explicit form of the constitutive relations for the ideal components is given by
\begin{align}
\begin{split}
\label{E:chargedideal}
	T_{id}^{\mu\nu} =& \left(\epsilon_0+\left(\rho_m+T \frac{\partial \rho_m}{\partial T}+\mu \frac{\partial \rho_m}{\partial \mu}\right) m_{\alpha}m^{\alpha} + \left(\rho_M+T \frac{\partial \rho_M}{\partial T} + \mu \frac{\partial \rho_M}{\partial \mu} \right)
	M_{\alpha\beta}M^{\alpha\beta} \right) u^{\mu}u^{\nu} \\
	&+ (P_0+\rho_m m_{\alpha}m^{\alpha} + \rho_M M^{\alpha\beta}M_{\alpha\beta}) \Delta^{\mu\nu} \\
	&+ u^{\mu}m_{\alpha}M^{\alpha\nu}\left(2 \rho_m - 4 \rho_M\right)  +\mathcal{O}(\nabla^4)\\
	S_{id}^{\lambda\mu\nu} =& u^{\lambda} \left(4 \rho_m m^{[\mu}u^{\nu]} - 4 \rho_M M^{\mu\nu} \right)  +\mathcal{O}(\nabla^4) \,,
\end{split}
\end{align}
where now
\begin{equation}
	\epsilon_0 = T \frac{\partial P_0}{\partial T} + \mu  \frac{\partial P_0}{\partial \mu}  - P_0\,.
\end{equation}
(Obtained by varying the pressure term in the generating function given in \eqref{E:chargedW}.)

The explicit form of the hydrostatic components of the stress tensor and current are given by
\begin{align}
\begin{split}
\label{E:chargedhydrostatic}
	S_{h\,BR}{}^{\lambda\mu\nu} = & 2 \chi^{(1)}_1 \Delta^{\lambda[\mu}u^{\nu]} \\
		 &- 2 \chi_1^{(2)} M^{\lambda[\mu}u^{\nu]} 
		 + 2  \chi_2^{(2)} u^{\lambda}  M^{\mu\nu} 
		 - 2 \chi_3^{(2)} u^{\lambda}  u^{[\mu}m^{\nu]}  
		 + 4 \chi_4^{(2)} \Delta^{\lambda[\mu}m^{\nu]}  \\
		 &
		 -2 \tilde{\chi}^{(2)}_2 u^{\lambda}u^{[\mu}E^{\nu]} 
		 + 4 \tilde{\chi}^{(2)}_3 \Delta^{\lambda[\mu}E^{\nu]}
		 + 2 \tilde{\chi}^{(2)}_5 u^{\lambda} B^{\mu\nu}
		 -2 \tilde{\chi}^{(2)}_6 B^{\lambda[\mu}u^{\nu]} \\
	S_{h\,nBR}{}^{\lambda\mu\nu} = &2 u^{\lambda} \left(-2 \tilde{\chi}^{(2)}_1 u^{[\mu}E^{\nu]} + \tilde{\chi}^{(2)}_4 B^{\mu\nu} \right) \\
\end{split}
\end{align}
and
\begin{align}
\begin{split}
	T_{h\,nBR}^{\mu\nu}  = &   0 \,.
\end{split}
\end{align}
(Obtained by varying the non pressure terms in the generating function given in \eqref{E:chargedW}.)
The non hydrostatic components of the stress tensor and current are given by
\begin{align}
\begin{split}
\label{E:nhcharged}
	S_{nh\,nBR}^{\lambda\mu\nu} &=  0 \,, \\
	S_{nh\,BR}^{\lambda\mu\nu} & = 2 \sigma_1 \sigma^{\lambda[\mu}u^{\nu]} + 2 \sigma_2 \theta \Delta^{\lambda[\mu}u^{\nu]} + 2 \sigma_3 u^{\lambda}u^{[\mu}\Delta^{\nu]\rho} \left( \mathring{\nabla}_{\rho} \left(\frac{\mu}{T}\right) - E_{\rho} \right)\,, \\
	T_{nh\,nBR}^{\mu\nu} &= - \zeta \theta \Delta^{\mu\nu} - \eta \sigma^{\mu\nu} + \sigma_m\left( \hat{m}^{[\mu}u^{\nu]} +  \tilde{\sigma}_A A^{[\mu} u^{\nu]} - \tilde{\sigma}_{e} u^{[\mu}\Delta^{\nu]\rho} \left(\mathring{\nabla}_{\rho} \left(\frac{\mu}{T}\right)-E_{\rho}\right)\right)   + \sigma_M \hat{M}^{\mu\nu}\,.
\end{split}
\end{align}
(Obtained using the entries in table \ref{T:chargecontribution}.) In writing the last equality we have replaced what we referred to in the uncharged case as $\sigma_A$ with $\sigma_m \tilde{\sigma_A}$ since the resulting equations of motion take a simpler form. Recall that if $\sigma_m=0$ we are in the dynamical spin limit and the constitutive relations change drastically. See \cite{Gallegos:2021bzp} or section \ref{S:discussion} for an extended discussion.

In addition to the energy momentum tensor and spin current we must also specify the constitutive relations for the charge current, $J^{\mu}$. Using \eqref{E:dWdmu} and the last entry in table \ref{T:chargecontribution} we find
\begin{equation}
\label{E:chargeducrrent}
	J^{\mu} = \frac{\partial P_0}{\partial \mu} u^{\mu} + \sigma_E \Delta^{\mu\rho} \left(T \mathring{\nabla}_{\rho} \left(\frac{\mu}{T}\right)-E_{\rho}\right)
\end{equation}
where we are working in a (Landau) frame where $J^{\mu}u_{\mu} = - \frac{\partial P_0}{\partial \mu}$.

\subsection{Conformal charged fluids in a torsionless background}
The constitutive relations for a conformal charged fluid (with $q=0$) in a flat torsionless background are of the form
\begin{align}
\begin{split}
	T_{id}^{\mu\nu} =& \left(p_0 (d-1) T^d  + r_M (d-1) T^{d-2} M_{\alpha\beta}M^{\alpha\beta} \right) u^{\mu}u^{\nu} \\
	&+ (p_0 T^d  + r_M T^{d-2} M^{\alpha\beta}M_{\alpha\beta}) \Delta^{\mu\nu} 
	- 4  u^{\mu}m_{\alpha}M^{\alpha\nu}T^{d-2}  r_M   +\mathcal{O}(\nabla^4)\\
	S_{id}^{\lambda\mu\nu} =& - 4 u^{\lambda} T^{d-2}  r_M M^{\mu\nu}   +\mathcal{O}(\nabla^4) \,,
\end{split}
\end{align}
where now $p_0$, $r_m$, and $r_M$ are functions of $mu/T$. (This relation can be obtained from \eqref{E:chargedideal} together with the constraint coming from the requirement that the $m_{\alpha}m^{\alpha}$ term be conformal invariant and that $P_0 = T^{d} p_0(\mu/T)$
and $\rho_M=T^{d-2} r_m(\mu/T)$.)
The explicit form of the hydrostatic components of the stress tensor and current are given by
\begin{align}
\begin{split}
	T^{-(d-2)} S_{h\,BR}{}^{\lambda\mu\nu} = & 2 x^{(1)}_1 T \Delta^{\lambda[\mu}u^{\nu]} 
		 - 2 x_1^{(2)} M^{\lambda[\mu}u^{\nu]} 
		 + 2  x_2^{(2)} u^{\lambda}  M^{\mu\nu} \\
		&
		 -2 \tilde{x}^{(2)}_2 u^{\lambda}u^{[\mu}E^{\nu]} 
		 + 4 \tilde{x}^{(2)}_3 \Delta^{\lambda[\mu}E^{\nu]}
		 + 2 \tilde{x}^{(2)}_5 u^{\lambda} B^{\mu\nu}
		 -2 \tilde{x}^{(2)}_6 B^{\lambda[\mu}u^{\nu]}
		  \\
	T^{-(d-2)} S_{h\,nBR}{}^{\lambda\mu\nu} = &2 u^{\lambda} \left(-2 \tilde{x}^{(2)}_1 u^{[\mu}E^{\nu]} + \tilde{x}^{(2)}_4 B^{\mu\nu} \right)
\end{split}
\end{align}
and
\begin{align}
\begin{split}
	 T_{h\,nBR}^{\mu\nu}  = & 0 \,.
\end{split}
\end{align}
(Obtained by removing non conformal terms from \eqref{E:chargedhydrostatic} (see \eqref{E:Wconformalcharged}) and setting $\chi^{(1)}_1 =T^{d-1} x_1^{(1)}(\mu/T)$ and $\chi^{(2)}_i =T^{d-1} x_i^{(2)}(\mu/T)$.)
The non hydrostatic components of the stress tensor and spin current are given by
\begin{align}
\begin{split}
	S_{nh\,nBR}^{\lambda\mu\nu} =&  0 \,, \\
	S_{nh\,BR}^{\lambda\mu\nu} =& 2 T^{d-2} s_1 \sigma^{\lambda[\mu}u^{\nu]}
		+ 2 T^{d-2} s_3 u^{\lambda}u^{[\mu}\Delta^{\nu]\rho} \left(T \mathring{\nabla}_{\rho} \left(\frac{\mu}{T}\right) - E_{\rho} \right)\,, \\
	T_{nh\,nBR}^{\mu\nu} =&  -h T^{d-1} \sigma^{\mu\nu}  + s_m T^{d-2} \left( \hat{m}^{[\mu}u^{\nu]} +  \tilde{s}_A A^{[\mu} u^{\nu]} - \tilde{s}_{e} u^{[\mu}\Delta^{\nu]\rho} \left(\mathring{\nabla}_{\rho} \left(\frac{\mu}{T}\right)-E_{\rho}\right)\right)   + s_M T^{d-2} \hat{M}^{\mu\nu}\,.
\end{split}
\end{align}
(Obtained by removing non conformal terms from \eqref{E:nhcharged} ($\sigma_2=0$ and $\zeta=0$ coming from the requirement of Weyl scaling, c.f., \eqref{E:deltaTS}) and scaling $\sigma_1$, $\sigma_3$, $\eta$, $\sigma_m$, $\sigma_M$, $\tilde{\sigma}_A$ and $\tilde{\sigma}_e$ by appropriate powers of the temperature multiplied by a function of the dimensionless ratio $\mu/T$.
Tthe charge current is given by
\begin{equation}
	J^{\mu} = p_0' T^{d-1} u^{\mu} + s_E T^{d-2} \Delta^{\mu\rho} \left(T \mathring{\nabla}_{\rho} \left(\frac{\mu}{T}\right)-E_{\rho}\right)
\end{equation}
(Obtained by scaling \eqref{E:chargeducrrent}.)

\section{Contributions to the constitutive relations which are quadratic in contorsion}
\label{A:quadratic}
As explained in the main text, in order to test conformal invariance (at non zero $q$) of the constitutive relations, beyond the hydrostatic limit, we need to know the full contributions of the scalars $S^{c}{}_{(1)}$ to $S^c_{(12)}$ to the constitutive relations, including the symmetric components of the stress tensor and contributions which are quadratic in the contorsion. In this appendix we have collected the important compoents of these contributions to the stress tensor for completeness. We denote the components of the stress tensor associated with $x^{(2)}_i$, which are second order in the contorsion, are not of the BR type, and also do not transform homogenously under Weyl rescalings by $\bar{T}_{{(i)}}^{\mu\nu}$. We have omitted an overall factor of $x^{(2)}_i T^{d-2}$ from these terms.
\begin{align}
\begin{split}
	\bar{T}_{(3)}^{[\mu\nu]} = & - 2 (1-q)^3 k_{\alpha} K^{\alpha[\mu}u^{\nu]} \\
	\bar{T}_{(4)}^{[\mu\nu]}  =& (1-q) \Bigg( 
		 2\left(K_A^{[\mu\nu]\alpha} +K_T^{[\mu\nu]\alpha} +(d-2) K^{\alpha[\mu}u^{\nu] }\right)k_{\alpha} 
		+\frac{1}{(d-2)} K_V^{[\mu}k^{\nu]} 
		+ \frac{2(d-2)}{d-1} \kappa u^{[\mu}k^{\nu]} \\
		&-k_{\alpha} \left(\kappa_A^{\alpha[\mu}u^{\nu]} + \kappa_S^{\alpha[\mu}u^{\nu]}\right) 
		\Bigg)
		\\
	\bar{T}_{(5)}^{[\mu\nu]} =& -\delta_{q\,0}\kappa  \left(K_{V}^{[\mu}u^{\nu]} + 2 \kappa_A^{\mu\nu}\right) \\
	\bar{T}_{(6)}^{[\mu\nu]} =& -4 u^{[\mu} K^{\nu]}{}_{\alpha} k^{\alpha}  \\
	\bar{T}_{(7)}^{[\mu\nu]} =& -4 K_V^{[\mu}k^{\nu]} + \frac{8(d-2)^2}{d-1} \kappa k^{[\mu}u^{\nu]} - \frac{4(d-2)}{d-1} \kappa K_V^{[\mu}u^{\nu]}  \\
	\bar{T}_{(8)}^{[\mu\nu]} =& \frac{1}{2(d-1)} K^{\mu\nu}\kappa + k_{\alpha}K^{\alpha[\mu}u^{\nu]} - 2 k_{\alpha}\kappa_A^{\alpha[\mu}u^{\nu]} \\
	\bar{T}_{(9)}^{[\mu\nu]} =& \frac{2}{d-1} \kappa \kappa_A^{\mu\nu} \kappa + \frac{1}{(d-2)} u^{[\mu} \kappa_A{}^{\nu]}{}_{\alpha} \mathcal{K}_{V}{}^{\alpha} \\
	\bar{T}_{(10)}^{[\mu\nu]} =&  - \frac{1}{(d-2)} u^{[\mu} \kappa_S{}^{\nu]}{}_{\alpha} \mathcal{K}_{V}{}^{\alpha} \\
	\bar{T}_{(11)}^{[\mu\nu]} =&  -\frac{2}{d-2} \mathcal{K}_A^{ \mu\nu \alpha} \mathcal{K}_{V\,\alpha}  \\
	\bar{T}_{(12)}^{[\mu\nu]} =& -\frac{1}{2(d-2)} \mathcal{K}_T^{[\mu\nu]\alpha}\mathcal{K}_{V\,\alpha}\,.
\end{split}
\end{align}
Note that when $q=0$ the non homogenous transformation of $\tilde{T}_{(5)}$ vanishes as it should. We have, nevertheless, included the explicit form of  $\tilde{T}_{(5)}$ for completeness.

\end{appendix}

\bibliographystyle{JHEP}
\bibliography{Spincurrentlong}

\end{document}